\documentclass[a4paper,11pt,final]{article}
\pdfoutput=1 

\usepackage{subfig}  
\usepackage{jheppub} 

\usepackage[T1]{fontenc}

\usepackage{commands_preprint}

\title{\boldmath Pseudoscalar mesons in a finite cubic volume with twisted boundary conditions}

\author{Gilberto Colangelo}
\author{and Alessio Vaghi}

\affiliation{Albert Einstein Center for Fundamental Physics, Institute for Theoretical Physics,\\ University of Bern, Sidlerstr. 5, 3012 Bern, Switzerland}

\emailAdd{gilberto@itp.unibe.ch}
\emailAdd{vaghial@itp.unibe.ch}

\abstract{%
  We study the effects of a finite cubic volume with twisted
  boundary conditions on pseudoscalar mesons. We first apply chiral
  perturbation theory in the $p$-regime and calculate the corrections for
  masses, decay constants, pseudoscalar coupling constants and form factors
  at next-to-leading order. We show that the Feynman-Hellmann theorem and
  the relevant Ward-Takahashi identity are satisfied. We then derive
  asymptotic formulae \emph{{\`a} la L{\"u}scher} for twisted boundary
  conditions. We show that chiral Ward identities for masses and decay
  constants are satisfied by the asymptotic formulae in finite volume as a
  consequence of infinite-volume Ward identities. Applying asymptotic
  formulae in combination with chiral perturbation theory we estimate
  corrections beyond next-to-leading order for twisted boundary
  conditions.}

\arxivnumber{XXXX.XXXX}

\graphicspath{{./figures/}}

\begin{document} 
\maketitle
\flushbottom

\section{Introduction}

The study of finite volume effects, besides a purely theoretical interest,
is also motivated by the need to correct results from lattice simulations.
These are necessarily performed in a volume of finite extent on which some
form of boundary conditions are imposed. If one chooses periodic boundary
conditions, momenta are discretized and can not take continuous values. To
overcome such limitation twisted boundary conditions were
proposed~\cite{deDiPeTa:04,deDiTa:04,GuMeSi:05,SaVi:05}. In this paper we
study the effects of a finite cubic volume with twisted boundary conditions
on observables related to pseudoscalar mesons by applying chiral
perturbation theory (ChPT). As is well known ChPT is the low-energy
effective field theory of QCD and can be formulated in finite volume
thereby providing a systematic tool to study finite-size effects on
observables calculated in lattice QCD.

The first analytical study of finite-volume effects with twisted boundary
conditions was published soon after the proposal to use these was 
made~\cite{SaVi:05}. This relied on one-loop ChPT. Further analytical
calculations have been made also by other 
groups~\cite{Ti:05,Ti:06,JiTi:07,JiTi:08}, 
and more recent ones have appeared in the last 
few years~\cite{BriDaLuuSav:13,BiRe:14}. 
All these studies rely on ChPT at one loop. 
In the case of periodic boundary conditions it has been
shown that the combined used of asymptotic formulae~\cite{Lu:85,CoHae:04}
and ChPT is the most efficient way to estimate higher orders in 
ChPT~\cite{CoDu:04,CoDuHae:05,CoFuLa:10}, with only tiny deviations from the
results of a full two-loop calculation~\cite{CoHae:06}.

Asymptotic formulae for twisted boundary conditions are not available in
the literature yet, so that all estimates of finite-volume effects have to
rely on one-loop ChPT. The latter estimates are known to suffer from large
two-loop corrections~\cite{CoDu:04,CoDuHae:05} (even though the absolute
size of finite-volume effects remains small). This paper fills this hole:
its main aim is the derivation and application of asymptotic formulae for
finite-volume effects with twisted boundary conditions for meson masses and
decay constants. We also apply and extend a suggestion by
H{\"a}feli~\cite{Hae:08} to use the Feynman-Hellmann 
theorem~\cite{Hell:37,Fey:39} to derive an asymptotic formula for the 
scalar form factor of the pion at zero momentum transfer. 
While for mesons most quantities have already been calculated at one loop in 
ChPT, we provide here for completeness also expressions for finite-volume effects 
for masses, decay constants and form factors. This gives us also the chance to
discuss issues like the very definition of masses and decay constants in
the presence of twisted boundary conditions --- since this is subject to a
certain degree of arbitrariness --- and the role of chiral Ward identities
in finite volume. We also clarify the meaning of the Ward-Takahashi identity
for the electromagnetic current in finite volume and discuss its violation
due to the breaking of Lorentz invariance.

With the help of these asymptotic formulae we make a numerical analysis of
finite-volume effects which goes beyond one-loop ChPT. As in the case of
periodic boundary conditions, also here two-loop effects can be very
sizable, whereas the effect of the twist tends to be, for small twisting
angles, which is the most relevant case, small. We trust that the results
presented here will allow a more reliable correction of finite-volume
effects in all lattice calculations using twisted boundary conditions.

This work is structured as follows.  In section~\ref{sec:ChPTFV} we present
ChPT in finite volume: we give an overview of periodic boundary conditions
and introduce the twisted ones.
Section~\ref{sec:FVC} focuses on finite-volume corrections at
next-to-leading order.  The corrections for masses, decay constants,
pseudoscalar coupling constants are recalculated and new results for pion
form factors are presented.  We show that the corrections of the pion
scalar form factor satisfy the Feynman-Hellman
theorem~\cite{Hell:37,Fey:39}.
In section~\ref{sec:AFs} we derive asymptotic formulae \emph{{\`a} la
  L{\"u}scher} for twisted boundary conditions.  We sketch the steps
necessary to generalize the original derivation of L{\"u}scher~\cite{Lu:85}
and present asymptotic formulae for masses, decay constants and
pseudoscalar coupling constants. These are related via chiral
Ward identities as we will show below. 
We also derive asymptotic formulae for
the pion scalar form factor at zero momentum transfer relying on the
Feynman-Hellmann theorem.
In section~\ref{sec:ApplicationAF} we apply the asymptotic formulae in
combination with ChPT.  We use the chiral representation at one loop to
express the amplitudes entering the formulae and present results beyond
next-to-leading order.
Section~\ref{sec:NumericalResults} contains the numerical analysis.  
Appendices give further details on analytical aspects.
Appendix~\ref{app:Sums} provides a list of useful results for the
evaluation of loop diagrams in finite volume.  Appendix~\ref{app:GITBC} is
devoted to the (electromagnetic) gauge symmetry in finite volume.  Therein,
we construct an effective theory for charged pions which is invariant under
gauge transformations and which reproduces results of
section~\ref{sec:ChPTFV}.  As the gauge symmetry is preserved, we show that
the Ward-Takahashi identity~\cite{Green:53,Taka:57a,Taka:57b} holds in
finite volume if the momentum transfer is discrete.  Finally
appendix~\ref{app:IntegralsS4} collects some long expressions related to
results presented in section~\ref{sec:ApplicationAF}.

\section{Chiral perturbation theory in finite volume}\label{sec:ChPTFV}

\subsection{Chiral perturbation theory}

QCD is the fundamental theory of the strong interaction~\cite{FritGellLe:73,Wei:73}.
It describes the dynamics of the strong interaction in terms of gluons and quarks.
The Lagrangian can be written as 
\begin{equation}\label{Eq:LaQCD}
 \mathcal{L}_{\textup{QCD}}=\bar{q}\,(\i\s{D}-\mathcal{M})\,q-\frac{1}{4}\,G_{\mu\nu,a}\,G_{a}^{\mu\nu},
\end{equation}
where $G_{a}^{\mu\nu}$ is the strength field tensor of gluon fields,
$q=q(x)$ represents the quark fields arranged as a vector of flavor space
and $\mathcal{M}$ is the matrix of the quark masses.
If quark masses are zero, $\mathcal{L}_{\textup{QCD}}$ exhibits a global
chiral symmetry. 
It is well known that chiral symmetry spontaneously breaks down and gives
rise to Goldstone bosons which can be identified with the lightest
pseudoscalar mesons. For $N_f=3$ quark flavors, the fields of Goldstone
bosons can be parametrized by a $3\times3$ unitary matrix, 
\begin{equation}\label{Eq:UMatrix}
    U(x)=\exp\left(\i\,\frac{\Phi(x)}{F_0}\right)\qquad\text{with}\qquad
 \Phi(x)=\begin{pmatrix}\vspace{0.3cm}
           \Pn+\frac{1}{\sqrt{3}}\eta    & \sqrt{2}\Pp                   & \sqrt{2}\Kp \\ \vspace{0.3cm}
           \sqrt{2}\Pm                   & -\Pn+\frac{1}{\sqrt{3}}\eta   & \sqrt{2}\Kn \\
           \sqrt{2}\Km                   & \sqrt{2}\Knb                  & -\frac{2}{\sqrt{3}}\eta
         \end{pmatrix}.
\end{equation}
The parameter $F_0$ is the decay constant in the limit of zero masses.

The effective chiral Lagrangian can be ordered as a series in powers of momenta
and quark masses
\begin{equation}\label{Eq:ChExpLa}
 \mathcal{L}_{\textup{eff}}=\mathcal{L}_2+\mathcal{L}_4+\dots,
\end{equation}
where each quark mass counts as a momentum square, i.e. $\Ord{m_f}\sim\Ord{p^2}$.
At leading order the effective Lagrangian consists of terms $\Ord{p^2}$ and
can be written as 
 \begin{equation}\label{Eq:La2}
  \mathcal{L}_2=\frac{F_{0}^2}{4}\braket{D_{\mu}U D^{\mu}U ^{\dagger}+\chi U^{\dagger}+U\chi^{\dagger}}.
 \end{equation}
The angular brackets $\braket{\,.\,}$ denote the trace in flavor space and 
\begin{equation}\label{Eq:Dder}
 D^{\mu}U=\partial^{\mu}U-\i\left[v^{\mu},U\right]-\i\left\{a^{\mu},U\right\},\qquad\chi=2 B_{0}(s+\i\, p),
\end{equation}
with $B_0$ a parameter of the effective Lagrangian.
We have introduced external fields $v^\mu$, $a^\mu$, $s$, $p$ as sources
for the chiral Noether currents and quark scalar and pseudoscalar
bilinears. They also allow one to include electromagnetism, semileptonic
weak interactions, as well as an explicit breaking of chiral symmetry via
quark masses. In particular, we work in the isospin limit (where
$\hat{m}:=m_u=m_d$) and include masses by setting
$s=\mathcal{M}=\diag(\hat{m},\hat{m},m_s)$.  

At next-to-leading order (NLO) the effective Lagrangian consists of terms
$\Ord{p^4}$ and can be compactly written as 
\begin{equation}\label{Eq:La4}
  \mathcal{L}_4=\sum_{j=1}^{12}L_{j}P_{j}. 
\end{equation}
The coupling constants $L_j$ contain the so-called low-energy constants
(LECs) and the monomials $P_j$ are contructed from $U$, $v^{\mu}$,
$a^{\mu}$, $s$, $p$. Their explicit expressions is well known and can be
found, e.g. in~\cite{GaLe:85a}. 

\subsection{Periodic boundary conditions}\label{subsec:PBC}

Numerical simulations of lattice QCD are by necessity performed in a volume
of finite extent. The volume is usually a spatial cubic box of the side
length $L$ on which boundary conditions are imposed. Mostly employed are
periodic boundary conditions (PBC) which require the periodicity of fields
within the cubic box, 
\begin{equation}\label{Eq:PBCquark}
 q(x+L\hat{e}_j)=q(x),\qquad j=1,2,3.
\end{equation} 
Here, $\hat{e}^{\mu}_j$ are unit Lorentz vectors pointing in the $j$-th spatial direction. 
In momentum space, the periodicity of
the fields corresponds to a discretization of the momenta. The spatial
components of momenta are discrete and read $\vec{p}=2\pi\vec{m}/L$,
$\vec{m}\in\Z{3}$. As a consequence, Lorentz invariance is broken.
Still, the subgroup of spatial rotations of $90^\circ$ (the so-called cubic
invariance) remains intact. 

The momentum discretization also introduces a new scale in the theory:
$1/L$. To apply ChPT one must consider momenta smaller than
$\Lambda_{\chi}=4\pi F_\pi$ --- where $F_\pi$ is the pion decay constant
--- and this provides the following quantitative condition~\cite{Co:04}: 
\begin{equation}\label{Eq:ChPTapp}
 \frac{2\pi}{L}\ll\ 4\pi F_\pi\quad\Longrightarrow\quad L\gg\frac{1}{2F_\pi}\approx\unit{1}{\fm}.
\end{equation}

In \Refs\cite{GaLe:86,GaLe:87:PL,GaLe:87:NP} Gasser and Leutwyler showed
how to apply ChPT in finite volume. They proved that for large enough
volume the effective chiral Lagrangian and the values of LECs remain the
same as in infinite volume. The most relevant change concerns the counting
scheme to be applied to the effective Lagrangian and the propagators. 
The counting scheme has to take into account also $1/L$ together with
momenta and quark masses, moreover propagators are modified by the discrete
momenta. For what concern the counting, there are two possible ones 
corresponding to different regimes: 
\begin{equation}
\begin{aligned}
 M_{\pi} L&\ll 1\qquad\text{$\epsilon$-regime},\\ 
 M_{\pi} L&\gg 1\qquad\text{$p$-regime}. 
\end{aligned}
\end{equation}
Since here we will work in the $p$-regime we do not discuss the
$\epsilon$-regime any further and refer the reader to
\cite{GaLe:87:PL,HaLe:89,Han:90,HanLe:90} for more details about it. 

In the $p$-regime, $M_{\pi}$ is larger than $1/L$: a pion fits well inside
the box and behaves almost as if it were in infinite volume.
Here, the counting scheme can be applied with the additional rule
$1/L\sim\Ord{p}$, see \Ref\cite{GaLe:87:NP}. The expressions for
propagators are similar to the infinite-volume ones, but integrals over
spatial components must be replaced by sums over discrete values.
This makes propagators periodic and dependent on $L$. Physical observables
can be then calculated as in infinite volume: tree graphs produce exactly
the same contributions and just the loop diagrams generate a finite-volume
dependence.

\subsection{Twisted boundary conditions}\label{subsec:TBC}

A serious limitation of PBC is the momentum discretization which makes it
difficult to access very small, finite momenta without using huge volumes.
Twisted boundary conditions (TBC), see
\Refs\cite{deDiPeTa:04,deDiTa:04,GuMeSi:05,SaVi:05}, have been introduced
to overcome this difficulty. They require that fields are periodic up to a
global symmetry transformation, 
\begin{equation}\label{Eq:TBCquark}
  q_{\scriptscriptstyle T}(x+L\hat{e}_j)=\mathcal{U}_jq_{\scriptscriptstyle T}(x),\qquad j=1,2,3.
\end{equation}
Here, the subscript $T$ indicates that fields satisfy TBC.
The transformation $\mathcal{U}_j$ has to be a symmetry of the action and
as such depends on the form of the Lagrangian~\cite{SaVi:05}.  
For QCD with $3$ light flavors one can consider 
\begin{equation}\label{Eq:transfj}
 \mathcal{U}_j=\e^{-\i L\hat{e}_{j}v_\vartheta}\in\mathrm{SU}(3)_V,
\end{equation}
where
$v_\vartheta^\mu=\diag(\vartheta^\mu_u,\vartheta^\mu_d,\vartheta^\mu_s)$ is
a traceless matrix commuting with $\mathcal{M}$. The Lorentz vectors, 
\begin{equation}
 \vartheta_f^\mu=\binom{0}{\vec{\vartheta}_f}, \qquad f=u,d,s,
\end{equation}
are called twisting angles and their spatial components can be arbitrarily
chosen. It is convenient to redefine quark fields as periodic ones by means
of 
\begin{equation}\label{Eq:TBCredefPBC}
 q_{\scriptscriptstyle T}(x)=\mathcal{V}(x)q(x), \qquad\mathcal{V}(x)=\e^{-\i v_{\vartheta}x}.
\end{equation}
The periodicity of $q(x)$ follows from the condition~\eqref{Eq:TBCquark}.
After this field redefinition the twisting angles appear in the Lagrangian
as a constant vector field,  
\begin{equation}
\begin{split}
 \mathcal{L}_{\textup{QCD}}&=\bar{q}_{\scriptscriptstyle T}(x)\left[\i\s{D}-\mathcal{M}\right]q_{\scriptscriptstyle T}(x) -\frac{1}{4}\,G_{\mu\nu,a}\,G_{a}^{\mu\nu}\\
                           &=\bar{q}(x)\big[\i\left(\s{D}-\i\s{v}_{\vartheta}\right)-\mathcal{M}\big]q(x) -\frac{1}{4}\,G_{\mu\nu,a}\,G_{a}^{\mu\nu}.
\end{split}
\end{equation}
The momentum of the flavor $f$ is shifted by the corresponding twisting
angle $\vartheta_f^\mu$, which is a free parameter and can therefore be
varied continuously. Note that one can either impose the
condition~\eqref{Eq:TBCquark} and work with the original form of the
Lagrangian or redefine quark fields as periodic and introduce the twisting
angles through the constant vector field $v_{\vartheta}^\mu$ in
$\mathcal{L}_{\textup{QCD}}$. The two approaches are equivalent.

Since they point in specific directions, twisting angles break various
symmetries.  In particular cubic invariance in momentum space is broken.
More generally, all symmetries whose generators do not commute with
$v_{\vartheta}^\mu$ are broken.  For three different twisting angles these
are: the vector symmetry $\mathrm{SU}(3)_V$ and the isospin symmetry.  Note
that the third isospin component $I_3$, the strangeness $S$ and the
electric charge $Q_e$ are still conserved quantities in this case. In
addition, the symmetry $\mathcal{U}_j$ induces a new one: at each vertex
the sum of incoming and outgoing twisting angles is conserved and equal to
zero, see \Ref\cite{SaVi:05}.

In the effective theory the condition~\eqref{Eq:TBCquark} implies that the
unitary matrix parametrizing the fields of pseudoscalar mesons satisfies  
\begin{equation}\label{Eq:TBCmatrix}
 U_{T}(x+L\hat{e}_j)=\mathcal{U}_jU_{T}(x)\mathcal{U}_j^{\dagger},\qquad j=1,2,3.
\end{equation}
Here, the repetition of $j$ does not imply any sum and $T$ specifies that fields satisfy TBC. 
Through a field redefinition the unitary matrix can be made periodic,
\begin{equation}\label{Eq:ReDefU}
 U(x)=\mathcal{V}^{\dagger}(x)U_{T}(x)\mathcal{V}(x),\qquad\mathcal{V}(x)=\e^{-\i v_{\vartheta}x}.
\end{equation}
The twisting angles enter the effective Lagrangian as a constant vector field $v_{\vartheta}^{\mu}$ where each derivative is replaced by $\partial^{\mu}.\mapsto\partial^{\mu}.-\i\left[v_{\vartheta}^{\mu},\,.\ \right]$.
At leading order the Lagrangian reads 
\begin{equation}\label{Eq:La2TBC}
  \mathcal{L}_2=\frac{F_{0}^2}{4}\braket{\hat{D}_{\mu}U \hat{D}^{\mu}U^{\dagger}+\chi U^{\dagger}+U\chi^{\dagger}},
\end{equation}
with $\hat{D}^{\mu}U:=D^{\mu}U -\i[v_{\vartheta}^{\mu},U]$, and $U(x)$ now
satisfying periodic boundary conditions. The commutator
$[v_{\vartheta}^{\mu},U]$ acts in different ways on the fields of
pseudoscalar mesons. Pseudoscalar mesons sitting in the diagonal of $U$
commute with $v_{\vartheta}^\mu$ and their momenta are
unshifted. Pseudoscalar mesons off the diagonal do not commute with
$v_{\vartheta}^\mu$ and their momenta are shifted by the twisting angles, 
\begin{equation}
 \begin{aligned}
  \tPp^\mu&=\vartheta_{u}^\mu-\vartheta_{d}^\mu,  &\qquad\qquad \tPm^\mu &=-\tPp^\mu,\label{Eq:tPc}\\
  \tKp^\mu&=\vartheta_{u}^\mu-\vartheta_{s}^\mu,  &\qquad\qquad \tKm^\mu&=-\tKp^\mu,\\
  \tKn^\mu&=\vartheta_{d}^\mu-\vartheta_{s}^\mu,  &\qquad\qquad \tKnb^\mu&=-\tKn^\mu.
 \end{aligned}
\end{equation}
Note that twisting angles reflect the flavor content of the particles.
A pseudoscalar meson with the flavor content $q_f\bar{q}_{f'}$ has the
twisting angle 
$\vartheta_{q_f\bar{q}_{f'}}^\mu=(\vartheta_{q_f}-\vartheta_{q_{f'}})^\mu$
whereas its antiparticle has a twisting angle of opposite sign.

As pointed out in \Ref\cite{SaVi:05} twisting angles enter the expressions
of external states and modify internal propagators. 
As an example we consider charged pions (kaons have similar expressions).
The field redefinition~\eqref{Eq:ReDefU} implies that the propagators read 
\begin{equation}\label{Eq:propTBCper}
  \Delta_{\Pc,L}(x)=\frac{1}{L^{3}}\sum_{
		  			\begin{subarray}{c}
		  			\vec{k}=\frac{2\pi}{L}\vec{m}\\
		  			\vec{m}\in\Z{3}
		  			\end{subarray}}
  			\int_{\R{}}\frac{\d{k_{0}}{}}{\left(2\pi\right)}\,\frac{\e^{-\i kx}}{M_\pi^2-(k+\tPc)^2-\i\epsilon}.
 \end{equation}
The propagators are still periodic but obey modified Klein-Gordon equations:\footnote{%
            Without the redefinition~\eqref{Eq:ReDefU} propagators are expressed as in \Eq\eqref{Eq:propTBCper} but twisting angles enter the exponential function of the numerator. 
            In that case, propagators are periodic up to a phase [e.g. $\Delta_{\Pc,L}(x+L\hat{e}_j)=\e^{-\i L\hat{e}_j\tPc}\Delta_{\Pc,L}(x)$ for $j=1,2,3$] and obey the usual Klein-Gordon equations. 
            }  
\begin{equation}
  \big[(\partial-\i\tPc)^2+M_\pi^2\big]\Delta_{\Pc,L}(x)=\delta^{(4)}(x).
\end{equation}
We observe that twisting angles shift the poles in the denominator of
propagators. Moreover, the substitution $k_0\mapsto -k_0$ and
$\vec{k}\mapsto -\vec{k}$ reverses the propagation direction and the sign
of twisting angles. Since antiparticles have twisting angles of opposite
sign, we conclude that the propagation of a positive pion with $\tPp^\mu$
in the forward direction of space-time is equivalent to a propagation of a
negative pion with $\tPm^\mu$ in the backward direction. 

To close this section, we remark that TBC are a generalization of PBC.
Setting all twisting angles to zero, the condition~\eqref{Eq:TBCquark}
reduces to \Eq\eqref{Eq:PBCquark}, which means that by taking this limit in
a TBC calculation and comparing the results with those for PBC, one gets a
non-trivial --- albeit partial --- check.

\section{Finite-volume corrections at NLO}\label{sec:FVC}

\subsection{Masses, decay constants and pseudoscalar coupling constants}\label{subsec:MassesDecayCouplConst}

In general, we define the corrections of an observable $X$ as 
\begin{equation} 
 \delta X=\frac{\Delta X}{X}, 
\end{equation}
where $\Delta X:=X(L)-X$ is the difference among the observable evaluated
in finite and in infinite volume. The corrections of masses and decay
constants of pseudoscalar mesons were first calculated in ChPT with TBC in
\Ref\cite{SaVi:05}. For what concerns the masses, before giving explicit
expressions for the finite-volume corrections, we need to define what we
mean by ``mass'' in the presence of TBC. Having introduced twisting angles
and with them a breaking of Lorentz symmetry, when we calculate the
self-energy of a particle we will get Lorentz non-invariant contributions
proportional to the twisting angles. In general, the self-energy of a meson
$P$ will have the form: 
\begin{equation}\label{Eq:selfE}
\Sigma_P(L)=A_P+B_P(p+\vartheta_P)^2+2(p+\vartheta_P)_\mu \Delta
\tSP^\mu \; .  
\end{equation}
The equation which determines the mass is:
\begin{equation}\label{Eq:massEq}
 M_P^2-(p+\vartheta_P)^2-\Sigma_P(L)=0 \; . 
\end{equation}
Since the solutions of the equation do not lie on a constant $p^2$
surface the mass would seem to depend on the direction of the momentum
$p^\mu$. A possible solution for this non-invariance of the pole position
can be obtained by completing the square and interpreting $\Delta\tSP^\mu$ 
as a renormalization of the twisting angle~\cite{JiTi:07}:
\begin{equation}\label{Eq:massEq2}
 M_P^2-\big[A_P-(1+B_P)\Delta\btSP^2\big]-(1+B_P)\big(p+\vartheta_P+\Delta\btSP\big)^2=0 \; ,
\end{equation}
where $\Delta\btSP^\mu:=\Delta\tSP^\mu/(1+B_P)$. The pole
position is then given by:
\begin{equation}\label{Eq:poleP}
 \big(p+\vartheta_P+\Delta\btSP\big)^2 =\frac{M_P^2-A_P}{1+B_P}+\Delta
  \btSP^2 =:M_P^2(L) \;,
\end{equation}
which is the mass definition we adopt, in agreement with
\Refs\cite{SaVi:05,JiTi:07}. In contrast, the authors of
\Ref\cite{BiRe:14} adopt a mass definition which is 
momentum-dependent and treat the terms proportional to $\Delta
\vartheta^\mu_{\Sigma_P}$ as part of the finite-volume correction to the
mass. 
We stress that our choice of the mass definition is consistent with the
idea to treat the twisting angle as part of the momentum --- which is the
basic point of TBC. In general the mass is defined as the energy of a
particle at zero spatial momentum, which is what is measured on the
lattice. It appears therefore natural to take as definition of the mass the
energy of a particle at zero total momentum (kinetic $+$ twisting
angle). The interpretation of $\Delta \vartheta^\mu_{\Sigma_P}$ is of
course somewhat arbitrary. As we will see below, the fact that at NLO
exactly the same contribution also appears as an additive correction to
$F_P(p+\vartheta_P)_\mu$ in the matrix element of the axial current
strongly suggests its interpretation as a renormalization of the twist.
The definition of mass given in \Eq(\ref{Eq:poleP}) then naturally
follows. 

Indeed a similar situation occurs for the decay constants: the matrix
element of the axial current is not just proportional to momentum, but
there is an extra shift, defined as follows:
\begin{equation}\label{Eq:AxME}
\bra{0} A^\mu_P(0)\ket{P(p+\vartheta_P)}_L = \i F_P(L)\big(p + \vartheta_P+\Delta\btAP\big)^\mu \; .
\end{equation}
Here too we consider $\Delta\btAP^\mu$ as a
twisting-angle renormalization and not as part of the finite-volume correction to
the decay constant (again in agreement with
\Refs\cite{SaVi:05,JiTi:07} and in contrast to \Ref\cite{BiRe:14}). 

In our notation, the results at NLO read
\begin{align}\label{Eq:Masses}
  \delta M_{\Pn}^2      &= \frac{\xi_{\pi}}{2}[2\g_1(\lambda_{\pi},\tPp)-\g_1(\lambda_{\pi})]-\frac{\xi_{\eta}}{6}\g_1(\lambda_{\eta}),\notag\\
  \delta M_{\Pc}^2      &= \frac{\xi_{\pi}}{2}\g_1(\lambda_{\pi})-\frac{\xi_{\eta}}{6}\g_1(\lambda_{\eta}),                            \notag\\ 
  \delta M^2_{\Kc(\Kn)} &= \frac{\xi_{\eta}}{3}\g_1(\lambda_{\eta}),\notag\\
  \delta M_\eta^2       &= -\xi_{\pi}\frac{x_{\pi\eta}}{6}[\g_1(\lambda_{\pi})+2\g_1(\lambda_{\pi},\tPp)]
                           +\frac{\xi_{K}}{2}\Big(1+\frac{x_{\pi\eta}}{3}\Big)[\g_1(\lambda_{K},\tKp)+\g_1(\lambda_{K},\tKn)]\notag\\
                        &\phantom{=}\
                           -\!\frac{2\xi_{\eta}}{3}\Big(1-\frac{x_{\pi\eta}}{4}\Big)\g_1(\lambda_{\eta}),
\end{align}
and 
\begin{equation}\label{Eq:DecayConst}
  \begin{aligned}
  \delta F_{\Pn}&=-\xi_{\pi}\g_1(\lambda_{\pi},\tPp)-\frac{\xi_{K}}{4}\left[\g_1(\lambda_{K},\tKp)+\g_1(\lambda_{K},\tKn)\right],\\ 
  \delta F_{\Pc}&=-\frac{\xi_{\pi}}{2}\left[\g_1(\lambda_{\pi})+\g_1(\lambda_{\pi},\tPp)\right]-\frac{\xi_{K}}{4}\left[\g_1(\lambda_{K},\tKp)+\g_1(\lambda_{K},\tKn)\right],\\
  \delta F_{\Kc}&=-\frac{\xi_{\pi}}{8}\left[\g_1(\lambda_{\pi})+2\g_1(\lambda_{\pi},\tPp)\right]\\
                &\phantom{=}\
                  -\!\frac{\xi_{K}}{4}\left[2\g_1(\lambda_{K},\tKp)+\g_1(\lambda_{K},\tKn)\right]-\frac{3}{8}\xi_{\eta}\g_1(\lambda_{\eta}),\\
  \delta F_{\Kn}&=-\frac{\xi_{\pi}}{8}\left[\g_1(\lambda_{\pi})+2\g_1(\lambda_{\pi},\tPp)\right]\\
                &\phantom{=}\
                  -\!\frac{\xi_{K}}{4}\left[\g_1(\lambda_{K},\tKp)+2\g_1(\lambda_{K},\tKn)\right]-\frac{3}{8}\xi_{\eta}\g_1(\lambda_{\eta}),\\
  \delta F_\eta &=-\frac{3}{4}\xi_{K}\left[\g_1(\lambda_{K},\tKp)+\g_1(\lambda_{K},\tKn)\right].
  \end{aligned}
\end{equation}
Here, $\xi_P\!=\!M_P^2/(4\pi F_\pi)^2$, $x_{PQ}\!=\!M_P^2/M_Q^2$ and $\lambda_P\!=\!M_PL$ for $P,Q\!=\!\pi,K,\eta$.
The function $\g_1(\lambda_{P},\vartheta)$ is defined in \Eq\eqref{Eq:gRfRhRTransferZero} for a generic twisting angle $\vartheta^\mu\!=\!\big(\begin{smallmatrix}0\\ \vec{\vartheta}\end{smallmatrix}\big)$.
Setting $\vartheta^\mu\!=\!0$ this function reduces to $\g_1(\lambda_{P})\!:=\!\g_1(\lambda_{P},0)$ which is the tadpole function for PBC, see e.g.~\cite{GaLe:86}.

The corrections decay exponentially in $\lambda_P$ and depend on twisting angles through phase factors.
Note that twisting angles can change the overall sign. 
This is a consequence of the breaking of the vector symmetry $\mathrm{SU}(3)_V$. 
For instance, mass corrections can turn negative whereas corrections of decay constants can turn posistive depending on $\tPp^\mu$, $\tKp^\mu$, $\tKn^\mu$. 
With an appropriate choice (or averaging over randomly chosen twisting angles) one can even suppress the corrections as discussed e.g. for nucleons in \Ref\cite{BriDaLuuSav:13}.

In addition, the evaluation of the corrections involves the terms which
remormalize the twisting angles. Such terms are not present for PBC and are
generated by the breaking of the cubic invariance, see \Ref\cite{JiTi:07}. 
In our notation, they read
\begin{equation}\label{Eq:DtPL}
\begin{aligned}
    \Delta\tSPc^\mu\phantom{\Big\{}&=\pm\left\{\xi_{\pi}\f_1^{\mu}(\lambda_{\pi},\tPp)+\frac{\xi_{K}}{2}\left[\f_1^{\mu}(\lambda_{K},\tKp)-\f_1^{\mu}(\lambda_{K},\tKn)\right]\right\},\\ 
    \Delta\tSKc^\mu\phantom{\Big\{}&=\pm\left\{\frac{\xi_{K}}{2}\left[2\f_1^{\mu}(\lambda_{K},\tKp)+\f_1^{\mu}(\lambda_{K},\tKn)\right]+\frac{\xi_{\pi}}{2}\f_1^{\mu}(\lambda_{\pi},\tPp)\right\},\\ 
    \left.
    \begin{aligned}
    \Delta\tSKn^\mu\\\Delta\tSKnb^\mu
    \end{aligned}          \right\}&=\pm\left\{\frac{\xi_{K}}{2}\left[\f_1^{\mu}(\lambda_{K},\tKp)+2\f_1^{\mu}(\lambda_{K},\tKn)\right]-\frac{\xi_{\pi}}{2}\f_1^{\mu}(\lambda_{\pi},\tPp)\right\}. 
    \end{aligned} 
\end{equation}
The function $\f_1^{\mu}(\lambda_{P},\vartheta)$ is defined in \Eq\eqref{Eq:gRfRhRTransferZero}.
Note that extra terms emerge only in the evaluation of corrections of charged pions and kaons.
They are non-vanishing in the directions where twisting angles are non-vanishing and disappear for $\tPp^\mu=\tKp^\mu=\tKn^\mu=0$. 
This intimately relates $\Delta\tSPc^\mu$, $\Delta\tSKc^\mu$, $\Delta\tSKn^\mu$ to twisting angles.

Expressions for $\Delta\tAPc^\mu$, $\Delta\tAKc^\mu$, $\Delta\tAKn^\mu$
would also be needed to complete the formulae at NLO, but at this order
they exactly coincide with the extra terms of \Eq\eqref{Eq:DtPL}. 

The corrections of the pseudoscalar coupling constants were first calculated in \Ref\cite{BiRe:14}.
At NLO the results read 
\begin{equation}\label{Eq:CouplConst}
 \begin{aligned}
  \delta G_{\Pn}&=-\frac{\xi_{\pi}}{2}\g_1(\lambda_{\pi})-\frac{\xi_{K}}{4}[\g_1(\lambda_{K},\tKp)+\g_1(\lambda_{K},\tKn)]-\frac{\xi_{\eta}}{6}\g_1(\lambda_{\eta}),\\
  \delta G_{\Pc}&=-\frac{\xi_{\pi}}{2}\g_1(\lambda_{\pi},\tPp)-\frac{\xi_{K}}{4}[\g_1(\lambda_{K},\tKp)+\g_1(\lambda_{K},\tKn)]-\frac{\xi_{\eta}}{6}\g_1(\lambda_{\eta}),\\
  \delta G_{\Kc}&=-\frac{\xi_{\pi}}{8}[\g_1(\lambda_{\pi})+2\g_1(\lambda_{\pi},\tPp)]\\
                &\phantom{=}\ 
                  -\!\frac{\xi_{K}}{4}[2\g_1(\lambda_{K},\tKp)+\g_1(\lambda_{K},\tKn)]-\frac{\xi_{\eta}}{24}\g_1(\lambda_{\eta}),\\
  \delta G_{\Kn}&=-\frac{\xi_{\pi}}{8}[\g_1(\lambda_{\pi})+2\g_1(\lambda_{\pi},\tPp)]\\
                &\phantom{=}\ 
                  -\!\frac{\xi_{K}}{4}[\g_1(\lambda_{K},\tKp)+2\g_1(\lambda_{K},\tKn)]-\frac{\xi_{\eta}}{24}\g_1(\lambda_{\eta}),\\
  \delta G_\eta &=-\frac{\xi_{\pi}}{6}[\g_1(\lambda_{\pi})+2\g_1(\lambda_{\pi},\tPp)]\\
                &\phantom{=}\ 
                  -\!\frac{\xi_{K}}{12}[\g_1(\lambda_{K},\tKp)+\g_1(\lambda_{K},\tKn)]-\frac{\xi_{\eta}}{2}\g_1(\lambda_{\eta}).
 \end{aligned}
\end{equation}
Note that $\delta G_{\Pn}$ (resp. $\delta G_\eta$) correspond to $\Delta^{V}G_{\Pn3}/G_\pi$ (resp. $\Delta^{V}G_{\eta8}/G_\eta$) 
of~\Ref\cite{BiRe:14}.
At this order, no extra terms like \Eq\eqref{Eq:DtPL} appear. 
We show that these results can be obtained with the mass definition~\cite{SaVi:05,JiTi:07} relying on chiral Ward identities.

We just illustrate the case of charged pions.
The relevant chiral Ward identities read
\begin{equation}\label{Eq:ChWIGPsL}
 \hat{m}\Bra{0}P_{1\mp\i2}(0)\Ket{\Pc(p+\tPc)}_L=\left(\partial-\i\tPc\right)_\mu\Bra{0}A^\mu_{1\mp\i2}(0)\Ket{\Pc(p+\tPc)}_L.
\end{equation}
Here, the subscript $L$ indicates that the matrix elements are evaluated in finite volume.
The operators are linear combinations of the pseudoscalar densities resp. axialvector currents: $P_{1\pm\i2}=(P_1\pm\i P_2)/\sqrt{2}$ resp. 
$A^\mu_{1\pm\i2}=(A_1\pm\i A_2)^\mu/\sqrt{2}$.
The matrix elements on the left- (resp. right-) hand side of the chiral Ward identities are proportional to the pseudoscalar coupling (resp. decay) constants.
Working out both sides and retaining terms up to $\Ord{p^6/F_\pi^3}$ we have  
\begin{equation}\label{Eq:GPcFPc}
  \hat{m} G_{\Pc}(L)=(p+\tPc)^2\,F_{\Pc}(L)+2F_\pi(p+\tPc)_\mu\Delta\tAPc^\mu.
\end{equation}
The mass definition~\cite{SaVi:05,JiTi:07} implies that charged pions lie on the following mass shells,
\begin{equation}\label{Eq:NLOMassPcFVCDef}
 \big(p+\tPc+\Delta\btSPc\big)^2=M_{\Pc}^2(L).
\end{equation}
Hence, the momentum squares on the right-hand side of \Eq\eqref{Eq:GPcFPc} value 
\begin{equation}\label{Eq:LimitOS}
 (p+\tPc)^2=M_{\Pc}^2(L)-2\,(p+\tPc)_\mu\Delta\tSPc^\mu+\Ord{p^6/F_\pi^4}, 
\end{equation}
and --- when multiplied with $F_{\Pc}(L)$ --- produce contributions that exactly cancel $2F_\pi(p+\tPc)_\mu\Delta\tAPc^\mu$ 
as at NLO, $\Delta\tAPc^\mu$ coincide with $\Delta\tSPc^\mu$.
Dividing by $\hat{m}G_\pi=M_\pi^2F_\pi$ we get 
\begin{equation}\label{Eq:dGPndGPc}
  \delta G_{\Pc}=\delta M_{\Pc}^2+\delta F_{\Pc}+\Ord{\xi_\pi^2}. 
\end{equation}
Thus, the corrections of pseudoscalar coupling constants are given by the sum of the corrections of masses and decay constants.\footnote{%
    For the eta meson, there is an additional term that must be considered as the relevant chiral Ward identity involves a matrix element with the pseudoscalar density $P_0$.
    } Inserting \Eqs~(\ref{Eq:Masses}, \ref{Eq:DecayConst}) one finds the results~\eqref{Eq:CouplConst}.

\subsection{Pion form factors}\label{subsec:FormFactors}

\begin{figure}[tbp]
\centering
  \hspace{\stretch{2}}
  \subfloat[][Tadpole diagram.]
  {\label{Fig:NLO_FFactors_Tadpole}\includegraphics[width=.2\columnwidth]{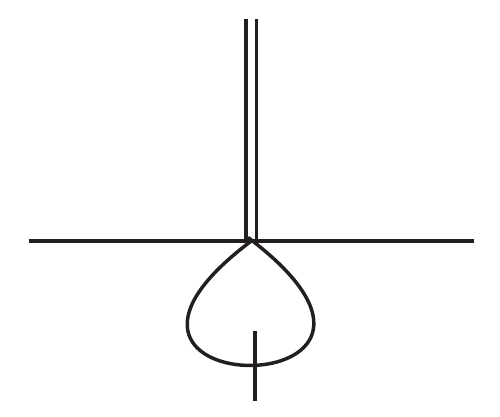}}\hspace{\stretch{2}}\phantom{.}
  \subfloat[][Fish diagram.]
  {\label{Fig:NLO_FFactors_Fish}\includegraphics[width=.2\columnwidth]{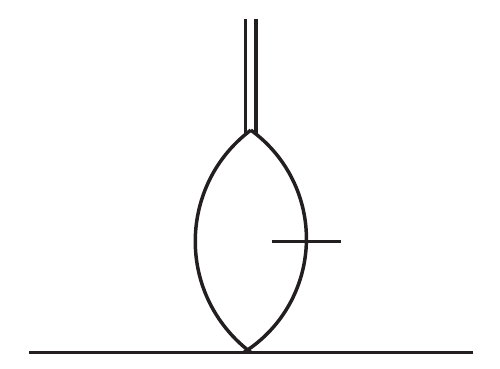}}\hspace{\stretch{2}}\phantom{.}
  \caption[Contributions to matrix elements of pion form factors.]
                                    {Contributions to matrix elements of pion form factors at NLO. 
                                     Single solid lines stand for pions while double solid lines represent the scalar densities (or the vector currents). 
                                     The spline on the loop indicates that the propagator is in finite volume.                                      
                                     }
 \label{Fig:NLO_FFactors}
\end{figure}

In infinite volume the pion form factors are defined by the matrix elements,
\begin{equation}\label{Eq:DefFFS}
\begin{aligned}
  \bra{\pi_b(\pp)}S_{0}\ket{\pi_a(p)}    &=\delta_{ab}\,F_S(q^2),\\ 
  \bra{\pi_b(\pp)}V^\mu_{3}\ket{\pi_a(p)}&=\i\epsilon_{ab3}(\pp+p)^{\mu}\,F_V(q^2).
\end{aligned}
\end{equation}
They depend on the square of the momentum transfer, $q^2=(\pp-p)^2$ and their expressions are known in ChPT at NLO~\cite{GaLe:84} and at NNLO~\cite{BiCoTa:98}.
At vanishing momentum transfer, they satisfy the relations
\begin{equation}\label{Eq:FHTEGI}
       F_S(0)=\frac{\partial M_\pi^2}{\partial\hat{m}}\qquad\text{and}\qquad 
       F_V(0)=1,
  \end{equation}  
which follow from the Feynman-Hellmann theorem \cite{Hell:37,Fey:39} and the Ward identity~\cite{Ward:50}. 

In finite volume the matrix elements of pion form factors receive additional corrections that can be defined as 
\begin{equation}\label{Eq:FVCMEFFS}
 \begin{aligned}
     \delta\GS^{ab}      &=\frac{\bra{\pi_b}S_{0}\ket{\pi_a}_L-\bra{\pi_b}S_{0}\ket{\pi_a}}{\bra{\pi_b}S_{0}\ket{\pi_a}_{q^2=0}},\\
  \i(\Delta\GV{ab})^{\mu}&=\bra{\pi_b}V^\mu_{3}\ket{\pi_a}_L-\bra{\pi_b}V^\mu_{3}\ket{\pi_a}.
 \end{aligned}
\end{equation}
Here, $L$ (resp. $q^2=0$) indicates that matrix elements are evaluated in
finite volume (resp. in infinite volume at vanishing momentum transfer).  
These corrections still depend on the momentum transfer. 
The twisting angles shift the momenta of charged pions but not necessarily
induces a continuous momentum transfer. If the incoming and outgoing pions
are the same, the twisting angles cancel out from the spatial components of
the momentum transfer and 
\begin{equation}\label{Eq:Transfer}
 \vec{q}=\left(\vpp+\vtPc\right)-\left(\vp+\vtPc\right)=\frac{2\pi}{L}\vec{l}\qquad\text{with}\qquad\vec{l}\in\Z{3}.
\end{equation}
We use this fact to work out the matrix elements in finite volume and
evaluate the corrections. However, keep in mind that ---
depending on the kinematics chosen --- the zeroth component $q^0$ may
contain twisting angles of external pions and hence, vary continuously. 

To study the pion form factors in finite volume we consider only $N_f=2$
light flavors. The corrections of masses, decay constants and pseudoscalar
coupling constants can be obtained from \Eqs(\ref{Eq:Masses},
\ref{Eq:DecayConst}, \ref{Eq:CouplConst}) discarding the contributions of
the virtual eta meson and kaons. Note that in this case, the extra terms read
\begin{equation}\label{Eq:DtL}
 \Delta\tSPc^\mu=\pm\xi_{\pi}\f_1^{\mu}(\lambda_{\pi},\tPp).
\end{equation} 

We first discuss the corrections of the matrix elements of the scalar form factor. 
At NLO the corrections can be evaluated from the loop diagrams of \Fig\ref{Fig:NLO_FFactors}.
The tadpole diagram generates corrections similar to those encountered before. 
The fish diagram generates additional corrections which can be calculated with the Feynman parametrization~\eqref{Eq:FeynmanPara}. 
Altogether, we find 
 \begin{align}\label{Eq:dGS}
  \delta\GSPn&=\frac{\xi_{\pi}}{2}\bigg\{2\g_1(\lambda_{\pi},\tPp)-\g_1(\lambda_{\pi})
              +\!\intz\!\big[M_\pi^2\g_2(\lambda_{z},q)+2\big(q^2-M_\pi^2\big)\g_2(\lambda_{z},q,\tPp)\big]\bigg\},\notag\\ 
  \delta\GSPc&=\frac{\xi_{\pi}}{2}\bigg\{\g_1(\lambda_{\pi})
                    +\intz\!\Big[\big(q^2-M_\pi^2\big)\g_2(\lambda_{z},q)+q^2\g_2(\lambda_{z},q,\tPp)\Big]\bigg\}+P_\mu\Delta\TPc^{\mu}, 
 \end{align}
with $\lambda_{z}=\lambda_{\pi}\sqrt{1+z(z-1)q^2/M_\pi^2}$.
The functions $\g_2(\lambda_{z},q,\vartheta)$, $\g_2(\lambda_{z},q):=\g_2(\lambda_{z},q,0)$ originate from the fish diagram and can be evaluated by means of the Poisson resummation formula~\eqref{Eq:Poisson}. 
Note that $\g_2(\lambda_{z},q,\vartheta)$ is even in the second and third argument. 
This is a consequence of the fact that the spatial components of the momentum transfer are discrete. 
The last term in \Eq\eqref{Eq:dGS} consists of the product among 
\begin{equation}\label{Eq:PDTPc} 
 \begin{aligned}
          P^\mu&=\left(\pp+\tPc\right)^\mu+\left(p+\tPc\right)^\mu,\\
 \Delta\TPc^\mu&=\pm\xi_{\pi}\intz\left[\f_2^{\mu}(\lambda_{z},q,\tPp)+ q^{\mu}\left(1/2-z\right)\g_2(\lambda_{z},q,\tPp)\right]. 
 \end{aligned}
\end{equation}
The Lorentz vectors $\Delta\TPc^\mu$ have non-vanishing components in the directions where both $\tPc^\mu$ and $q^\mu$ are non-vanishing.
They disappear for $\tPc^\mu=0$.
The function $\f_2^{\mu}(\lambda_{z},q,\vartheta)$ is defined in \Eq\eqref{Eq:FishSums} and originates from the fish diagram. 
Note that as $\vec{q}$ is discrete, the function $\f_2^{\mu}(\lambda_{z},q,\vartheta)$ is even in the second argument and odd in the third one. 

The corrections of the matrix elements of the scalar form factor decay exponentially in $\lambda_{\pi}=M_{\pi} L$ and disappear for $L\to\infty$.
As a check we set $\tPc^\mu=0$ and find the result for PBC~\cite{Hae:08,JLQCD:09}.
In that case, the corrections are negative.
For small twisting angles, the corrections stay also negative as the dependence on twisting angles is roughly a phase factor.
They may turn positive for large twisting angles.
Note that $P_\mu\Delta\TPc^{\mu}$ depend linearly on $\tPc^\mu$.
This dependence increases $\delta\GSPc$ at large twisting angles. 
Thus, in order to keep the corrections under control, it is important to employ small twisting angles, e.g. $\abs{\vtPc}<\pi/L$. 

At vanishing momentum transfer, the corrections reduce to
\begin{equation}\label{Eq:dGSPnPc0}
 \begin{aligned}
  \delta\GSPn\big|_{q^2=0}&=\frac{\xi_{\pi}}{2}\left\{2\g_1(\lambda_{\pi},\tPp)-\g_1(\lambda_{\pi})+M_\pi^2\left[\g_2(\lambda_{\pi})-2\g_2(\lambda_{\pi},\tPp)\right]\right\},\\
  \delta\GSPc\big|_{q^2=0}&=\frac{\xi_{\pi}}{2}\left[\g_1(\lambda_{\pi})-M_\pi^2\g_2(\lambda_{\pi})\right]\pm2\xi_{\pi}\left(p+\tPc\right)_\mu\f_2^\mu(\lambda_{\pi},\tPp).
 \end{aligned}
\end{equation}
The functions $\g_2(\lambda_{\pi},\tPp):=\g_2(\lambda_{\pi},0,\tPp)$, $\g_2(\lambda_{\pi}):=\g_2(\lambda_{\pi},0,0)$ and $\f_2^\mu(\lambda_{\pi},\tPp):=\f_2^\mu(\lambda_{\pi},0,\tPp)$ are defined in \Eq\eqref{Eq:gRfRhRTransferZero}. 
At vanishing momentum transfer the Feynman-Hellman
theorem~\cite{Hell:37,Fey:39} relates the scalar form factor with the
derivative of the pion mass, see \Eq\eqref{Eq:FHTEGI}. 
This relation can be extended to finite volume.
However, one must make some specification.
As pointed out in~\Ref\cite{GaLe:84} the Feynman-Hellmann theorem states
that the expectation value $\bra{\pi_b}S_0\ket{\pi_a}_{q^2=0}$ is related
to the derivative of the energy level describing the pion eigenstate.  
In finite volume the energy levels are additionally shifted by twisting
angles and by the corrections of self-energies: 
\begin{equation}\label{Eq:EPc2}
 E_{\Pc}^2(L)=M_\pi^2+\left(\vp+\vtPc\right)^2-\Delta\Sigma_{\Pc}.
\end{equation}
Here, $\Delta\Sigma_{\Pc}:=\Sigma_{\Pc}(L)-\Sigma_{\Pc}$ just contain corrections in finite volume.
Taking the derivative $\partial_{\hat{m}}=(\partial M_\pi^2/\partial\hat{m})\,\partial_{M_\pi^2}$, 
and retaining only contributions in finite volume, we obtain
\begin{equation}\label{Eq:FHTPc}
 \delta\GSPc\big|_{q^2=0}=\frac{\partial}{M_\pi^2}\left[-\Delta\Sigma_{\Pc}\right]. 
\end{equation}
This relation extends the statement of the Feynman-Hellmann theorem in finite volume: 
at $q^2=0$ the corrections of the matrix elements of the scalar form factor
are related with the derivative of the self-energies with the respect to
the mass. One can show that deriving the expressions of the self-energies
at NLO one obtains the expressions~\eqref{Eq:dGSPnPc0}. 

The corrections of the matrix elements of the vector form factor can be
similarly evaluated from the loop diagrams of~\Fig\ref{Fig:NLO_FFactors}. 
In this case, some attention must be paid as the evaluation involves
tensors in finite volume.
At NLO we find 
\begin{equation}\label{Eq:DGVPc}
 \begin{aligned}
  \DGVPc{\mu}&=Q_e\xi_\pi\intz\Big\{P^\mu\big[\g_1(\lambda_{z},q,\tPp)-\g_1(\lambda_{\pi},\tPp)\big]+2 P_\nu\h_2^{\mu\nu}(\lambda_{z},q,\tPp)\Big\}\\
             &\quad-Q_e\left\{\xi_\pi\intz q^\mu\left(1-2z\right)\left[P_\nu\f_2^{\nu}(\lambda_{z},q,\tPp)\right]-2\Delta\tGammaPc^\mu+q^2\Delta\TPc^\mu\right\}. 
 \end{aligned}
\end{equation}
Here, $Q_e=\pm1$ represents the electric charge of $\Pc$ in elementary units and the functions $\g_1(\lambda_{z},q,\vartheta)$, $\h_2^{\mu\nu}(\lambda_{z},q,\vartheta)$ are defined in \Eq\eqref{Eq:FishSums}.
At this order, $\Delta\tGammaPc^\mu$ exactly coincide with the extra terms of \Eq\eqref{Eq:DtL}. 

The corrections of the matrix elements of the vector form factor decay exponentially in $\lambda_{\pi}=M_{\pi} L$ and disappear for $L\to\infty$. 
As a check we set $\tPc^\mu=0$ and find the result for PBC obtained by H{\"a}feli~\cite{Hae:08}.
This result differs from the expression published in \Ref\cite{BuJiTi:06} by a term proportional to $q^\mu$.
We stress that such term contributes for PBC and only disappears when the momentum transfer is zero, as well.
In this sense, we disagree with \Ref\cite{JiTi:07} where it is claimed that the contribution~$\boldsymbol{G}_{FV}^{\textup{rot}}$ disappears for vanishing twisting angles.
Actually, such contribution disappears when both the momentum transfer and the twisting angles are zero. 
In \Ref\cite{BiRe:14} the corrections were calculated in $\mathrm{SU}(3)$ ChPT with TBC. 
Their results coincide with~\Eq\eqref{Eq:DGVPc} if contributions of the virtual eta meson and kaons are discarded.
In general, the corrections roughly depend on twisting angles through a phase factor. 
For small twisting angles the corrections stay negative (resp. positive) for positive (resp. negative) pions. 
Note that in~\eqref{Eq:DGVPc} there are terms linear in $\tPc^\mu$. 
This increases the absolute value of $\DGVPc{\mu}$ at large twisting angles. 

At vanishing momentum transfer the corrections reduce to 
\begin{equation}\label{Eq:DGVPc0}
\DGVPc{\mu}_{q^2=0}=Q_e\left\{4\,\xi_\pi (p+\tPc)_\nu\h_2^{\mu\nu}(\lambda_{\pi},\tPp)+2\,\Delta\tGammaPc^\mu\right\},
\end{equation}
and disappear only for $L\to\infty$. 
Here, $\h_2^{\mu\nu}(\lambda_{\pi},\tPp):=\h_2^{\mu\nu}(\lambda_{\pi},0,\tPp)$, see \Eq\eqref{Eq:gRfRhRTransferZero}. 
Setting $\tPc^\mu=0$ the corrections~\eqref{Eq:DGVPc0} reduce to the result obtained for PBC~\cite{HuJiTi:07}.
In general, the fact that $\DGVPc{\mu}_{q^2=0}$ differ from zero indicates that the (electromagnetic) gauge symmetry or Lorentz invariance are broken.
In infinite volume, the vector form factor equals unity at vanishing momentum transfer, see~\Eq\eqref{Eq:FHTEGI}.
This result follows from the Ward identity~\cite{Ward:50} which relies on both the gauge symmetry as well as Lorentz invariance and can be derived from the Ward-Takahashi identity~\cite{Green:53,Taka:57a,Taka:57b}.
The derivation implies a continuous limit on the momentum transfer which in finite volume can not be taken due to the discretization of spatial components.
This invalidates the Ward identity in finite volume.
It turns out that the corrections~\eqref{Eq:DGVPc0} are consequences of the breaking of Lorentz invariance.
In~\Ref\cite{HuJiTi:07} these considerations were presented for PBC.
The authors demonstrated that at vanishing momentum transfer the corrections respect the gauge symmetry and that the Ward-Takahashi identity holds for PBC.
In appendix~\ref{app:GITBC} we generalize their derivations for TBC. 
In particular, we construct an effective theory invariant under gauge transformations which reproduces \Eq\eqref{Eq:DGVPc0} 
and we show that the Ward-Takahashi identity holds for TBC as long as the spatial components of the transfer momentum are discrete.

\section{Asymptotic formulae for TBC}\label{sec:AFs}

Asymptotic formulae represent another method to estimate finite-volume
corrections. They relate the corrections of a given physical quantity to
an integral of a specific amplitude, evaluated in infinite volume. The
method was introduced by L{\"u}scher~\cite{Lu:83} and it has been widely
applied in combination with ChPT as it allows one to get a chiral order
almost for free at the price of neglecting exponentially suppressed
contributions. Currently, there are asymptotic formulae for
pseudoscalar
mesons~\cite{CoDuSo:02,CoDu:04,CoHae:04,CoDuHae:05,CoWenWu:10},
nucleons~\cite{Ali:03,Koma:04,Koma:05a,Koma:05b,CoFuHae:05,CoFuLa:10} and
heavy mesons~\cite{CoFuLa:10}.  These formulae are valid in the $p$-regime
and for PBC.  Here, we generalize the method to TBC and derive asymptotic formulae for small twisting
angles estimating the corrections for masses, decay constants, pseudoscalar coupling constants
and scalar form factors of pseudoscalar mesons.

\subsection{Masses, decay and pseudoscalar coupling constants} 

\begin{figure}[tbp]
   \centering
   \includegraphics[width=.30\columnwidth]{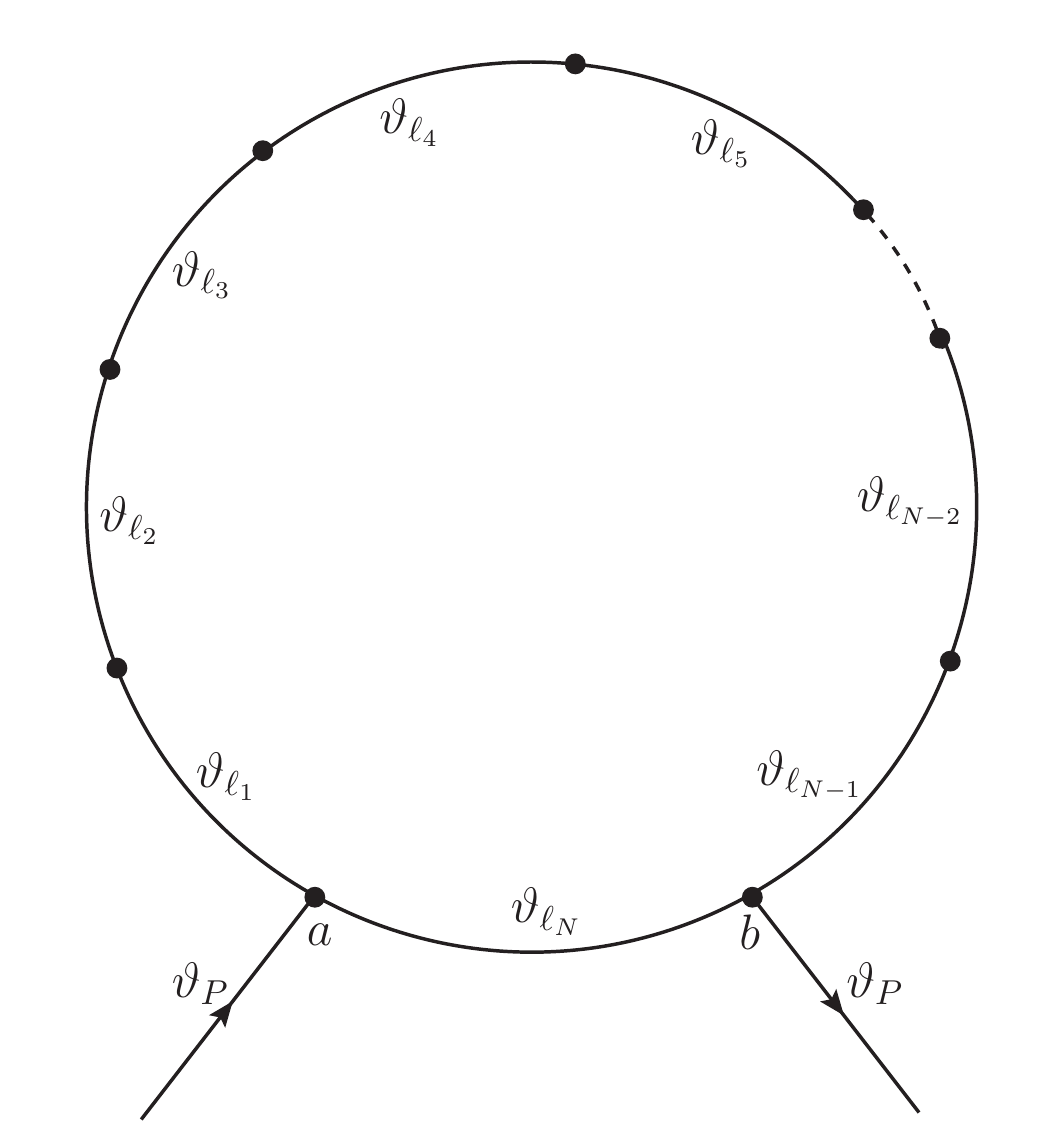}
   \caption[Auxiliary graph.]{%
              Auxiliary graph for the sum of internal twisting angles.
              The endpoints of each line correspond to vertices of the non-empty finite set $\Graph{V}$ of the abstract graph $\Graph{G}$. 
              Along the internal lines $\ell_1,\dots,\ell_N$ flow twisting angles $\vartheta_{\ell_1}^\mu,\dots,\vartheta_{\ell_N}^\mu$. 
              Along external lines flows the twisting angle $\vartheta_{P}^\mu$ of the particle $P$ under consideration.
            }
  \label{Fig:Twist_Conservation}
\end{figure}

\subsubsection{Generalization of L\"uscher's derivation}\label{subsubsec:GeneralizationDerivation} 

The derivation of the asymptotic formulae for TBC can be led back to the
original derivation of L{\"u}scher~\cite{Lu:85}. In the following we
outline the necessary steps to generalize the L{\"u}scher's derivation to
TBC and refer the reader to his paper for details. 

\begin{figure}[tbp]
   \centering
   \includegraphics[width=.2\columnwidth]{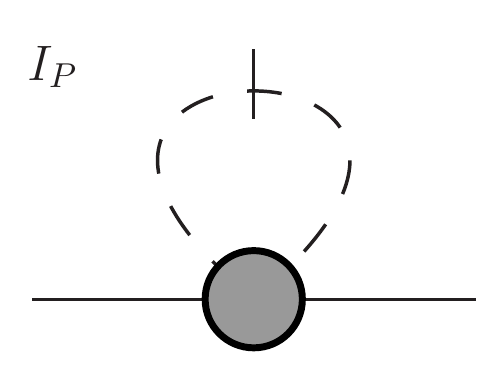}
   \caption[Skeleton diagram contributing to asymptotic formulae for masses.]{%
            Skeleton diagram contributing to asymptotic formulae for masses. 
            Solid lines stand for a generic pseudoscalar meson $P$ and dashed lines for virtual pions.
            The spline indicates that the pion propagator is in finite volume and accounts for integer vectors $\vec{n}\in\Z{3}$ with $\abs{\vec{n}}=1$.
            The blob corresponds to the vertex functions defined by the one-particle irreducible part of the amputated four-point function, see \Ref\cite{Lu:85}.
            }
  \label{Fig:I3}
\end{figure}

In the first part of the proof, L{\"u}scher showed by means of abstract
graph theory that, for a generic loop diagram contributing to the self-energy, the dominant corrections are obtained if one takes all propagators in
infinite volume but the ones of the lightest particles (in this case,
pions) are taken in finite volume.\footnote{%
This concerns only propagators which are contained in at least one loop, cfr. \Ref\cite{Lu:85}. 
}
For TBC the propagators depend also on twisting angles.
In that case, the twisting angles flowing along internal lines of a generic
loop diagram add up to match the total external twisting angles
entering the diagram. The situation can be illustrated by the graph
in \Fig\ref{Fig:Twist_Conservation}. 
This is a consequence of the conservation law pointed out in
\Ref\cite{SaVi:05}: at each vertex the sum of twisting angles is conserved
and equal zero. In position space the contribution
$\Graph{J}\!(\Graph{D},L)$ of the diagram $\Graph{D}$ to the self-energy
takes then the general form, 
\begin{equation}
 \begin{aligned}
  \Graph{J}\!(\Graph{D},L)  &=\!\sum_{[n]}\Graph{J}\!(\Graph{D},n,L),\\
  \Graph{J}\!(\Graph{D},n,L)&=\!\!\prod_{v\in\Graph{V}'}\int_{\R{4}}\!\!\!\d{x(v)}{4}\mathbb{V}
                            \bigg\{\e^{-\i(p+\vartheta_{P})[x(b)-x(a)]}\!\prod_{\ell\in\Graph{L}}\!G_{\pi}\big(\Delta x_\ell+Ln(\ell)\big)\e^{-\i Ln(\ell)\vartheta_{\ell}}\bigg\}.
 \end{aligned}
\end{equation}
Here, $x^\mu(v)$ is the space-time coordinates of the vertex
$v\in\Graph{V}':=\Graph{V}\setminus\set{b}$ and $\Delta x^\mu_\ell$ is the
difference among the final and initial vertex of the line $\ell$. 
The quantity $\mathbb{V}$ is a product of differential operators and
generates the vertex functions of the diagram $\Graph{D}$. 
The pion propagators $G_{\pi}\big(\Delta x_\ell+Ln(\ell)\big)$ are in
infinite volume and can be expressed, e.g., with the heat kernel
representation, see \Ref\cite{Lu:85}. 
For every line $\ell$ we have assigned an integer Lorentz vector $n^\mu(\ell)=\big(\begin{smallmatrix}0\\ \vec{n}(\ell)\end{smallmatrix}\big)$.
Note that the summation over all possible sets of integers $[n]$ 
for all internal propagators is a well-defined operation also in the case of TBC. 
The term $[n]=[0]$ corresponds to the contribution in infinite volume with external momenta
shifted by $\vartheta_{P}^\mu$. Discarding this term one finds that 
$\abs{\Graph{J}\!(\Graph{D},L)}$ is exponentially bound so that
the self-energy decays as $\Ord{\e^{-\sqrt{3}\,\lambda_{\pi}/2}}$ at
asymptotically large $L$. The dominant corrections of the self-energy
is then given by the contribution where for only one propagator $\abs{\vec{n}(\ell^{*})}=1$ and all others $\abs{\vec{n}(\ell\neq\ell^*)}=0$.
This is represented by the skeleton diagram of \Fig\ref{Fig:I3}.  

The second part of the derivation consists in showing that by modifying the
integration countour in the complex plane, the dominant corrections can be
written as an integral of the forward $P\pi$-scattering amplitude evaluated
in Minkowski space and analytically continued to complex values of its
arguments. For TBC the pion propagators as well as the vertex functions
depend on twisting angles. Through an integration shift one can express the
dependence on twisting angles of virtual particles as a phase factor
multiplying the vertex functions. The dependence on external twisting
angles may be worked out expanding the vertex functions around small
twisting angles. This is hardly a limitation since the main goal of the
introduction of TBC is precisely to be able to access small momenta, which
requires the use of small twisting angles.  
The first term of the expansion contributes to the dominant corrections of
masses whereas a part of the second term provides the dominant contribution
to the extra terms of the self-energy. The results are asymptotic formulae
which, in the case of the neutral pion and the eta meson are valid for
arbitrary twisting angles, and in the case 
of charged pions and kaons are valid for small external twisting angles
only. Note that similar argumentations can be extended to the derivation of
asymptotic formulae for decay constants and pseudoscalar coupling constants. 

\subsubsection{Analytical results}\label{subsubsec:AnalyticalResults} 

From the generalization of the L{\"u}scher's derivation we obtain the following asymptotic formulae for the masses of pseudoscalar mesons,
\begin{subequations}\label{Eq:AFMPs} 
\begin{align}
  \delta M_{\Pn}     &=\frac{-1}{2(4\pi)^{2}\lambda_{\pi}}\!\sum_{%
				  \begin{subarray}{c}
				  \vec{n}\in\Z{3}\\
				  \abs{\vec{n}}\neq0
				  \end{subarray}
				  }%
				  \int_{\R{}}\frac{\d{y}{}}{\abs{\vec{n}}}\,\e^{-\lambda_{\pi}\abs{\vec{n}}\sqrt{1+y^2}}\mathcal{F}_{\Pn}(\i y,\tPp)
				  +\Ord{\e^{-\bar{\lambda}}},\notag\\ 
				  \\
  \delta M_{\Pc}     &=\frac{-1}{2(4\pi)^{2}\lambda_{\pi}}\!\sum_{%
				  \begin{subarray}{c}
				  \vec{n}\in\Z{3}\\
				  \abs{\vec{n}}\neq0
				  \end{subarray}
				  }%
				  \int_{\R{}}\frac{\d{y}{}}{\abs{\vec{n}}}\,\e^{-\lambda_{\pi}\abs{\vec{n}}\sqrt{1+y^2}}\bigg(\!1+y\,\frac{D_{\Pc}}{M_{\pi}}\,\frac{\partial}{\partial y}\bigg)\mathcal{F}_{\Pc}(\i y,\tPp)+\Ord{\e^{-\bar{\lambda}}},\notag
\end{align}

\begin{align}                                   
  \delta M_{\Kc(\Kn)}&=\frac{-1}{2(4\pi)^{2}\lambda_{K}}\frac{M_{\pi}}{M_K}\!\sum_{%
				  \begin{subarray}{c}
				  \vec{n}\in\Z{3}\\
				  \abs{\vec{n}}\neq0
				  \end{subarray}
				  }%
				  \int_{\R{}}\frac{\d{y}{}}{\abs{\vec{n}}}\,\e^{-\lambda_{\pi}\abs{\vec{n}}\sqrt{1+y^2}}\notag\\
				  &\qquad\qquad\qquad\qquad\qquad\quad\ \,\times\bigg(\!1+y\,\frac{D_{\Kc(\Kn)}}{M_K}\,\frac{\partial}{\partial y}\bigg)\,\mathcal{F}_{\Kc(\Kn)}(\i y,\tPp)
				  +\Ord{\e^{-\bar{\lambda}}},\notag\\ 
				  \notag\\
  \delta
  M_\eta             &=\frac{-1}{2(4\pi)^{2}\lambda_{\eta}}\,\frac{M_{\pi}}{M_\eta}\!\sum_{%
				  \begin{subarray}{c}
				  \vec{n}\in\Z{3}\\
				  \abs{\vec{n}}\neq0
				  \end{subarray}
				  }%
				  \int_{\R{}}\frac{\d{y}{}}{\abs{\vec{n}}}\,\e^{-\lambda_{\pi}\abs{\vec{n}}\sqrt{1+y^2}}\mathcal{F}_{\eta}(\i y,\tPp)
				  +\Ord{\e^{-\bar{\lambda}}}. 
\end{align}
\end{subequations}
Here, we display the resummed version of the formulae for which
$\bar{\lambda}=\Mb L$ with $\Mb=(\sqrt{3}+1)\,M_{\pi}/\sqrt{2}$, see
\Refs\cite{Co:04,CoDuHae:05}. Each symbol on the right-hand side refers to
a quantity in infinite volume. The amplitudes are all defined in similar way. 
For instance, 
\begin{equation}\label{Eq:AmpFKc}
 \begin{aligned}
 \mathcal{F}_{\Kc}(\tilde{\nu},\tPp)&=T_{\Kc\Pn}(0,-4M_K\nu)\\
                                    &+[T_{\Kc\Pp}(0,-4M_K\nu)+T_{\Kc\Pm}(0,-4M_K\nu)]\e^{\i L\vec{n}\vtPp},
 \end{aligned}
\end{equation}
where $\nu=(s-u)/(4M_K)$, $\tilde{\nu}=\nu/M_{\pi}=\i y$ and $s$, $t$, $u$
are Mandelstam variables in Minkowski space. The functions
$T_{\Kc\Pn}(t,u-s)$, $T_{\Kc\Pp}(t,u-s)$, $T_{\Kc\Pm}(t,u-s)$ represent the
isospin components of the $K\pi$-scattering in the $t$-channel with zero
isospin~\cite{CoDuHae:05}:  
\begin{equation}\label{Eq:TKpiI0}
  T_{K\pi}^{I=0}(t,u-s)=T_{\Kc\Pn}(t,u-s)+T_{\Kc\Pp}(t,u-s)+T_{\Kc\Pm}(t,u-s).
\end{equation}
In general, the asymptotic formulae depend on twisting angles through the
phase factor $\exp(\i L\vec{n}\vtPp)$ in the amplitudes. 
The formulae for $\delta M_{\Pc}$, $\delta M_{\Kc}$, $\delta M_{\Kn}$
additionally depend on external twisting angles through  
 \begin{equation}\label{Eq:DPcDKcDKn} 
  D_{\Pc}=\sqrt{M_\pi^2+\abs{\vtPc}^2}-M_{\pi}\qquad\text{and}\qquad
  D_{\Kc(\Kn)}=\sqrt{M_K^2+\abs{\vec{\vartheta}_{\Kc(\Kn)}}^2}-M_K.
 \end{equation}
We stress that the latter formulae are only valid for small external
twisting angles. At tree level the chiral representation of
$\mathcal{F}_{\Pc}(\tilde{\nu},\tPp)=-M_\pi^2/F_\pi^2$,
$\mathcal{F}_{\Kc(\Kn)}(\tilde{\nu},\tPp)=0$
does not depend on twisting angles and if inserted in the asymptotic
formulae provides the results~\eqref{Eq:Masses} obtained with ChPT at
NLO. The formulae for $\delta M_{\Pn}$, 
$\delta M_{\eta}$ are valid for arbitrary twisting angles. Inserting the
chiral representation at tree level of
$\mathcal{F}_{\Pn}(\tilde{\nu},\tPp)=M_\pi^2\big(1-2\,\e^{\i
  L\vec{n}\vtPp}\big)/F_\pi^2$ and
$\mathcal{F}_{\eta}(\tilde{\nu},\tPp)=M_\pi^2\big(1+2\e^{\i
  L\vec{n}\vtPp}\big)/\big(3F_\pi^2\big)$, one recovers the results
obtained with ChPT at NLO. While the asymptotic formulae are in principle
valid up to terms $\Ord{\e^{-\bar{\lambda}}}$, at this chiral order they
give the full result. Note that setting all twisting angles to zero, the
asymptotic formulae reduce to the formulae valid for PBC~\cite{CoDuHae:05}. 

Along with the formulae for mass corrections, we also derive asymptotic formulae for the extra terms breaking the Lorentz invariance of the self-energies. 
These formulae read 
\begin{equation}\label{Eq:AFDtPL}
 \begin{aligned}
  \Delta\vtSPc&\!=\frac{-M_{\pi}}{2(4\pi)^{2}}\!
                                  \sum_{%
				  \begin{subarray}{c}
				  \vec{n}\in\Z{3}\\
				  \abs{\vec{n}}\neq0
				  \end{subarray}
				  }%
				  \frac{\vec{n}}{\abs{\vec{n}}}\!\int_{\R{}}\!\!\!\d{y}{}\e^{-\lambda_{\pi}\abs{\vec{n}}\sqrt{1+y^2}}y\,\mathcal{G}_{\Pc}(\i y,\tPp)+\Ord{\e^{-\bar{\lambda}}},\\ 
  \Delta\vtSKcKn\!&=\frac{-1}{2(4\pi)^{2}}\frac{M^2_\pi}{M_K}\!\sum_{%
				  \begin{subarray}{c}
				  \vec{n}\in\Z{3}\\
				  \abs{\vec{n}}\neq0
				  \end{subarray}
				  }%
				  \frac{\vec{n}}{\abs{\vec{n}}}\!\int_{\R{}}\!\!\!\d{y}{}\e^{-\lambda_{\pi}\abs{\vec{n}}\sqrt{1+y^2}}y\,\mathcal{G}_{\Kc(\Kn)}(\i y,\tPp)+\Ord{\e^{-\bar{\lambda}}}. 
 \end{aligned}
\end{equation}
In that case, the amplitudes are given by differences of isospin components. 
For instance, 
\begin{equation}\label{Eq:AmpGKc}
  \mathcal{G}_{\Kc}(\tilde{\nu},\tPp)=\left[T_{\Kc\Pp}(0,-4M_K\nu)-T_{\Kc\Pm}(0,-4M_K\nu)\right]\e^{\i L\vec{n}\vtPp},
\end{equation}
and similarly for other pseudoscalar mesons.
These asymptotic formulae are valid for small external twisting angles and
provide the results~\eqref{Eq:DtPL} obtained with ChPT at NLO 
if one inserts the tree-level chiral representation of the amplitudes,
namely $\mathcal{G}_{\Pc}(\tilde{\nu},\tPp)=\mp(4M^2_\pi\,\tilde{\nu}\,\e^{\i L\vec{n}\vtPp})/F_\pi^2$ and $\mathcal{G}_{\Kc(\Kn)}(\tilde{\nu},\tPp)=\mp\ppp{}{}(2 M_K M_\pi\,\tilde{\nu}\,\e^{\i L\vec{n}\vtPp})/F_\pi^2$.
Note that for $\tPp^\mu=0$ the sums in~\Eq\eqref{Eq:AFDtPL} are odd in
$\vec{n}$ and hence, vanish --- as they should for PBC. 

The asymptotic formulae for decay constants are similar to those for masses,
 \begin{align}\label{Eq:AFFPs}
  \delta
  F_{\Pn}&=\frac{1}{(4\pi)^2\lambda_{\pi}}\frac{M_{\pi}}{F_\pi}\!\sum_{%
				  \begin{subarray}{c}
				  \vec{n}\in\Z{3}\\
				  \abs{\vec{n}}\neq0
				  \end{subarray}
				  }%
                                  \int_{\R{}}\frac{\d{y}{}}{\abs{\vec{n}}}\,\e^{-\lambda_{\pi}\abs{\vec{n}}\sqrt{1+y^2}}\mathcal{N}_{\Pn}(\i y,\tPp)
                                  +\Ord{\e^{-\bar{\lambda}}},\notag\\ 
                                  \notag\\
   \delta
   F_{\Pc}&=\frac{1}{(4\pi)^2\lambda_{\pi}}\frac{M_{\pi}}{F_\pi}\!\sum_{%
				  \begin{subarray}{c}
				  \vec{n}\in\Z{3}\\
				  \abs{\vec{n}}\neq0
				  \end{subarray}
				  }%
				  \int_{\R{}}\frac{\d{y}{}}{\abs{\vec{n}}}\,\e^{-\lambda_{\pi}\abs{\vec{n}}\sqrt{1+y^2}}\bigg(\!1+y\frac{D_{\Pc}}{M_{\pi}}\frac{\partial}{\partial y}\bigg)\mathcal{N}_{\Pc}(\i y,\tPp)+\Ord{\e^{-\bar{\lambda}}},\notag\\
				  \notag\\
   \delta
   F_{\Kc(\Kn)}&=\frac{1}{(4\pi)^{2}\lambda_{K}}\frac{M_{\pi}}{F_K}\!\sum_{%
				  \begin{subarray}{c}
				  \vec{n}\in\Z{3}\\
				  \abs{\vec{n}}\neq0
				  \end{subarray}
				  }%
				  \int_{\R{}}\frac{\d{y}{}}{\abs{\vec{n}}}\,\e^{-\lambda_{\pi}\abs{\vec{n}}\sqrt{1+y^2}}\\
				  &\qquad\qquad\qquad\qquad\qquad\qquad\ \times\bigg(\!1+y\frac{D_{\Kc(\Kn)}}{M_K}\frac{\partial}{\partial y}\bigg)\mathcal{N}_{\Kc(\Kn)}(\i y,\tPp)
				  +\Ord{\e^{-\bar{\lambda}}},\notag\\ 
				  \notag\\
  \delta
  F_\eta&=\frac{1}{(4\pi)^{2}\lambda_{\eta}}\frac{M_{\pi}}{F_\eta}\!\sum_{%
				  \begin{subarray}{c}
				  \vec{n}\in\Z{3}\\
				  \abs{\vec{n}}\neq0
				  \end{subarray}
				  }%
				  \int_{\R{}}\frac{\d{y}{}}{\abs{\vec{n}}}\,\e^{-\lambda_{\pi}\abs{\vec{n}}\sqrt{1+y^2}}\mathcal{N}_{\eta}(\i y,\tPp)
				  +\Ord{\e^{-\bar{\lambda}}},\notag
 \end{align}
where $D_{\Pc}$, $D_{\Kc(\Kn)}$ are given in \Eq\eqref{Eq:DPcDKcDKn}.
The amplitudes are defined from the matrix elements of the axialvector
decay after a pole subtraction described in
\Refs\cite{CoHae:04,CoDuHae:05}. For instance, the amplitudes of charged
kaons read
\begin{equation}\label{Eq:AmpNKc}
 \begin{aligned}
     \mathcal{N}_{\Kc}(\tilde{\nu},\tPp)&=-\i\Big\{\bar{A}_{\Kc\Pn}(0,-4M_K\nu)\Big.\\
                                        &\phantom{-\i\Big\{\Big\}}\ \, 
                                                   \Big.+\left[\bar{A}_{\Kc\Pp}(0,-4M_K\nu)+\bar{A}_{\Kc\Pm}(0,-4M_K\nu)\right]\e^{\i L\vec{n}\vtPp}\Big\},  
 \end{aligned}
\end{equation}
where $\nu=(u-t)/(4M_K)$ and $\tilde{\nu}=\nu/M_{\pi}=\i y$. 
Here, $\bar{A}_{\Kc\Pn}(s,t-u)$, $\bar{A}_{\Kc\Pp}(s,t-u)$,
$\bar{A}_{\Kc\Pm}(s,t-u)$ are the isospin components of the matrix elements
describing the kaon decay into a two-pion state with zero
isospin~\cite{CoDuHae:05}: 
\begin{equation}
 \bar{A}_K^{I=0}(s,t-u)=\bar{A}_{\Kc\Pn}(s,t-u)+\bar{A}_{\Kc\Pp}(s,t-u)+\bar{A}_{\Kc\Pm}(s,t-u).
\end{equation} 
Inserting the chiral representation at tree level of the amplitudes, the
asymptotic formulae provide the results~\eqref{Eq:DecayConst} obtained with ChPT at
NLO and if all twisting angles are set to zero,  
they reduce to the formulae valid for PBC~\cite{CoDuHae:05}.

We also derive asymptotic formulae for the extra terms arising in the matrix elements of the axial vector current. 
These read 
\begin{equation}\label{Eq:AFFDtPLPsc}
 \begin{aligned}
  \Delta\vtAPc\!&=\frac{-1}{2(4\pi)^{2}}\frac{M_\pi^2}{F_\pi}\!\sum_{%
				  \begin{subarray}{c}
				  \vec{n}\in\Z{3}\\
				  \abs{\vec{n}}\neq0
				  \end{subarray}
				  }%
				  \frac{\vec{n}}{\abs{\vec{n}}}\!\int_{\R{}}\!\!\!\d{y}{}\e^{-\lambda_{\pi}\abs{\vec{n}}\sqrt{1+y^2}}\,y\,\mathcal{H}_{\Pc}(\i y,\tPp)+\Ord{\e^{-\bar{\lambda}}}, \\ 
  \Delta\vtAKcKn\!&=\frac{-1}{2(4\pi)^{2}}\frac{M_\pi^2}{F_K}\!\sum_{%
				  \begin{subarray}{c}
				  \vec{n}\in\Z{3}\\
				  \abs{\vec{n}}\neq0
				  \end{subarray}
				  }%
				  \frac{\vec{n}}{\abs{\vec{n}}}\!\int_{\R{}}\!\!\!\d{y}{}\e^{-\lambda_{\pi}\abs{\vec{n}}\sqrt{1+y^2}}\,y\,\mathcal{H}_{\Kc(\Kn)}(\i y,\tPp)+\Ord{\e^{-\bar{\lambda}}}. 
 \end{aligned}
\end{equation}
The amplitudes are given by differences of isospin components, e.g. 
\begin{equation}\label{Eq:AmpHKc}
 \mathcal{H}_{\Kc}(\tilde{\nu},\tPp)=-\i\left[\bar{A}_{\Kc\Pp}(0,-4M_K\nu)-\bar{A}_{\Kc\Pm}(0,-4M_K\nu)\right]\e^{\i L\vec{n}\vtPp}.
\end{equation}
For $\tPp^\mu=0$ the sums in~\Eq\eqref{Eq:AFFDtPLPsc} are odd in $\vec{n}$
and hence vanish, as expected for PBC.

The asymptotic formulae for the pseudoscalar coupling constants are 
\begin{align}\label{Eq:AFGPs}
    \delta G_{\Pn}&=\frac{1}{(4\pi)^2\lambda_{\pi}}\,\frac{M_\pi^2}{G_\pi}\!\sum_{%
				  \begin{subarray}{c}
				  \vec{n}\in\Z{3}\\
				  \abs{\vec{n}}\neq0
				  \end{subarray}
				  }%
                                  \int_{\R{}}\frac{\d{y}{}}{\abs{\vec{n}}}\,\e^{-\lambda_{\pi}\abs{\vec{n}}\sqrt{1+y^2}}\mathcal{C}_{\Pn}(\i y,\tPp)
                                  +\Ord{\e^{-\bar{\lambda}}},\notag\\ 
                                  \notag\\
   \delta G_{\Pc}&=\frac{1}{(4\pi)^2\lambda_{\pi}}\,\frac{M_\pi^2}{G_\pi}\!\sum_{%
				  \begin{subarray}{c}
				  \vec{n}\in\Z{3}\\
				  \abs{\vec{n}}\neq0
				  \end{subarray}
				  }%
				  \int_{\R{}}\frac{\d{y}{}}{\abs{\vec{n}}}\,\e^{-\lambda_{\pi}\abs{\vec{n}}\sqrt{1+y^2}}\bigg(\!1+y\frac{D_{\Pc}}{M_{\pi}}\frac{\partial}{\partial y}\bigg)\mathcal{C}_{\Pc}(\i y,\tPp)+\Ord{\e^{-\bar{\lambda}}},\notag\\
				  \notag\\
   \delta G_{\Kc(\Kn)}&=\frac{1}{(4\pi)^{2}\lambda_{K}}\frac{M_{\pi}M_K}{G_K}\!\sum_{%
				  \begin{subarray}{c}
				  \vec{n}\in\Z{3}\\
				  \abs{\vec{n}}\neq0
				  \end{subarray}
				  }%
				  \int_{\R{}}\frac{\d{y}{}}{\abs{\vec{n}}}\,\e^{-\lambda_{\pi}\abs{\vec{n}}\sqrt{1+y^2}}\\
				  &\qquad\qquad\qquad\qquad\qquad\qquad\ \times\bigg(\!1+y\frac{D_{\Kc(\Kn)}}{M_K}\frac{\partial}{\partial y}\bigg)\mathcal{C}_{\Kc(\Kn)}(\i y,\tPp)
				  +\Ord{\e^{-\bar{\lambda}}},\notag\\ 
				  \notag\\
  \delta G_\eta&=\frac{1}{(4\pi)^{2}\lambda_{\eta}}\frac{M_{\pi}M_\eta}{G_\eta}\!\sum_{%
				  \begin{subarray}{c}
				  \vec{n}\in\Z{3}\\
				  \abs{\vec{n}}\neq0
				  \end{subarray}
				  }%
				  \int_{\R{}}\frac{\d{y}{}}{\abs{\vec{n}}}\,\e^{-\lambda_{\pi}\abs{\vec{n}}\sqrt{1+y^2}}\mathcal{C}_{\eta}(\i y,\tPp)
				  +\Ord{\e^{-\bar{\lambda}}}. \notag
 \end{align}
The amplitudes are all defined in similar way for the various pseudoscalar mesons.
For charged kaons, they are 
\begin{equation}\label{Eq:AmpCKc}
 \begin{aligned}
   \mathcal{C}_{\Kc}(\tilde{\nu},\tPp)&=\bar{C}_{\Kc\Pn}(0,-4M_K\nu)\\
                                      &+[\bar{C}_{\Kc\Pp}(0,-4M_K\nu)+\bar{C}_{\Kc\Pm}(0,-4M_K\nu)]\,\e^{\i L\vec{n}\vtPp},
 \end{aligned}
\end{equation}
where the isospin components $\bar{C}_{\Kc\Pn}(s,t-u)$, $\bar{C}_{\Kc\Pp}(s,t-u)$, can be determined from the matrix elements, 
\begin{equation}
 \begin{aligned}
  C_{\Kc\Pn}&=\Bra{\Pn(p_1)\Pn(p_2)}P_{4\mp\i5}(0)\Ket{\Kc(p_3)},\\
  C_{\Kc\Pp}&=\Bra{\Pp(p_1)\Pm(p_2)}P_{4\mp\i5}(0)\Ket{\Kc(p_3)},
 \end{aligned}
\end{equation}
after the pole subtraction, 
\begin{equation}\label{Eq:barCKcPion}
 \begin{aligned}
  \bar{C}_{\Kc\Pn}(s,t-u)&=C_{\Kc\Pn}-G_K\frac{T_{\Kc\Pn}(s,t-u)}{M_K^2-\tilde{Q}^2},\\
  \bar{C}_{\Kc\Pp}(s,t-u)&=C_{\Kc\Pp}-G_K\frac{T_{\Kc\Pp}(s,t-u)}{M_K^2-\tilde{Q}^2}.
 \end{aligned}
\end{equation}
Here, $\tilde{Q}^\mu=(p_3-p_1-p_2)^\mu$ and $T_{\Kc\Pn}(s,t-u)$, $T_{\Kc\Pp}(s,t-u)$ are the isospin components of the $K\pi$-scattering in the $s$-channel, see \Eq\eqref{Eq:TKpiI0}. 
Note that the isospin component $\bar{C}_{\Kc\Pm}(s,t-u)$ can be determined in a similar way from $C_{\Kc\Pp}$ by exchanging $p_1^\mu\leftrightarrow p_2^\mu$.

Furthermore, we derive asymptotic formulae for extra terms of the matrix elements
\begin{equation}
 \begin{aligned}
  \mathscr{G}_{\Pc}&=\Bra{0}P_{1\mp\i2}(0)\Ket{\Pc(p+\tPc)}_L,\\
  \mathscr{G}_{\Kc}&=\Bra{0}P_{4\mp\i5}(0)\Ket{\Kc(p+\tKc)}_L,\\
  \mathscr{G}_{\Kn}&=\Bra{0}P_{6-\i7}(0)\Ket{\Kn(p+\tKn)}_L.
 \end{aligned}
\end{equation}
These formulae are valid for small external twisting angles and read
 \begin{align}\label{Eq:AFGDtPLPsc}
  \Delta\vtGPc\!&=\frac{-1}{2(4\pi)^{2}}\frac{M_{\pi}}{G_{\pi}}\!\sum_{%
				  \begin{subarray}{c}
				  \vec{n}\in\Z{3}\\
				  \abs{\vec{n}}\neq0
				  \end{subarray}
				  }%
				  \frac{\vec{n}}{\abs{\vec{n}}}\!\int_{\R{}}\!\!\!\d{y}{}\e^{-\lambda_{\pi}\abs{\vec{n}}\sqrt{1+y^2}}y\,\mathcal{K}_{\Pc}(\i y,\tPp)+\Ord{\e^{-\bar{\lambda}}}, 
				  \notag\\[-0.5cm]\\
  \Delta\vtGKcKn\!&=\frac{-1}{2(4\pi)^{2}}\frac{M_\pi^2}{G_KM_K}\!\sum_{%
				  \begin{subarray}{c}
				  \vec{n}\in\Z{3}\\
				  \abs{\vec{n}}\neq0
				  \end{subarray}
				  }%
				  \frac{\vec{n}}{\abs{\vec{n}}}\!\int_{\R{}}\!\!\!\d{y}{}\e^{-\lambda_{\pi}\abs{\vec{n}}\sqrt{1+y^2}}y\,\mathcal{K}_{\Kc(\Kn)}(\i y,\tPp)+\Ord{\e^{-\bar{\lambda}}}.\notag 
 \end{align}
The amplitudes are given by the difference of the isospin components, e.g. 
\begin{equation}\label{Eq:AmpKKc}
 \mathcal{K}_{\Kc}(\tilde{\nu},\tPp)=\left[\bar{C}_{\Kc\Pp}(0,-4M_K\nu)-\bar{C}_{\Kc\Pm}(0,-4M_K\nu)\right]\e^{\i L\vec{n}\vtPp}.
\end{equation}
Setting $\tPp^\mu=0$ the sums in~\Eq\eqref{Eq:AFGDtPLPsc} vanish, as expected for PBC.

\subsubsection{Chiral Ward identities}\label{subsubsec:ChWI}

In \Ref\cite{CoHae:04} it was pointed out that the asymptotic formulae for
masses, decay constants and pseudoscalar coupling constants are related by
means of chiral Ward identities. 
We are now going to show that the relation can be generalized to TBC.

For convenience, we only illustrate the case of charged pions.
We start from the relevant chiral Ward identities which in momentum space read
\begin{equation}\label{Eq:ChWIAFGPc}
 -\i\left(\tilde{Q}+\tPc\right)_\mu\big(A_{\Pc\pi_c}(\tPc)\big)^\mu=\hat{m}\,C_{\Pc\pi_c}(\tPc),
\end{equation}
where  
\begin{equation}\label{Eq:APcpicCPcpi}
\begin{aligned}
 \big(A_{\Pc\pi_c}(\tPc)\big)^\mu&:=\Bra{\pi_c(p_1)\pi_c(p_2)}A^\mu_{1\mp\i2}(0)\Ket{\Pc(p_3+\tPc)},\\
               C_{\Pc\pi_c}(\tPc)&:=\Bra{\pi_c(p_1)\pi_c(p_2)}P_{1\mp\i2}(0)\Ket{\Pc(p_3+\tPc)}.
\end{aligned}
\end{equation}
The matrix elements~\eqref{Eq:APcpicCPcpi} are in infinite volume though
momenta are shifted by a twisting angle. Note that
$\Bra{\pi_c(p_1)\pi_c(p_2)}$ is a two-pion state with zero isospin. 
We can leave out the twisting angles of that state as they will appear as
phase factors after a shift of the loop momentum, see
\Eq\eqref{Eq:Omegac}. This is why the twisting angle just appears in the initial
states. 

According to \Ref\cite{CoFiUr:96} the matrix elements
$\big(A_{\Pc\pi_c}(\tPc)\big)^\mu$ have a pole that does not enter the
amplitudes in the asymptotic formulae. We subtract that pole expanding the
matrix elements around $(\tilde{Q}+\tPc)^2=M_\pi^2$, 
\begin{equation}\label{Eq:APcpic}
 \big(A_{\Pc\pi_c}(\tPc)\big)^\mu=\big(\bar{A}_{\Pc\pi_c}(\tPc)\big)^\mu+\i F_\pi(\tilde{Q}+\tPc)^\mu\frac{T_{\Pc\pi_c}\big(s,t(\tPc)-u(\tPc)\big)}{M_\pi^2-(\tilde{Q}+\tPc)^2}.
\end{equation}
Here, $T_{\Pc\pi_c}\big(s,t(\tPc)-u(\tPc)\big)$ correspond to the isospin
components of the $\pi\pi$-scattering in the $s$-channel with zero isospin
and 
\begin{equation}\label{Eq:Mandelstamtw}
  s=(\tilde{Q}-p_3)^2,\qquad t(\tPc)=(\tilde{Q}+\tPc+p_2)^2,\qquad u(\tPc)=(\tilde{Q}+\tPc+p_1)^2,
\end{equation}
are Mandelstam variables shifted by the twisting angles of $p_3$ and $\tilde{Q}$.
Note that the first variable does not depend on $\tPc^\mu$ as twisting
angles exactly cancel out. The bar of
$\big(\bar{A}_{\Pc\pi_c}(\tPc)\big)^\mu$ indicates that the pole has been
subtracted from the matrix elements. 

Similarly, the pole in the matrix elements $C_{\Pc\pi_c}(\tPc)$ should also
be removed, see~\cite{CoHae:04}. We do this by expanding the matrix
elements around $(\tilde{Q}+\tPc)^2=M_\pi^2$, 
\begin{equation}\label{Eq:CPcpic}
 C_{\Pc\pi_c}(\tPc)=\bar{C}_{\Pc\pi_c}(\tPc)+G_\pi\frac{T_{\Pc\pi_c}\big(s,t(\tPc)-u(\tPc)\big)}{M_\pi^2-(\tilde{Q}+\tPc)^2}.
\end{equation}

We insert (\ref{Eq:APcpic}, \ref{Eq:CPcpic}) in the
identities~\eqref{Eq:ChWIAFGPc} and divide by $M_{\pi}$.  
Using the relation $\hat{m}G_\pi=M_\pi^2F_\pi$, we find
\begin{equation}
 \frac{-\i}{M_{\pi}}\big(\tilde{Q}+\tPc\big)_\mu\big(\bar{A}_{\Pc\pi_c}(\tPc)\big)^\mu=\frac{\hat{m}}{M_{\pi}}\,\bar{C}_{\Pc\pi_c}(\tPc)+\frac{F_\pi}{M_{\pi}}\,T_{\Pc\pi_c}\big(s_3,s_1(\tPc)-s_2(\tPc)\big).
\end{equation}
The last term can be brought on the left-hand side: we can set
$p_1^\mu=-p_2^\mu=k^\mu$ and rewrite the momenta
$(p_3+\tPc)^\mu=\big(\hat{p}+\htPc\big)^\mu$ with
$\hat{p}^\mu=\Big(\begin{smallmatrix}M_{\pi}\\\vec{0}\end{smallmatrix}\Big)$
and
$\htPc^\mu=\Big(\begin{smallmatrix}D_{\Pc}\\\vtPc\end{smallmatrix}\Big)$. 
We obtain 
\begin{equation}
 \frac{-\i}{M_{\pi}}\Big(\hat{p}+\htPc\Big)_\mu\Big(\bar{A}_{\Pc\pi_c}\big(\htPc\big)\Big)^\mu-\frac{F_\pi}{M_{\pi}}\,T_{\Pc\pi_c}\big(0,-4M_{\pi}\nu_{\pm}\big)=\frac{\hat{m}}{M_{\pi}}\,\bar{C}_{\Pc\pi_c}\big(\htPc\big),
\end{equation}
where $\nu_{\pm}=\big[u\big(\htPc\big)-t\big(\htPc\big)\big]/(4M_{\pi})=\nu+k_\mu\htPc^\mu/M_{\pi}$.
Now, we expand for small external twisting angles (i.e. around
$\htPc^\mu=0$ or 
$\nu_{\pm}=\nu$) and multiply both
sides by
\begin{equation}\label{Eq:Omegac}
 \frac{1}{2}\sum_{%
		\begin{subarray}{c}
		\vec{n}\in\Z{3}\\
		\abs{\vec{n}}\neq0
		\end{subarray}
		}%
		\int_{\R{4}}\frac{\d{k}{4}}{\left(2\pi\right)^4}\,\e^{\i L\vec{n}\vec{k}}\,\frac{\Omega_c}{M_\pi^2+k^2}\qquad\text{with}\qquad
 \Omega_c=\begin{cases}
           \e^{-\i L\vec{n}\vtPp} &\text{for}\quad c=1\\
           \e^{-\i L\vec{n}\vtPm} &\text{for}\quad c=2\\
           1                      &\text{for}\quad c=3
          \end{cases}.
\end{equation}
The integration over $\vec{k}$ can be performed to an accuracy of
$\Ord{\e^{-\bar{\lambda}}}$ and what remains is the integral over $y=-\i 
k_0/M_{\pi}$ appearing in the asymptotic formulae. 
This provides us with 
\begin{equation}\label{Eq:ChWIAFPc}
\begin{aligned}
 \delta G_{\Pc}&=2\,\delta M_{\Pc}+\delta F_{\Pc}+\Ord{\e^{-\bar{\lambda}}},\\
   \Delta\vtGPc&=\frac{1}{M_\pi^2}\big(\Delta\vtAPc-\Delta\vtSPc\big)+\Ord{\e^{-\bar{\lambda}}},
\end{aligned}
\end{equation}
where $\delta M_{\Pc}$, $\delta F_{\Pc}$, $\delta G_{\Pc}$
resp. $\Delta\vtSPc$, $\Delta\vtAPc$, $\Delta\vtGPc$ are given in terms of
\Eqs(\ref{Eq:AFMPs}, \ref{Eq:AFFPs}, \ref{Eq:AFGPs})
resp. \Eqs(\ref{Eq:AFDtPL}, \ref{Eq:AFFDtPLPsc}, \ref{Eq:AFGDtPLPsc}). 
These relations hold if the amplitudes entering the asymptotic formulae satisfy
\begin{equation}\label{Eq:CondAmpPc}
 \begin{aligned}
  \frac{\hat{m}}{M_{\pi}}\,\mathcal{C}_{\Pc}(\tilde{\nu},\tPp)&=\mathcal{N}_{\Pc}(\tilde{\nu},\tPp)-\frac{F_\pi}{M_{\pi}}\,\mathcal{F}_{\Pc}(\tilde{\nu},\tPp),\\
  \frac{\hat{m}}{M_{\pi}}\,\mathcal{K}_{\Pc}(\tilde{\nu},\tPp)&=\mathcal{H}_{\Pc}(\tilde{\nu},\tPp)-\frac{F_\pi}{M_{\pi}}\,\mathcal{G}_{\Pc}(\tilde{\nu},\tPp) ,
 \end{aligned}
\end{equation} 
which they actually do in general. As a check one can insert the chiral
representation of these amplitudes provided below and explicitly verify
that both relations hold.

The relations for other pseudoscalar mesons can be proved in an analogous
way and give conditions on the amplitudes similar to
\Eq\eqref{Eq:CondAmpPc}, see \Ref\cite{Va:15}. From these conditions it is
possible to determine unknown amplitudes starting from the explicit
representation of the related amplitudes. For instance, one can determine
$\mathcal{N}_{\Pn}$ from the chiral representation of $\mathcal{F}_{\Pn}$,
$\mathcal{C}_{\Pn}$ (see \Refs\cite{BiCoEcGaSa:97,GaLe:84}) or one can
determine $\mathcal{C}_{\Kc}$ (resp. $\mathcal{K}_{\Kc}$) from
$\mathcal{F}_{\Kc}$, $\mathcal{N}_{\Kc}$ (resp. $\mathcal{G}_{\Kc}$,
$\mathcal{H}_{\Kc}$), see \Refs\cite{BeKaiMei:90,BiCoGa:94}. Later, we will
use this fact to work out the asymptotic formulae for the decay constant of
the neutral pion and for the pseudoscalar coupling constants of charged
kaons.

\subsection{Pion form factors}\label{subsec:AFpionsFFS}

As proposed by H{\"a}feli~\cite{Hae:08} one can rely on the Feynman-Hellmann theorem to derive asymptotic formulae for the matrix elements of the scalar form factor.
In finite volume the Feynman-Hellmann theorem relates the corrections of the matrix elements of the scalar form factor with the derivative of the self-energies, see \Eq\eqref{Eq:FHTPc}. 
Starting from such relation we can derive asymptotic formulae valid at vanishing momentum transfer via
\begin{equation}\label{Eq:DSigmaDMassPion}
 \begin{aligned}
  \Delta\Sigma_{\Pn}&=-2M_{\pi}^2\,\delta M_{\Pn}+\Ord{\e^{-\bar{\lambda}}},\\
  \Delta\Sigma_{\Pc}&=-2M_{\pi}^2\,\delta M_{\Pc}-2\vtPc\Delta\vtSPc+\Ord{\e^{-\bar{\lambda}}}.
 \end{aligned}
\end{equation}
Taking the derivative of the asymptotic formulae $\delta M_{\Pn}$, $\delta M_{\Pc}$, $\Delta\vtSPc$ one finds
\begin{subequations}\label{Eq:AFGSPnPc0}
\begin{align}
  \delta\GSPn\big|_{q^2=0}&=\frac{-1}{2\left(4\pi\right)^{2}\lambda_{\pi}}\!\sum_{%
			  \begin{subarray}{c}
			  \vec{n}\in\Z{3}\\
			  \abs{\vec{n}}\neq0
			  \end{subarray}
			  }%
			  \int_{\R{}}\frac{\d{y}{}}{\abs{\vec{n}}}\,\e^{-\lambda_{\pi}\abs{\vec{n}}\sqrt{1+y^2}}\notag\\
			  &\qquad\qquad\quad\times\Big(\!1-\lambda_{\pi}\abs{\vec{n}}{\textstyle\sqrt{1+y^2}}+2M_{\pi}^2\partial_{M_\pi^2}\Big)\mathcal{F}_{\Pn}(\i y,\tPp)
			  +\Ord{\e^{-\bar{\lambda}}},
\end{align}

\begin{align}
  \delta\GSPc\big|_{q^2=0}&=\frac{-1}{2\left(4\pi\right)^{2}\lambda_{\pi}}\!\sum_{%
				  \begin{subarray}{c}
				  \vec{n}\in\Z{3}\\
				  \abs{\vec{n}}\neq0
				  \end{subarray}
				  }%
				  \int_{\R{}}\frac{\d{y}{}}{\abs{\vec{n}}}\,\e^{-\lambda_{\pi}\abs{\vec{n}}\sqrt{1+y^2}}\notag\\
				  &\times\Bigg[\bigg(1-\lambda_{\pi}\abs{\vec{n}}{\textstyle\sqrt{1+y^2}}+2M_\pi^2\partial_{M_\pi^2}\bigg)\mathcal{F}_{\Pc}(\i y,\tPp)\\
				  &\qquad-y\,\frac{D_{\Pc}}{M_{\pi}}\bigg(\lambda_{\pi}\abs{\vec{n}}{\textstyle\sqrt{1+y^2}}+\frac{M_{\pi}}{M_{\pi}+D_{\Pc}}-2M_\pi^2\partial_{M_\pi^2}\bigg)\partial_y\mathcal{F}_{\Pc}(\i y,\tPp)\notag\\
				  &\qquad+y\,L\vec{n}\vtPc\bigg(\!1-\lambda_{\pi}\abs{\vec{n}}{\textstyle\sqrt{1+y^2}}+2M_\pi^2\partial_{M_\pi^2}\bigg)\mathcal{G}_{\Pc}(\i y,\tPp)\Bigg]\!
				  +\Ord{\e^{-\bar{\lambda}}}.\notag
 \end{align}
\end{subequations}
These asymptotic formulae depend on the amplitudes 
entering the expressions for
$\delta M_{\Pn}$, $\delta M_{\Pc}$, $\Delta\vtSPc$. Here, the dependence on
the twist is threefold. The formulae depend on the twisting angle of the
virtual positive pion through the phase factor $\exp(\i L\vec{n}\vtPp)$ in
the amplitudes. Furthermore, they depend on the external twisting angles
through the parameter $D_{\Pc}$ and through the product $\vec{n}\vtPc$ in
the last line of \Eq\eqref{Eq:AFGSPnPc0}. Note that the formulae for
charged pions are only valid for small external twisting angles.  
The formula for the neutral pion is valid for arbitrary twisting angles.

We have checked the above asymptotic formulae in two ways.
First, we have set all twisting angles to zero and we have recovered the formula valid
for PBC, originally proposed by H{\"a}feli~\cite{Hae:08}. 
Second, we have inserted the tree-level chiral representation of $\mathcal{F}_{\Pn}$, $\mathcal{F}_{\Pc}$, $\mathcal{G}_{\Pc}$ and we have obtained
the results~\eqref{Eq:dGSPnPc0} found at NLO with $\mathrm{SU}(2)$ ChPT. 

The general derivation of asymptotic formulae for form factors is
complicated by the presence of a non-zero momentum transfer. 
We anticipate that in this case we were not able to derive asymptotic formulae.
The first part of the derivation is similar to the one outlined in
section~\ref{subsubsec:GeneralizationDerivation}. 
In this case, there is an additional external line due to the insertion of
the pseudoscalar densities (resp. vector currents). 
Taking appropriate modifications one can still show that the argumentation
holds and the matrix elements of form factors decay as
$\Ord{\e^{-\sqrt{3}\,\lambda_{\pi}/2}}$ at asymptotically large $L$. 
In the second part of the derivation, complications arise due to the
injected non-zero momentum transfer, which implies that the integration
over the loop momentum can not be performed as for the self-energy. 
Note that these complications should not arise in matrix elements with two
different external states, as e.g. $\Bra{\Kp}S_{4+\i 5}\Ket{\Pn}$. 
In such matrix elements, both external particles can be taken in the rest
frame and the momentum transfer remains non-zero. The vertex functions could
be then expanded around small external angles as outlined in
section~\ref{subsubsec:GeneralizationDerivation} and the integration 
performed in a similar way as for the self-energy~\cite{CoVa:16b}.

\section{Application of asymptotic formulae}\label{sec:ApplicationAF}

\subsection{Amplitudes in ChPT}

\begin{table}[tbp]
 \centering
 \setlength{\fboxrule}{1.5pt}
 \framebox[\width]{ 
 \begin{tabular}{LLll} 
  \text{Asymptotic formula}                                                                   & \text{Amplitude}                                                               & Process                               & \Ref               \\
  \midrule 
  \delta M_{\Pn},\ \delta M_{\Pc},\ \Delta\vec{\vartheta}_{\Sigma_{\Pc}}                      & \mathcal{F}_{\Pn},\ \mathcal{F}_{\Pc},\ \mathcal{G}_{\Pc}                      & $\pi\pi$-scattering                   & \cite{BiCoEcGaSa:97}\\
  \delta M_{\Kc(\Kn)},\ \Delta\vec{\vartheta}_{\Sigma_{\Kc(\Kn)}}                             & \mathcal{F}_{\Kc(\Kn)},\ \mathcal{G}_{\Kc(\Kn)}                                & $K\pi$-scattering                     & \cite{BeKaiMei:90}  \\
  \delta M_\eta                                                                               & \mathcal{F}_{\eta}                                                             & $\eta\pi$-scattering                  & \cite{BeKaiMei:91}  \\
  \midrule 
  \delta F_{\Pc},\ \Delta\vec{\vartheta}_{\mathscr{A}_{\Pc}}                                  & \mathcal{N}_{\Pc},\ \mathcal{H}_{\Pc}                                          & $\tau$-decay                          & \cite{CoFiUr:96}    \\
  \delta F_{\Kc},\ \Delta\vec{\vartheta}_{\mathscr{A}_{\Kc}}                                  & \mathcal{N}_{\Kc},\ \mathcal{H}_{\Kc}                                          & $K_{\ell4}$-decay                     & \cite{BiCoGa:94}    \\
  \delta F_{\Pn},\ \delta F_{\Kn},\ \Delta\vec{\vartheta}_{\mathscr{A}_{\Kn}},\ \delta F_\eta & \mathcal{N}_{\Pn},\ \mathcal{N}_{\Kn},\ \mathcal{H}_{\Kn},\ \mathcal{N}_{\eta} &                                       &                     \\
  \midrule 
  \delta G_{\Pn},\ \delta G_{\Pc},\ \Delta\vec{\vartheta}_{\mathscr{G}_{\Pc}}                 & \mathcal{C}_{\Pn},\ \mathcal{C}_{\Pc},\ \mathcal{K}_{\Pc}                      & $P_a\longrightarrow\pi_a(\pi_c\pi_c)$ & \cite{GaLe:84}      \\
  \delta G_{\Kc(\Kn)},\ \Delta\vec{\vartheta}_{\mathscr{G}_{\Kc(\Kn)}},\ \delta G_\eta        & \mathcal{C}_{\Kc(\Kn)},\ \mathcal{K}_{\Kc(\Kn)},\ \mathcal{C}_{\eta}           &                                       &                     \\
  \midrule 
  \delta\Gamma_{S}^{\Pn}|_{q^2=0},\ \delta\Gamma_{S}^{\Pc}|_{q^2=0}                           & \mathcal{F}_{\Pn},\ \mathcal{F}_{\Pc},\ \mathcal{G}_{\Pc}                      & $\pi\pi$-scattering                   & \cite{BiCoEcGaSa:97}\\
 \end{tabular}
 }
  \caption{We summarize the quantities for which asymptotic formulae have
    been derived in this work, the corresponding infinite-volume amplitudes needed
    and the processes from which the chiral representation can be
    determined. We also provide references where the results for the
    amplitudes at one loop can be found. Here,
    $P_a\longrightarrow\pi_a(\pi_c\pi_c)$ denotes the pseudoscalar decay
    into three pions in which two pions form a state with zero isospin. 
          }
 \label{Tab:Amplitudes}         
\end{table}

We apply the asymptotic formulae 
in combination with ChPT and estimate finite-volume corrections beyond NLO.
We use the chiral representation of the amplitudes at one loop which is
known for most of the amplitudes derived in this work.  In
\Tab\ref{Tab:Amplitudes} we summarize all quantities for which we have
derived an asymptotic formula and the infinite-volume amplitudes needed
therein. If available we give the reference where the chiral representation
at one loop can be found. We also list the process from which the chiral
representation can be determined. Note that currently, $\mathcal{N}_{\Pn}$,
$\mathcal{C}_{\Kc}$, $\mathcal{K}_{\Kc}$ are unknown in ChPT but they can
be determined from $\mathcal{F}_{\Pn},\mathcal{C}_{\Pn}$, $\
\mathcal{F}_{\Kc},\mathcal{N}_{\Kc}\ $ and $\
\mathcal{G}_{\Kc},\mathcal{H}_{\Kc}$ relying on chiral Ward identities.

\subsubsection{Chiral representation at one loop}\label{subsubsec:ChRepr1L}

\paragraph{Pions}

The amplitudes $\mathcal{F}_{\Pn}$, $\mathcal{F}_{\Pc}$, $\mathcal{G}_{\Pc}$ are defined similarly to \Eqs(\ref{Eq:AmpFKc}, \ref{Eq:AmpGKc}).
Their chiral representation can be determined from the $\pi\pi$-scattering, $ \pi(p_1)+\pi(p_2)\longrightarrow\pi(p_3)+\pi(p_4)$, for forward kinematics: 
\begin{equation}\label{Eq:fwdpipi}
   s=(p_1+p_2)^2\!=2M_{\pi}(M_{\pi}+\nu),\quad\
   t=(p_1-p_3)^2\!=0,\quad\
   u=(p_1-p_4)^2\!=2M_{\pi}(M_{\pi}-\nu).
\end{equation}
All isospin components of the $\pi\pi$-scattering can be given in terms of
the invariant amplitude $A_{stu}:=A(s,t,u)$ which is known up to
NNLO in ChPT~\cite{BiCoEcGaSa:97}. 
In terms of $A_{stu}$ the isospin components of $\mathcal{F}_{\Pn}$,
$\mathcal{F}_{\Pc}$, $\mathcal{G}_{\Pc}$ read  
\begin{equation}\label{Eq:IsoTPnTPc}
 \begin{aligned}
   T_{\Pn\Pn}(t,u-s)&=A_{stu}+A_{tus}+A_{ust},                                       \\
   T_{\Pn\Pp}(t,u-s)&=T_{\Pn\Pm}(t,u-s)=T_{\Pp\Pn}(t,u-s)=T_{\Pm\Pn}(t,u-s)=A_{tus}, \\
   T_{\Pp\Pp}(t,u-s)&=T_{\Pm\Pm}(t,u-s)=A_{tus}+A_{ust},                             \\
   T_{\Pp\Pm}(t,u-s)&=T_{\Pm\Pp}(t,u-s)=A_{stu}+A_{tus}. 
 \end{aligned}
\end{equation}
Inserting the expression of $A_{stu}$ at NLO~\cite{GaLe:84} and evaluating
the isospin components in the kinematics~\eqref{Eq:fwdpipi} one obtains the
chiral representation of $\mathcal{F}_{\Pn}$, $\mathcal{F}_{\Pc}$,
$\mathcal{G}_{\Pc}$ at one loop.  

The amplitudes $\mathcal{N}_{\Pc}$, $\mathcal{H}_{\Pc}$ have similar
definitions as \Eqs(\ref{Eq:AmpNKc}, \ref{Eq:AmpHKc}). Their chiral
representation can be determined from the matrix elements of the
$\tau$-decay,
$\tau(p_\tau)\longrightarrow\pi(p_1)+\pi(p_2)+\pi(p_3)+\nu_{\tau}(p_\tau-Q)$,
for forward kinematics: 
\begin{equation} 
 s_1\!=(p_2+p_3)^2\!=2M_{\pi}(M_{\pi}-\nu),\quad 
 s_2\!=(p_1+p_3)^2\!=2M_{\pi}(M_{\pi}+\nu),\quad
 s_3\!=(p_1+p_2)^2\!=0.
\end{equation}
In ChPT the matrix elements of this $\tau$-decay are known at
NLO~\cite{CoFiUr:96} and can be written as 
\begin{equation}
 \begin{aligned}
  (A_{\Pm\Pn})^\mu&:=\Bra{\Pn(p_1)\Pn(p_2)\Pm(p_3)}A^\mu_{1-\i2}(0)\Ket{0},\\
  (A_{\Pm\Pm})^\mu&:=\Bra{\Pm(p_1)\Pp(p_2)\Pm(p_3)}A^\mu_{1-\i2}(0)\Ket{0}.
 \end{aligned}
\end{equation}
According to \cite{CoFiUr:96} these matrix elements can be decomposed as\footnote{%
       The factors $F_{123},G_{123},H_{123}$ are $1/\sqrt{2}$ times smaller than those of \Ref\cite{CoFiUr:96}.
       } %
\begin{equation}\label{Eq:MEPiPiPmAm}
 \begin{aligned}
  (A_{\Pm\Pn})^\mu&=F_{123}\,p_3^\mu+G_{123}\,(p_1+p_2)^\mu+H_{123}\,(p_1-p_2)^\mu,\\
  (A_{\Pm\Pm})^\mu&=F_{123}^{[-]}\,p_3^\mu+G_{123}^{[-]}(p_1+p_2)^\mu+H_{123}^{[-]}(p_1-p_2)^\mu.
 \end{aligned}
\end{equation}
The scalar functions $F_{123}=F(s_1,s_2,s_3)$, $G_{123}=G(s_1,s_2,s_3)$, $H_{123}=H(s_1,s_2,s_3)$ are related to $F_{123}^{[-]}=F^{[-]}(s_1,s_2,s_3)$, $G_{123}^{[-]}=G^{[-]}(s_1,s_2,s_3)$, $H_{123}^{[-]}=H^{[-]}(s_1,s_2,s_3)$ through 
\begin{equation}
\begin{aligned}
 F_{123}^{[-]}&=F_{123}+G_{231}-H_{231},\\
 G_{123}^{[-]}&=G_{123}+\frac{1}{2}[F_{231}+G_{231}+H_{231}],\\
 H_{123}^{[-]}&=H_{123}+\frac{1}{2}[F_{231}-G_{231}-H_{231}].
\end{aligned} 
\end{equation}

As described in \Ref\cite{CoHae:04} the matrix
elements~\eqref{Eq:MEPiPiPmAm} have a pole that does not contribute to the
amplitudes of asymptotic formulae and must be subtracted. The subtraction
occurs expanding the matrix elements around $Q^2=(p_1+p_2+p_3)^2=M_\pi^2$
and isolating the pole appearing in $F_{123},G_{123}$
resp. $F_{123}^{[-]},G_{123}^{[-]}$.   
One gets
\begin{equation}\label{Eq:DefIsoCompNPm}
 \begin{aligned}
  \big(\bar{A}_{\Pm\Pn}\big)^\mu&=(A_{\Pm\Pn})^\mu-\i F_\pi Q^\mu\frac{T_{\Pm\Pn}(s_3,s_1-s_2)}{M_\pi^2-Q^2},\\
  \big(\bar{A}_{\Pm\Pm}\big)^\mu&=(A_{\Pm\Pm})^\mu-\i F_\pi Q^\mu\frac{T_{\Pm\Pm}(s_3,s_1-s_2)}{M_\pi^2-Q^2},
 \end{aligned} 
\end{equation}
where $T_{\Pm\Pn}(s_3,s_1-s_2)$, $T_{\Pm\Pm}(s_3,s_1-s_2)$ are isospin
components of the $\pi\pi$-scattering amplitude in the $s_3$-channel, see
\Eq\eqref{Eq:IsoTPnTPc}. The bar indicates that the pole has been
subtracted from the matrix elements. The amplitudes can be obtained
contracting with $p_3^\mu/M_{\pi}$ and setting the momenta to
$p_2^\mu=-p_1^\mu$
resp.
$p_3^\mu=\big(\begin{smallmatrix}M_{\pi}\\\vec{0}\end{smallmatrix}\big)$. 
For instance, the isospin components of $\mathcal{N}_{\Pm}$, $\mathcal{H}_{\Pm}$ read
\begin{equation}\label{Eq:bAPm}
  \bar{A}_{\Pm\Pn}(s_3,s_1-s_2)=\frac{p_3^\mu}{M_{\pi}}\big(\bar{A}_{\Pm\Pn}\big)_\mu\quad\text{and}\quad
  \bar{A}_{\Pm\Pm}(s_3,s_1-s_2)=\frac{p_3^\mu}{M_{\pi}}\big(\bar{A}_{\Pm\Pm}\big)_\mu. 
\end{equation} 
The isospin component $\bar{A}_{\Pm\Pp}(s_3,s_1-s_2)$ can be obtained from
$\big(A_{\Pm\Pm}\big)^\mu$ exchanging $p_1^\mu\leftrightarrow
p_2^\mu$. Note that isospin symmetry relates the amplitudes of the negative
pion to those of the positive pion via 
\begin{equation}
  \mathcal{N}_{\Pp}(\tilde{\nu},\tPp)=\mathcal{N}_{\Pm}(\tilde{\nu},\tPp)\qquad\text{and}\qquad 
  \mathcal{H}_{\Pp}(\tilde{\nu},\tPp)=-\mathcal{H}_{\Pm}(\tilde{\nu},\tPp). 
\end{equation}

The amplitudes $\mathcal{C}_{\Pn}$, $\mathcal{C}_{\Pc}$,
$\mathcal{K}_{\Pc}$ are defined similarly to \Eqs(\ref{Eq:AmpCKc},
\ref{Eq:AmpKKc}). 
Their representation can be determined from the point function containing
four pseudoscalar densities, where three of them serve as interpolating
fields for pions. In ChPT such point function can be calculated from
\Eq(16.2) of \Ref\cite{GaLe:84}: one must take the off-shell amplitude
$A(s,t,u;p_1^2,p_2^2,p_3^2,p_4^2)$ and set the momenta to
$p_1^2=p_2^2=p_3^2=M_\pi^2$ resp. $p_4^2=\tilde{Q}^2$ with
$\tilde{Q}^\mu=(p_3-p_1-p_2)^\mu$. 
Defining $C_{stu}:=A(s,t,u;M_\pi^2,M_\pi^2,M_\pi^2,\tilde{Q}^2)$, the
isospin components of $\mathcal{C}_{\Pn}$, $\mathcal{C}_{\Pc}$,
$\mathcal{K}_{\Pc}$ can be expressed as  
 \begin{align}\label{Eq:IsoCPnCPc}
   C_{\Pn\Pn}(s,t-u)&=\frac{G_\pi}{M_\pi^2-\tilde{Q}^2}[C_{stu}+C_{tus}+C_{ust}],                                       \notag\\
   C_{\Pn\Pp}(s,t-u)&=C_{\Pn\Pm}(s,t-u)=C_{\Pp\Pn}(s,t-u)=C_{\Pm\Pn}(s,t-u)=\frac{G_\pi}{M_\pi^2-\tilde{Q}^2}\,C_{stu}, \notag\\
   C_{\Pp\Pp}(s,t-u)&=C_{\Pm\Pm}(s,t-u)=\frac{G_\pi}{M_\pi^2-\tilde{Q}^2}[C_{stu}+C_{tus}],                             \notag\\
   C_{\Pp\Pm}(s,t-u)&=C_{\Pm\Pp}(s,t-u)=\frac{G_\pi}{M_\pi^2-\tilde{Q}^2}[C_{stu}+C_{ust}],
 \end{align}
after having subtracted the pole at $\tilde{Q}^2=M_\pi^2$.
In section~\ref{subsubsec:AnalyticalResults} we describe how to subtract
this pole and how to define the isospin components, see
\Eq\eqref{Eq:barCKcPion}. From that definition, one obtains the chiral
representation evaluating the isospin components in the forward kinematics:
\begin{equation}
   s=0,\qquad t=2M_{\pi}(M_{\pi}-\nu),\qquad u=2M_{\pi}(M_{\pi}+\nu).
\end{equation}

In section~\ref{subsec:AFpionsFFS} we have derived asymptotic formulae for
the matrix elements of the scalar form factor of pions at vanishing
momentum transfer. The amplitudes entering those formulae are
$\mathcal{F}_{\Pn}$, $\mathcal{F}_{\Pc}$, $\mathcal{G}_{\Pc}$. The chiral
representation is given by the isospin components~\eqref{Eq:IsoTPnTPc}
evaluated in the forward kinematics. In this case, the asymptotic formulae
also contain $\partial_{M_\pi^2}\mathcal{F}_{\Pn}$,
$\partial_{M_\pi^2}\mathcal{F}_{\Pc}$,
$\partial_{M_\pi^2}\mathcal{G}_{\Pc}$, for which attention must be paid during their calculation.

\paragraph{Kaons}

The amplitudes $\mathcal{F}_{\Kc(\Kn)}$, $\mathcal{G}_{\Kc(\Kn)}$ can be
determined from the $K\pi$-scattering,
$\Pp(p_1)+\Kp(p_2)\longrightarrow\Pp(p_3)+\Kp(p_4)$, in the case of forward
kinematics:
\begin{equation}\label{Eq:fwdPpKp}
   s=M_K^2+M_\pi^2+2M_K\nu,\qquad t=0,\qquad u=M_K^2+M_\pi^2-2M_K\nu.
\end{equation}
In ChPT this scattering is given by the amplitude
$T^{3/2}_{stu}:=T^{3/2}(s,t,u)$ which is known at NLO from
\Ref\cite{BeKaiMei:90}.\footnote{%
    As noted in~\cite{CoDuHae:05} there are two misprints in \Eq(3.16) of \cite{BeKaiMei:90}. 
    The prefactor of $[M^{r}_{\pi K}(u)+M^{r}_{K\eta}(u)]$ should read $(M_K^2-M_\pi^2)^2$ and the factor of $\frac{3}{8}J^{r}_{K\eta}(u)$ should read $[u-\frac{2}{3}(M_\pi^2+M_K^2)]^2$. 
} %
In terms of $T^{3/2}_{stu}$ the isospin components of
$\mathcal{F}_{\Kc(\Kn)}$, $\mathcal{G}_{\Kc(\Kn)}$ read  
\begin{equation}\label{Eq:IsoTKcTKn}
 \begin{aligned}
      T_{\Kp\Pn}(t,u-s)&=T_{\Km\Pn}(t,u-s)=T_{\Kn\Pn}(t,u-s)=\frac{1}{2}\left[T^{3/2}_{stu}+T^{3/2}_{uts}\right],\\
      T_{\Kp\Pp}(t,u-s)&=T_{\Km\Pm}(t,u-s)=T_{\Kn\Pm}(t,u-s)=T^{3/2}_{stu},\\
      T_{\Kp\Pm}(t,u-s)&=T_{\Km\Pp}(t,u-s)=T_{\Kn\Pp}(t,u-s)=T^{3/2}_{uts}.
 \end{aligned}
\end{equation}
The chiral representation of $\mathcal{F}_{\Kc(\Kn)}$,
$\mathcal{G}_{\Kc(\Kn)}$ can be then obtained evaluating these expressions
in the kinematics~\eqref{Eq:fwdPpKp}. 
Note that from \Eq\eqref{Eq:IsoTKcTKn} it turns out that the isospin
components of the negative kaon are equal to those of the neutral and hence,   
\begin{equation}\label{Eq:AmpFKnFKmGKnGKm}
  \mathcal{F}_{\Kn}(\tilde{\nu},\tPp)=\mathcal{F}_{\Km}(\tilde{\nu},\tPp)\qquad\text{and}\qquad
  \mathcal{G}_{\Kn}(\tilde{\nu},\tPp)=\mathcal{G}_{\Km}(\tilde{\nu},\tPp).
\end{equation}

The amplitudes $\mathcal{N}_{\Kc}$, $\mathcal{H}_{\Kc}$ are defined in
\Eqs(\ref{Eq:AmpNKc}, \ref{Eq:AmpHKc}).  
Their chiral representation can be determined from the matrix elements of
the $K_{\ell4}$-decay,
$\Kp(p_3)\longrightarrow\pi(p_1)+\pi(p_2)+\ell^{+}(p_{\ell})+\bar{\nu}_\ell(\tilde{Q}-p_\ell)$,
for forward kinematics:
\begin{equation}
   s=0,\qquad t=M_K^2+M_\pi^2-2M_K\nu,\qquad u=M_K^2+M_\pi^2+2M_K\nu.
\end{equation}
In ChPT the matrix elements of the $K_{\ell4}$-decay are known at
NLO~\cite{BiCoGa:94} and read
\begin{equation} 
 \begin{aligned}
  (A_{\Kp\Pn})^\mu&=\Bra{\Pn(p_1)\Pn(p_2)}A^\mu_{4-\i 5}(0)\Ket{\Kp(p_3)},\\
  (A_{\Kp\Pp})^\mu&=\Bra{\Pp(p_1)\Pm(p_2)}A^\mu_{4-\i 5}(0)\Ket{\Kp(p_3)}.
 \end{aligned}
\end{equation}
According to \cite{BiCoGa:94} these matrix elements can be decomposed as\footnote{%
       The factors $F_{stu},G_{stu},R_{stu}$ are $1/\sqrt{2}$ times smaller than those of \Ref\cite{BiCoGa:94}.
       }%
\begin{equation}\label{Eq:MEPiPiAmKp}
 \begin{aligned}
  (A_{\Kp\Pn})^\mu&=\frac{-\i}{M_K}\left[F_{stu}^{+}\,(p_1+p_2)^\mu+G_{stu}^{-}\,(p_1-p_2)^\mu+R_{stu}^{+}\,\tilde{Q}^\mu\right],\\ 
  (A_{\Kp\Pp})^\mu&=\frac{-\i}{M_K}\left[F_{stu}\,(p_1+p_2)^\mu+G_{stu}\,(p_1-p_2)^\mu+R_{stu}\,\tilde{Q}^\mu\right],
 \end{aligned}
\end{equation}
where $\tilde{Q}^\mu=(p_3-p_1-p_2)^\mu$.
The scalar functions $F_{stu}^{+}=F^{+}(s,t,u)$, $G_{stu}^{-}=G^{-}(s,t,u)$, $R_{stu}^{+}=R^{+}(s,t,u)$ are related to $F_{stu}=F(s,t,u)$, $G_{stu}=G(s,t,u)$, $R_{stu}=R(s,t,u)$ via
\begin{equation}
 F_{stu}^{+}=\frac{1}{2}[F_{stu}+F_{sut}],\qquad
 G_{stu}^{-}=\frac{1}{2}[G_{stu}-G_{sut}],\qquad
 R_{stu}^{+}=\frac{1}{2}[R_{stu}+R_{sut}].
\end{equation}

As described in \Ref\cite{CoDuHae:05} the matrix
elements~\eqref{Eq:MEPiPiAmKp} have a pole that must be subtracted. 
Expanding around  $\tilde{Q}^2=M_K^2$ the pole appears in $R_{stu}^{+}$
resp. $R_{stu}$ and can be expressed in terms of the isospin components of
the $K\pi$-scattering amplitude.
One gets 
\begin{equation}\label{Eq:DefIsoCompNKp}
 \begin{aligned}
  \big(\bar{A}_{\Kp\Pn}\big)^\mu&=(A_{\Kp\Pn})^\mu-\i F_K\tilde{Q}^\mu\frac{T_{\Kp\Pn}(s,t-u)}{M_K^2-\tilde{Q}^2},\\
  \big(\bar{A}_{\Kp\Pp}\big)^\mu&=(A_{\Kp\Pp})^\mu-\i F_K\tilde{Q}^\mu\frac{T_{\Kp\Pp}(s,t-u)}{M_K^2-\tilde{Q}^2}.
 \end{aligned} 
\end{equation}
The amplitudes of asymptotic formulae can be obtained contracting with
$p_3^\mu/M_K$ and setting the momenta to $p_2^\mu=-p_1^\mu$
resp. $p_3^\mu=\Big(\begin{smallmatrix}M_K\\\vec{0}\end{smallmatrix}\Big)$. 
Thus, the isospin components of $\mathcal{N}_{\Kp}$,
$\mathcal{H}_{\Kp}$ read
\begin{equation}\label{Eq:bAKp}
  \bar{A}_{\Kp\Pn}(s,t-u)=\frac{p^\mu}{M_K}\big(\bar{A}_{\Kp\Pn}\big)_\mu,\qquad
  \bar{A}_{\Kp\Pp}(s,t-u)=\frac{p^\mu}{M_K}\big(\bar{A}_{\Kp\Pp}\big)_\mu, 
\end{equation} 
and $\bar{A}_{\Kp\Pm}(s,t-u)$ can be obtained from
$\big(\bar{A}_{\Kp\Pp}\big)^\mu$ exchanging $p_1^\mu\!\leftrightarrow\!
p_2^\mu$. Note that isospin symmetry relates the amplitudes of the
positive kaon to those of the negative one as 
\begin{equation}
  \mathcal{N}_{\Km}(\tilde{\nu},\tPp)=\mathcal{N}_{\Kp}(\tilde{\nu},\tPp)\qquad\text{and}\qquad 
  \mathcal{H}_{\Km}(\tilde{\nu},\tPp)=-\mathcal{H}_{\Kp}(\tilde{\nu},\tPp).
\end{equation}

\paragraph{Eta meson}

The amplitude $\mathcal{F}_{\eta}$ is defined similarly to
\Eq\eqref{Eq:AmpFKc}. Its chiral representation can be determined from the
$\eta\pi$-scattering,
$\pi(p_1)+\eta(p_2)\longrightarrow\pi(p_3)+\eta(p_4)$, for forward
kinematics: 
\begin{equation}\label{Eq:fwdetapi}
   s=M_\eta^2+M_\pi^2+2M_\eta\nu,\qquad t=0,\qquad u=M_\eta^2+M_\pi^2-2M_\eta\nu.
\end{equation}
In ChPT the $\eta\pi$-scattering is given by the invariant amplitude $T_{\pi\eta}(s,t,u)$.
In terms of $T_{\pi\eta}(s,t,u)$ the isospin components of
$\mathcal{F}_{\eta}$ are all the same,
\begin{equation}\label{Eq:IsoTetaPnTetaPc}
  T_{\eta\Pn}(t,u-s)=T_{\eta\Pp}(t,u-s)
                    =T_{\eta\Pm}(t,u-s)
                    =T_{\pi\eta}(s,t,u).
\end{equation}
Inserting the expression of $T_{\pi\eta}(s,t,u)$ at NLO~\cite{BeKaiMei:91}
and evaluating the isospin components for forward kinematics one obtains
the chiral representation of $\mathcal{F}_{\eta}$ at one loop.

\subsubsection{Chiral expansion}

We now apply the asymptotic formulae of section~\ref{sec:AFs}.
The results are presented in sections
\ref{subsec:AFpions}--\ref{subsec:AFeta} with long expressions
relegated to appendix~\ref{app:IntegralsS4}. 
To better organize these results we follow~\Ref\cite{CoDuHae:05} and make
use of the chiral expansion. 

We first keep the discussion general and consider a pseudoscalar meson $P$
with the mass $M_P$ and the twisting angle $\vartheta_{P}^\mu$. 
The amplitudes of \Tab\ref{Tab:Amplitudes} can be expressed in the generic forms,
\begin{equation}
 \begin{aligned}
 \mathcal{X}_{P}(\tilde{\nu},\tPp)&=Z_{P\Pn}(0,-4M_P\nu)+[Z_{P\Pp}(0,-4M_P\nu)+Z_{P\Pm}(0,-4M_P\nu)]\e^{\i L\vec{n}\vtPp},\\
 \mathcal{Y}_{P}(\tilde{\nu},\tPp)&=[Z_{P\Pp}(0,-4M_P\nu)-Z_{P\Pm}(0,-4M_P\nu)]\e^{\i L\vec{n}\vtPp},
 \end{aligned}
\end{equation}
where $\nu=(s-u)/(4M_P)$ and $\tilde{\nu}=\nu/M_{\pi}$.
The functions $Z_{P\Pn}(t,u-s)$, $Z_{P\Pp}(t,u-s)$, $Z_{P\Pm}(t,u-s)$ are
isospin components in infinite volume. Let us assume that $\mathcal{X}_{P}(\tilde{\nu},\tPp)$ enters the asymptotic formula for
the observable $X_{P}$ and so, provides an estimate for the corrections $\delta X_{P}$. 
The asymptotic formula has then the form 
\begin{equation}\label{Eq:AFXPs}
\begin{aligned}
 \delta X_{P}&=R(X_{P})+\Ord{\e^{-\bar{\lambda}}},\\
     R(X_{P})&=\frac{1}{(4\pi)^{2}\lambda_{P}}\,\frac{M_{\pi}}{X_{P}}\!\sum_{%
				  \begin{subarray}{c}
				  \vec{n}\in\Z{3}\\
				  \abs{\vec{n}}\neq0
				  \end{subarray}
				  }%
                                  \int_{\R{}}\frac{\d{y}{}}{\abs{\vec{n}}}\,\e^{-\lambda_{\pi}\abs{\vec{n}}\sqrt{1+y^2}}
                                  \bigg(\!1+y\frac{D_{P}}{M_P}\frac{\partial}{\partial y}\bigg)\mathcal{X}_{P}(\i y,\tPp),
\end{aligned}
\end{equation}
where $\lambda_{P}=M_P L$ and $D_{P}=\sqrt{M^2_P+\abs{\vec{\vartheta}_{P}}^2}-M_P$. 
Analogously, assume that the amplitude $\mathcal{Y}_{P}(\tilde{\nu},\tPp)$ enters the
asymptotic formula for the extra term $\Delta\vartheta_{\mathscr{X}_{P}}^\mu$ which has the form 
\begin{equation}\label{Eq:AFDvtwXPs}
\begin{aligned}
           \Delta\vt_{\mathscr{X}_{P}}&=\vec{R}(\vartheta_{\mathscr{X}_{P}})+\Ord{\e^{-\bar{\lambda}}},\\
  \vec{R}(\vartheta_{\mathscr{X}_{P}})&=\frac{-1}{2(4\pi)^{2}}\,\frac{M_\pi^2}{X_{P}}\!\sum_{%
					\begin{subarray}{c}
					\vec{n}\in\Z{3}\\
					\abs{\vec{n}}\neq0
					\end{subarray}
					}%
					\frac{\vec{n}}{\abs{\vec{n}}}\int_{\R{}}\!\!\!\d{y}{}\e^{-\lambda_{\pi}\abs{\vec{n}}\sqrt{1+y^2}}y\,\mathcal{Y}_{P}(\i y,\tPp).
\end{aligned}
\end{equation}
For convenience, we rewrite the amplitudes as 
\begin{equation}\label{Eq:DecompoXPsYPs}
 \begin{aligned}
 \mathcal{X}_{P}(\tilde{\nu},\tPp)&=\mathcal{X}_{P}(X_{P},\Pn)+\mathcal{X}_{P}(X_{P},\Pc)\,\e^{\i L\vec{n}\vtPp},\\
 \mathcal{Y}_{P}(\tilde{\nu},\tPp)&=\mathcal{Y}_{P}(\vartheta_{\mathscr{X}_{P}})\,\e^{\i L\vec{n}\vtPp},
 \end{aligned}
\end{equation}
where we collect the isospin components in
\begin{equation}
 \begin{aligned}
                    \mathcal{X}_{P}(X_{P},\Pn)&:=Z_{P\Pn}(0,-4M_P\nu),\\
                    \mathcal{X}_{P}(X_{P},\Pc)&:=Z_{P\Pp}(0,-4M_P\nu)+Z_{P\Pm}(0,-4M_P\nu),\\
  \mathcal{Y}_{P}(\vartheta_{\mathscr{X}_{P}})&:=Z_{P\Pp}(0,-4M_P\nu)-Z_{P\Pm}(0,-4M_P\nu).
 \end{aligned}
\end{equation}

In ChPT the amplitudes $\mathcal{X}_{P}(\tilde{\nu},\tPp)$,
$\mathcal{Y}_{P}(\tilde{\nu},\tPp)$ can be developed according to the
chiral expansion in powers of $\xi_P=M_P^2/(4\pi F_\pi)^2$. 
Expanding the amplitudes one gets 
\begin{equation}\label{Eq:XPsYPsChiral}
 \begin{aligned}
  \mathcal{X}_{P}(\tilde{\nu},\tPp)&=\mathcal{X}_{P}^{(2)}(\tilde{\nu},\tPp)+\xi_{P}\mathcal{X}_{P}^{(4)}(\tilde{\nu},\tPp)+\Ord{\xi_P^2},\\
  \mathcal{Y}_{P}(\tilde{\nu},\tPp)&=\mathcal{Y}_{P}^{(2)}(\tilde{\nu},\tPp)+\xi_{P}\mathcal{Y}_{P}^{(4)}(\tilde{\nu},\tPp)+\Ord{\xi_P^2}.
 \end{aligned}
\end{equation}
At each order the terms can be written by means of the definitions~\eqref{Eq:DecompoXPsYPs} as
\begin{equation}\label{Eq:XYPSj}
 \begin{aligned}
 \mathcal{X}_{P}^{(j)}(\tilde{\nu},\tPp)&=\mathcal{X}_{P}^{(j)}(X_{P},\Pn)+\mathcal{X}_{P}^{(j)}(X_{P},\Pc)\e^{\i L\vec{n}\vtPp},\\
 \mathcal{Y}_{P}^{(j)}(\tilde{\nu},\tPp)&=\mathcal{Y}_{P}^{(j)}(\vartheta_{\mathscr{X}_{P}})\e^{\i L\vec{n}\vtPp},
 \end{aligned}
\end{equation}
with $j=2,4,\dots$
This allows one to factorize order by order the phase factor $\exp(\i
L\vec{n}\vtPp)$ within the chiral expansion.  
The amplitudes become 
\begin{subequations}
\begin{align}
 \mathcal{X}_{P}(\tilde{\nu},\tPp)&=\mathcal{X}_{P}^{(2)}(X_{P},\Pn)+\xi_{P}\mathcal{X}_{P}^{(4)}(X_{P},\Pn)+\Ord{\xi_P^2}\notag\\
                                  &\phantom{=}\ 
                                   +\!\Big[\mathcal{X}_{P}^{(2)}(X_{P},\Pc)+\xi_{P}\mathcal{X}_{P}^{(4)}(X_{P},\Pc)+\Ord{\xi_P^2}\Big]\e^{\i L\vec{n}\vtPp},\label{Eq:XPs}\\                                
 \mathcal{Y}_{P}(\tilde{\nu},\tPp)&=\Big[\mathcal{Y}_{P}^{(2)}(\vartheta_{\mathscr{X}_{P}})+\xi_{P}\mathcal{Y}_{P}^{(4)}(\vartheta_{\mathscr{X}_{P}})+\Ord{\xi_P^2}\Big]\e^{\i L\vec{n}\vtPp}.\label{Eq:YPs}
 \end{align}
\end{subequations}
These expressions induce a similar expansion in the asymptotic formulae.  
The asymptotic formula~\eqref{Eq:AFXPs} exhibit four contributions which in general take the form
\begin{equation}\label{Eq:RXPs}
 R(X_{P})=R(X_{P},\Pn)+R(X_{P},\Pc)+R_D(X_{P},\Pn)+R_D(X_{P},\Pc),
\end{equation}
with
\begin{align}\label{Eq:RXPsPion}
   R(X_{P},\Pn)&=\frac{\xi_\pi}{\lambda_{\pi}}\frac{X_\pi}{X_P}\sum_{%
				  \begin{subarray}{c}
				  \vec{n}\in\Z{3}\\
				  \abs{\vec{n}}\neq0
				  \end{subarray}
				  }%
				  \frac{1}{\abs{\vec{n}}}\Big[I^{(2)}(X_P,\Pn)+\xi_P I^{(4)}(X_P,\Pn)+\Ord{\xi_P^2}\Big],\notag\\
   R(X_{P},\Pc)&=\frac{\xi_\pi}{\lambda_{\pi}}\frac{X_\pi}{X_P}\!\sum_{%
				  \begin{subarray}{c}
				  \vec{n}\in\Z{3}\\
				  \abs{\vec{n}}\neq0
				  \end{subarray}
				  }%
				  \frac{1}{\abs{\vec{n}}}\Big[I^{(2)}(X_P,\Pc)+\xi_P I^{(4)}(X_P,\Pc)+\Ord{\xi_P^2}\Big]\e^{\i L\vec{n}\vtPp},\notag\\[-0.5cm]
				  \\
 R_D(X_{P},\Pn)&=\frac{\xi_\pi}{\lambda_{\pi}}\frac{X_\pi}{X_P}\frac{D_{P}}{M_P}\sum_{%
				  \begin{subarray}{c}
				  \vec{n}\in\Z{3}\\
				  \abs{\vec{n}}\neq0
				  \end{subarray}
				  }%
                                  \frac{1}{\abs{\vec{n}}}\Big[I_D^{(2)}(X_P,\Pn)+\xi_P I_D^{(4)}(X_P,\Pn)+\Ord{\xi_P^2}\Big],\notag\\
 R_D(X_{P},\Pc)&=\frac{\xi_\pi}{\lambda_{\pi}}\frac{X_\pi}{X_P}\frac{D_{P}}{M_P}\sum_{%
				  \begin{subarray}{c}
				  \vec{n}\in\Z{3}\\
				  \abs{\vec{n}}\neq0
				  \end{subarray}
				  }%
				  \frac{1}{\abs{\vec{n}}}\Big[I_D^{(2)}(X_P,\Pc)+\xi_P I_D^{(4)}(X_P,\Pc)+\Ord{\xi_P^2}\Big]\e^{\i L\vec{n}\vtPp}.\notag
\end{align}
Each contribution is rescaled by
$X_\pi/X_P$ where $X_\pi$ denotes the observable $X_{P}$ in infinite volume
for $P=\pi$. The integrals $I^{(j)}(X_P,\Pn)$, $\ldots,$
$I_D^{(j)}(X_P,\Pc)$ can be determined from the terms
$\mathcal{X}_{P}^{(j)}(X_{P},\Pn)$, $\mathcal{X}_{P}^{(j)}(X_{P},\Pc)$ of
\Eq\eqref{Eq:XYPSj}. Note that the contributions $R_D(X_{P},\Pn)$,
$R_D(X_{P},\Pc)$ are proportional to the parameter $D_{P}$ and are not
present if the particle $P$ has no external twisting angle (as e.g. for the
neutral pion and the eta meson).  

The asymptotic formula~\eqref{Eq:AFDvtwXPs} can be expanded in a similar way as 
\begin{equation}\label{Eq:RtwXPs}
  \vec{R}(\vartheta_{\mathscr{X}_{P}})=-\frac{\xi_\pi M_P}{2}\frac{X_\pi}{X_P}\sum_{%
				  \begin{subarray}{c}
				  \vec{n}\in\Z{3}\\
				  \abs{\vec{n}}\neq0
				  \end{subarray}
				  }%
				  \frac{\i\vec{n}}{\abs{\vec{n}}}\Big[I^{(2)}(\vartheta_{\mathscr{X}_{P}})+\xi_P I^{(4)}(\vartheta_{\mathscr{X}_{P}})+\Ord{\xi_P^2}\Big]\e^{\i L\vec{n}\vtPp}.
\end{equation}
The integrals $I^{(j)}(\vartheta_{\mathscr{X}_{P}})$ can be determined from the term $\mathcal{Y}_{P}^{(j)}(\vartheta_{\mathscr{X}_{P}})$ of \Eq\eqref{Eq:XYPSj}.
Note that $\vec{R}(\vartheta_{\mathscr{X}_{P}})$ is not present if the particle $P$ has no twisting angle (as e.g. for $\Pn$ and $\eta$) and disappears for $\tPp^\mu=0$.

\subsection{Pions}\label{subsec:AFpions}

\subsubsection{Masses}

We start with the asymptotic formulae for pion masses. 
In the generic form of \Eq\eqref{Eq:RXPs} the formula of the neutral pion exhibits two contributions: $R(M_{\Pn})=R(M_{\Pn},\Pn)+R(M_{\Pn},\Pc)$.
At one loop, $R(M_{\Pn},\Pn)$ and $R(M_{\Pn},\Pc)$ are given by the first two expressions of \Eq\eqref{Eq:RXPsPion} multiplied by $(-1/2)$ and replacing $X_P=X_\pi=M_P=M_\pi$.
The integrals $I^{(j)}(M_{\Pn},\Pn)$ resp. $I^{(j)}(M_{\Pn},\Pc)$ 
can be determined from the chiral representation of
$\mathcal{F}_{\Pn}$ and read
\begin{align}\label{Eq:IMPn}
  I^{(2)}(M_{\Pn},\Pn)&=-\frac{1}{2}I^{(2)}(M_{\Pn},\Pc)= B^0,\notag\\
  I^{(4)}(M_{\Pn},\Pn)&=-B^0\!\bigg[\frac{9}{2}-\frac{4}{3}\elb_1-\frac{8}{3}\elb_2+\frac{3}{2}\elb_3-2\elb_4\bigg]+B^2\!\bigg[8-\frac{8}{3}\elb_1-\frac{16}{3}\elb_2\bigg]+S^{(4)}(M_{\Pn},\Pn),\notag\\
  I^{(4)}(M_{\Pn},\Pc)&= B^0\!\bigg[\frac{13}{9}+\frac{8}{3}\elb_1-\elb_3-4\elb_4\bigg]+B^2\!\bigg[\frac{40}{9}-\frac{16}{3}\elb_2\bigg]+S^{(4)}(M_{\Pn},\Pc). 
\end{align}
The functions $B^{2k}=B^{2k}(\lambda_{\pi}\abs{\vec{n}})$ were defined in~\Ref\cite{CoDu:04} and can be evaluated analytically 
\begin{equation}\label{Eq:B2k}
 B^{2k}=\int_\R{}\d{y}{}y^{2k}\,\e^{-\lambda_{\pi}\abs{\vec{n}}\sqrt{1+y^2}}
       =\frac{\Gamma(k+1/2)}{\Gamma(3/2)}\left[\frac{2}{\lambda_{\pi}\abs{\vec{n}}}\right]^k\KB_{k+1}(\lambda_{\pi}\abs{\vec{n}}),
\end{equation}
where $\KB_{r}(x)$ are modified Bessel functions of the second kind.
The constants $\elb_j$ were originally introduced in \Ref\cite{GaLe:84} and depend logarithmically on the pion mass,
\begin{equation}\label{Eq:elbj}
 \elb_j=\elbPhys_j+2\log\bigg(\frac{\MpionPhys}{M_{\pi}}\bigg).
\end{equation}
Here, $\MpionPhys=\unit{0.140}{\GeV}$ and $\elbPhys_j$ are listed in \Tab\ref{Tab:LECnum}. 
The terms $S^{(4)}(M_{\Pn},\Pn)$, $S^{(4)}(M_{\Pn},\Pc)$ contain integrals
that can not be evaluated analytically but just numerically.  
Their explicit expressions are given in \Eq\eqref{Eq:S4MPn}.

The asymptotic formulae for masses of charged pions
exhibit four contributions, namely 
$R(M_{\Pc})=R(M_{\Pc},\Pn)+R(M_{\Pc},\Pc)+R_D(M_{\Pc},\Pn)+R_D(M_{\Pc},\Pc)$. 
At one loop, these contributions are given by the expressions of \Eq\eqref{Eq:RXPsPion} multiplied by $(-1/2)$ and replacing $X_P=X_\pi=M_P=M_\pi$ as well as $D_P=D_{\Pc}$, see \Eq\eqref{Eq:DPcDKcDKn}.
The integrals $I^{(j)}(M_{\Pc},\Pn)$ resp. $I^{(j)}(M_{\Pc},\Pc)$ are related to to those of~\Eq\eqref{Eq:IMPn} through
\begin{equation}\label{Eq:IMPc}
 \begin{aligned}
  I^{(j)}(M_{\Pc},\Pn)&=\frac{1}{2}\,I^{(j)}(M_{\Pn},\Pc),\\
  I^{(j)}(M_{\Pc},\Pc)&=I^{(j)}(M_{\Pn},\Pn)+\frac{1}{2}\,I^{(j)}(M_{\Pn},\Pc),\qquad j=2,4.
 \end{aligned} 
\end{equation}
Such relations follow from the representation of isospin components in ChPT, see~\Eq\eqref{Eq:IsoTPnTPc}.
The integrals $I_D^{(j)}(M_{\Pc},\Pn)$ resp. $I_D^{(j)}(M_{\Pc},\Pc)$ can be evaluated from the derivative of the chiral representation $\partial_y\mathcal{F}_{\Pc}$. 
For $j=2,4$, we find
\begin{equation}\label{Eq:IDMPc}
 \begin{aligned}
  I_D^{(2)}(M_{\Pc},\Pn)&=I_D^{(2)}(M_{\Pc},\Pc)=0,\\
  I_D^{(4)}(M_{\Pc},\Pn)&=B^2\bigg[\frac{40}{9}-\frac{16}{3}\elb_2\bigg]+S_D^{(4)}(M_{\Pc},\Pn),\\
  I_D^{(4)}(M_{\Pc},\Pc)&=B^2\bigg[\frac{184}{9}-\frac{16}{3}\elb_1-16\elb_2\bigg]+S_D^{(4)}(M_{\Pc},\Pc),
 \end{aligned}
\end{equation}
where $S_D^{(4)}(M_{\Pc},\Pn)$, $S_D^{(4)}(M_{\Pc},\Pc)$ are given in \Eq\eqref{Eq:SD4MPc}.

The asymptotic formulae for the extra terms $\Delta\tSPc^\mu$ can be expressed in the form of \Eq\eqref{Eq:RtwXPs} replacing $\vartheta_{\mathscr{X}_{P}}=\vartheta_{\Sigma_{\Pc}}$ and $X_P=X_\pi=M_P=M_\pi$.
The integrals of $\vec{R}(\tSPc)$ can be determined from the chiral representation of $\mathcal{G}_{\Pc}$ and read
\begin{equation}\label{Eq:IMvtwPc}
 \begin{aligned}
  I^{(2)}(\tSPc)&=\pm\big\{-4B^2\big\}, \\
  I^{(4)}(\tSPc)&=\pm\Big\{8B^2\big[1-\elb_4\big]+S^{(4)}(\tSPp)\Big\},
 \end{aligned}
\end{equation}
where $S^{(4)}(\tSPp)$ is given in \Eq\eqref{Eq:SM4vtwPc}.

\subsubsection{Decay constants}

The asymptotic formula for the decay constant of the neutral pion is given in \Eq\eqref{Eq:AFFPs}.
In section~\ref{subsubsec:ChWI} we have mentioned that $\mathcal{N}_{\Pn}$ is related to $\mathcal{F}_{\Pn}$, $\mathcal{C}_{\Pn}$ by virtue chiral Ward identities. 
We use this relation to determine the chiral representation of $\mathcal{N}_{\Pn}$ at one loop.
Expanding the chiral representation according to \Eq\eqref{Eq:XPs} the asymptotic formula exhibits two contributions: $R(F_{\Pn})=R(F_{\Pn},\Pn)+R(F_{\Pn},\Pc)$.
At one loop, $R(F_{\Pn},\Pn)$ and $R(F_{\Pn},\Pc)$ are given by the first two expressions of \Eq\eqref{Eq:RXPsPion} where $X_P=X_\pi=F_\pi$ and $M_P=M_\pi$.
The integrals $I^{(j)}(F_{\Pn},\Pn)$ resp. $I^{(j)}(F_{\Pn},\Pc)$ are related to those of the mass and of the pseudoscalar coupling constant through 
\begin{equation}\label{Eq:IFPn}
 \begin{aligned}
  I^{(j)}(F_{\Pn},\Pn)&=I^{(j)}(M_{\Pn},\Pn)+I^{(j)}(G_{\Pn},\Pn),\\
  I^{(j)}(F_{\Pn},\Pc)&=I^{(j)}(M_{\Pn},\Pc)+I^{(j)}(G_{\Pn},\Pc),\qquad j=2,4,
 \end{aligned}
\end{equation}
and can be explicitly evaluated from \Eqs(\ref{Eq:IMPn}, \ref{Eq:IGPn}).

The asymptotic formulae for the decay constants of charged pions exhibit four contributions:
$R(F_{\Pc})=R(F_{\Pc},\Pn)+R(F_{\Pc},\Pc)+R_D(F_{\Pc},\Pn)+R_D(F_{\Pc},\Pc)$.
Their expressions at one loop are given by \Eq\eqref{Eq:RXPsPion} where $X_P=X_\pi=F_\pi$, $M_P=M_\pi$ and $D_P=D_{\Pc}$.
The integrals $I^{(j)}(F_{\Pc},\Pn)$ resp. $I^{(j)}(F_{\Pc},\Pc)$ can be evaluated from the chiral representation of $\mathcal{N}_{\Pc}$ and are related to the integrals $I^{(j)}_{F,0}$ resp. $I^{(j)}_{F,\pm}$ of \Ref\cite{CoWenWu:10} through
\begin{equation}\label{Eq:IFPc}
 \begin{aligned}
  I^{(j)}(F_{\Pc},\Pn)&=I^{(j)}_{F,0},\\
  I^{(j)}(F_{\Pc},\Pc)&=I^{(j)}_{F,\pm},\qquad j=2,4.
 \end{aligned}
\end{equation}
The integrals $I_D^{(j)}(F_{\Pc},\Pn)$ resp. $I_D^{(j)}(F_{\Pc},\Pc)$ can
be determined from the derivative of the chiral representation
$\partial_y\mathcal{N}_{\Pc}$ and read 
\begin{equation}\label{Eq:IDFPc}
 \begin{aligned}
  I_D^{(2)}(F_{\Pc},\Pn)&=I_D^{(2)}(F_{\Pc},\Pc)=0,\\
  I_D^{(4)}(F_{\Pc},\Pn)&=B^2\bigg[\frac{40}{9}-\frac{16}{3}\elb_2\bigg]+S^{(4)}_D(F_{\Pc},\Pn),\\
  I_D^{(4)}(F_{\Pc},\Pc)&=B^2\bigg[\frac{184}{9}-\frac{16}{3}\elb_1-16\elb_2\bigg]+S^{(4)}_D(F_{\Pc},\Pc),
 \end{aligned}
\end{equation}
where $S_D^{(4)}(F_{\Pc},\Pn)$, $S_D^{(4)}(F_{\Pc},\Pc)$ are given in \Eq\eqref{Eq:SD4FPc}. 

The asymptotic formulae for the extra terms $\Delta\tAPc^\mu$ can be expressed in the form of \Eq\eqref{Eq:RtwXPs} replacing $\vartheta_{\mathscr{X}_{P}}=\vartheta_{\mathscr{A}_{\Pc}}$, $X_P=X_\pi=F_\pi$ and $M_P=M_\pi$.
The integrals of $\vec{R}(\tAPc)$ can be evaluated from the chiral representation of $\mathcal{H}_{\Pc}$ and read
\begin{equation}\label{Eq:IFvtwPc}
 \begin{aligned}
  I^{(2)}(\tAPc)&=\pm\big\{-4B^2\big\}, \\
  I^{(4)}(\tAPc)&=\pm\Big\{4B^2\big[1-\elb_4\big]+S^{(4)}(\tAPp)\Big\},
 \end{aligned}
\end{equation}
where $S^{(4)}(\tAPp)$ is given in \Eq\eqref{Eq:SF4vtwPp}.

\subsubsection{Pseudoscalar coupling constants}

The asymptotic formula for the pseudoscalar coupling constant of the neutral pion exhibits two contributions:
$R(G_{\Pn})=R(G_{\Pn},\Pn)+R(G_{\Pn},\Pc)$.
At one loop, $R(G_{\Pn},\Pn)$ and $R(G_{\Pn},\Pc)$ are given by the first two expressions of \Eq\eqref{Eq:RXPsPion} where $X_P=X_\pi=G_\pi$ and $M_P=M_\pi$.
The integrals $I^{(j)}(G_{\Pn},\Pn)$ resp. $I^{(j)}(G_{\Pn},\Pc)$ can be evaluated from the chiral representation of $\mathcal{C}_{\Pn}$ and read
\begin{equation}\label{Eq:IGPn}
 \begin{aligned}
  I^{(2)}(G_{\Pn},\Pn)&=-B^0,\\ 
  I^{(2)}(G_{\Pn},\Pc)&=0,\\
  I^{(4)}(G_{\Pn},\Pn)&=B^0\bigg[\frac{7}{2}-\frac{2}{3}\elb_1-\frac{4}{3}\elb_2+\frac{3}{2}\elb_3-3\elb_4\bigg]+S^{(4)}(G_{\Pn},\Pn),\\
  I^{(4)}(G_{\Pn},\Pc)&=-B^0\bigg[\frac{11}{9}+\frac{4}{3}\elb_1-\elb_3-2\elb_4\bigg]+S^{(4)}(G_{\Pn},\Pc).
 \end{aligned}
\end{equation}
The terms $S^{(4)}(G_{\Pn},\Pn)$, $S^{(4)}(G_{\Pn},\Pc)$ are explicitly given in \Eq\eqref{Eq:S4GPn}. 

The asymptotic formulae for pseudoscalar coupling constants of charged pions exhibit four contributions:
$R(G_{\Pc})\!=\!R(G_{\Pc},\Pn)\!+\!R(G_{\Pc},\Pc)\!+\!R_D(G_{\Pc},\Pn)\!+\!R_D(G_{\Pc},\Pc)$.
Their expressions at one loop are given by \Eq\eqref{Eq:RXPsPion} where $X_P=X_\pi=G_\pi$, $M_P=M_\pi$ and $D_P=D_{\Pc}$.
The integrals $I^{(j)}(G_{\Pc},\Pn)$ resp. $I^{(j)}(G_{\Pc},\Pc)$ can be evaluated from the chiral representation of $\mathcal{C}_{\Pc}$ and are related to those of \Eq\eqref{Eq:IGPn} by means of 
\begin{equation}\label{Eq:IGPc}
 \begin{aligned}
  I^{(j)}(G_{\Pc},\Pn)&=\frac{1}{2}\,I^{(j)}(G_{\Pn},\Pc),\\
  I^{(j)}(G_{\Pc},\Pc)&=I^{(j)}(G_{\Pn},\Pn)+\frac{1}{2}\,I^{(j)}(G_{\Pn},\Pc),\qquad j=2,4.
 \end{aligned}
\end{equation}
These relations follow from~\Eq\eqref{Eq:IsoCPnCPc} once the pole has been subtracted.
The integrals $I_D^{(j)}(G_{\Pc},\Pn)$ resp. $I_D^{(j)}(G_{\Pc},\Pc)$ can be determined from the derivative of the chiral representation $\partial_y\mathcal{C}_{\Pc}$ and are related to those of masses and of decay constants through
\begin{equation}
 \begin{aligned}
  I_D^{(j)}(G_{\Pc},\Pn)&=I_D^{(j)}(F_{\Pc},\Pn)-I_D^{(j)}(M_{\Pc},\Pn),\\ 
  I_D^{(j)}(G_{\Pc},\Pc)&=I_D^{(j)}(F_{\Pc},\Pc)-I_D^{(j)}(M_{\Pc},\Pc),\qquad j=2,4.
 \end{aligned}
\end{equation}
Inserting \Eqs(\ref{Eq:IDMPc}, \ref{Eq:IDFPc}) one can obtain the explicit expressions.

The asymptotic formulae for the extra terms $\Delta\tGPc^\mu$ can be expressed in the form of \Eq\eqref{Eq:RtwXPs} replacing $\vartheta_{\mathscr{X}_{P}}=\vartheta_{\mathscr{G}_{\Pc}}$, $X_P=X_\pi=G_\pi$ and $M_P=M_\pi$.
In this case, the equation must be divided by $\sMpion$.
The integrals of $\vec{R}(\tGPc)$ can be evaluated from the chiral representation of $\mathcal{K}_{\Pc}$ and are related to those of \Eqs(\ref{Eq:IMvtwPc}, \ref{Eq:IFvtwPc}) through
\begin{equation}
 I^{(j)}(\tGPc)=I^{(j)}(\tAPc)-I^{(j)}(\tSPc),\qquad j=2,4.
\end{equation}

\subsubsection{Scalar form factors at vanishing momentum transfer}\label{subsubsec:AFpionsSFF0}
 
In section~\ref{subsec:AFpionsFFS} we have presented asymptotic formulae for the matrix elements of the scalar form factor valid at vanishing momentum transfer.
The formula for the neutral pion is given in \Eq\eqref{Eq:AFGSPnPc0}.
If we insert the chiral representation of $\mathcal{F}_{\Pn}$ and expand according to \Eq\eqref{Eq:XPs}, the formula exhibits two contributions, $R(\GSPn)=R(\GSPn,\Pn)+R(\GSPn,\Pc)$.
At one loop, $R(\GSPn,\Pn)$ and $R(\GSPn,\Pc)$ are given by the first two expressions of \Eq\eqref{Eq:RXPsPion} multiplied by $(-1/2)$ and replacing $X_P=X_\pi=F_S(0)$ as well as $M_P=M_\pi$.
The integrals $I^{(j)}(\GSPn,\Pn)$ resp. $I^{(j)}(\GSPn,\Pc)$ can be evaluated from the chiral representation of $\mathcal{F}_{\Pn}$.
The evaluation involves terms with ${\textstyle\sqrt{1+y^2}}\mathcal{F}_{\Pn}$ and we make use of 
\begin{equation}
 \begin{split}
  \int_\R{}\d{y}{}y^{2k}\lambda_{\pi}\abs{\vec{n}}\sqrt{1+y^2}\,\e^{-\lambda_{\pi}\abs{\vec{n}}\sqrt{1+y^2}}=(2k-1)B^{2k-2}+(2k+1)B^{2k}, 
 \end{split}
\end{equation}
where $B^{2k}$ are defined in \Eq\eqref{Eq:B2k}.
The derivative $\partial_{M_\pi^2}\mathcal{F}_{\Pn}$ must be evaluated with care.
The operator $\partial_{M_\pi^2}$ acts on all quantities depending on the pion mass.
In particular, it acts on the decay constant~$F_\pi$ and on the constants $\elb_j$ of~\Eq\eqref{Eq:elbj}.
This leads to supplementary terms which must be integrated and added according to their chiral order to the corresponding integral.
Altogether, we find
\begin{align}\label{Eq:IGSPn}
  I^{(2)}(\GSPn,\Pn)&=-\frac{1}{2}\,I^{(2)}(\GSPn,\Pc)=B^{-2}+2B^0,  \notag\\ 
                   \notag\\
  I^{(4)}(\GSPn,\Pn)&=-B^{-2}\bigg[\frac{9}{2}-\frac{4}{3}\elb_1-\frac{8}{3}\elb_2+\frac{3}{2}\elb_3-2\elb_4\bigg]
                      -B^0\bigg[31-8\elb_1-16\elb_2+6\elb_3-4\elb_4\bigg],\notag\\
                    &\phantom{=}\ 
                      +\!B^2\bigg[32-\frac{16}{3}\elb_1-\frac{32}{3}\elb_2\bigg]+S^{(4)}(\GSPn,\Pn),\notag\\
                   \notag\\
  I^{(4)}(\GSPn,\Pc)&=B^{-2}\bigg[\frac{13}{9}+\frac{8}{3}\elb_1-\elb_3-4\elb_4\bigg]-B^0\bigg[2-\frac{32}{3}\elb_1-\frac{16}{3}\elb_2+4\elb_3+8\elb_4\bigg]\notag\\
                    &\phantom{=}\
                     +\!B^2\bigg[\frac{176}{9}-\frac{32}{3}\elb_2\bigg]+S^{(4)}(\GSPn,\Pc),
\end{align}
where $S^{(4)}(\GSPn,\Pn)$, $S^{(4)}(\GSPn,\Pc)$ are explicitly given in~\Eq\eqref{Eq:S4GSPn}. 

The asymptotic formulae for the matrix elements of the scalar form factor of charged pions are given in \Eq\eqref{Eq:AFGSPnPc0}.
If we insert the chiral representation of $\mathcal{F}_{\Pc}$, $\mathcal{G}_{\Pc}$ and expand according to~\Eq\eqref{Eq:XPs} the formulae exhibit five contributions which can be written as 
$R(\GSPc)=R(\GSPc,\Pn)+R(\GSPc,\Pc)+R_D(\GSPc,\Pn)+R_D(\GSPc,\Pc)+2\vtPc\vec{R}(\TPc)$.
At one loop, the first four contributions are given by the expressions of \Eq\eqref{Eq:RXPsPion} multiplied by $(-1/2)$ where $X_P=X_\pi=F_S(0)$, $M_P=M_\pi$ and $D_P=D_{\Pc}$.
The integrals $I^{(j)}(\GSPc,\Pn)$ resp. $I^{(j)}(\GSPc,\Pc)$ can be evaluated from the first group of terms in the square brackets of \Eq\eqref{Eq:AFGSPnPc0}.
Using the chiral representation of $\mathcal{F}_{\Pc}$ we find
\begin{equation}\label{Eq:IGSPc}
 \begin{aligned}
  I^{(2)}(\GSPc,\Pn)&=-[B^{-2}+2B^0], \\
                   \\
  I^{(2)}(\GSPc,\Pc)&=0,\\  
                   \\
  I^{(4)}(\GSPc,\Pn)&=B^{-2}\bigg[\frac{13}{18}+\frac{4}{3}\elb_1-\frac{1}{2}\elb_3-2\elb_4\bigg]-B^0\bigg[1-\frac{16}{3}\elb_1-\frac{8}{3}\elb_2+2\elb_3+4\elb_4\bigg]\\
                    &\phantom{=}\ 
                      +\!B^2\bigg[\frac{88}{9}-\frac{16}{3}\elb_2\bigg]+S^{(4)}(\GSPc,\Pn),\\  
                   \\
  I^{(4)}(\GSPc,\Pc)&=-B^{-2}\bigg[\frac{34}{9}-\frac{8}{3}\elb_1-\frac{8}{3}\elb_2+2\elb_3\bigg]-B^0\bigg[32-\frac{40}{3}\elb_1-\frac{56}{3}\elb_2+8\elb_3\bigg]\\
                    &{}\phantom{=}\
                      +\!B^2\bigg[\frac{376}{9}-\frac{16}{3}\elb_1-16\elb_2\bigg]+S^{(4)}(\GSPc,\Pc),
 \end{aligned}
\end{equation}
where $S^{(4)}(\GSPc,\Pn),S^{(4)}(\GSPc,\Pc)$ are given in \Eq\eqref{Eq:S4GSPcSD4GSPcSGS4vTwPc}. 
The integrals $I_D^{(j)}(\GSPc,\Pn)$ resp. $I_D^{(j)}(\GSPc,\Pc)$ can be determined from the second group of terms in the square brackets of \Eq\eqref{Eq:AFGSPnPc0} after having taken the derivative of the chiral representation $\partial_y\mathcal{F}_{\Pc}$. 
We find
\begin{align}\label{Eq:IDGSPc}
  I_D^{(2)}(\GSPc,\Pn)&=I_D^{(2)}(\GSPc,\Pc)=0,\notag\\
  I_D^{(4)}(\GSPc,\Pn)&=-B^0\!\bigg[\frac{40}{9}-\frac{16}{3}\elb_2\bigg]+B^2\!\bigg[\frac{32}{3}+(1-C_{\Pc})\bigg(\frac{40}{9}-\frac{16}{3}\elb_2\bigg)\bigg]+S_D^{(4)}(\GSPc,\Pn),\notag\\
  I_D^{(4)}(\GSPc,\Pc)&=-B^0\!\bigg[\frac{184}{9}-\frac{16}{3}\elb_1-16\elb_2\bigg]\notag\\
                      &{}\phantom{=}\
                        +\!B^2\!\bigg[\frac{128}{3}+(1-C_{\Pc})\bigg(\frac{184}{9}-\frac{16}{3}\elb_1-16\elb_2\bigg)\bigg]+S_D^{(4)}(\GSPc,\Pc),                   
\end{align}
where $C_{\Pc}=M_{\pi}/(M_{\pi}+D_{\Pc})$ with $D_{\Pc}$ defined by \Eq\eqref{Eq:DPcDKcDKn}.
The terms $S_D^{(4)}(\GSPc,\Pn)$, $S_D^{(4)}(\GSPc,\Pc)$ are explicitly given in \Eq\eqref{Eq:S4GSPcSD4GSPcSGS4vTwPc}. 

The fifth contribution $\vec{R}(\TPc)$ can be expressed in the form of \Eq\eqref{Eq:RtwXPs} replacing $\vartheta_{\mathscr{X}_{P}}=\TPc$, $X_P=X_\pi=F_S(0)$ and $M_P=M_\pi$.
In this case, the equation must be divided by $2\sMpion$.
The integrals of $\vec{R}(\TPc)$ can be evaluated from the last group of terms in the square brackets of \Eq\eqref{Eq:AFGSPnPc0}.
Using the chiral representation of $\mathcal{G}_{\Pc}$ we find
\begin{equation}\label{Eq:IGSvTwPc}
 \begin{aligned}
  I^{(2)}(\TPc)&=\pm\big\{4B^0\big\}, \\
  I^{(4)}(\TPc)&=\pm\Big\{-8B^0\big[1-\elb_4\big]+16B^2+S^{(4)}(\TPp)\Big\},
 \end{aligned}
\end{equation}
where $S^{(4)}(\TPp)$ can be found in \Eq\eqref{Eq:S4GSPcSD4GSPcSGS4vTwPc}.

\subsection{Kaons}\label{subsec:AFkaons}

\subsubsection{Masses}

The asymptotic formulae for the kaon masses are given in \Eq\eqref{Eq:AFMPs}. 
They differ in terms of the parameters $D_{\Kc}$, $D_{\Kn}$ and in terms of the amplitudes $\mathcal{F}_{\Kc}$, $\mathcal{F}_{\Kn}$.
In section~\ref{subsubsec:ChRepr1L} we have seen that the chiral representation of $\mathcal{F}_{\Kn}$ is equal to that of $\mathcal{F}_{\Km}$, see~\Eq\eqref{Eq:AmpFKnFKmGKnGKm}.
Hence, we can just consider the asymptotic formulae of charged kaons as the results of the neutral kaon can be obtained replacing $D_{\Kc}$ with $D_{\Kn}$.

We insert the chiral representation of $\mathcal{F}_{\Kc}$ and expand according to \Eq\eqref{Eq:XPs}.
The asymptotic formulae of charged kaons exhibit four contributions which can be written as 
$R(M_{\Kc})=R(M_{\Kc},\Pn)+R(M_{\Kc},\Pc)+R_D(M_{\Kc},\Pn)+R_D(M_{\Kc},\Pc)$.
At one loop, these contributions are given by \Eq\eqref{Eq:RXPsPion} multiplied by $(-1/2)$ where $X_P=X_\pi=M_P=M_K$ and $D_P=D_{\Kc}$. 
The integrals $I^{(j)}(M_{\Kc},\Pn)$ resp. $I^{(j)}(M_{\Kc},\Pc)$ can be evaluated from the chiral representation of $\mathcal{F}_{\Kc}$ and are related to the integrals $I^{(j)}_{M_{K}}$ of \Ref\cite{CoDuHae:05} through
\begin{equation}\label{Eq:IMKc}
 \begin{aligned}
  I^{(j)}(M_{\Kc},\Pn)&=\frac{x_{\pi K}^{-1/2}}{3}\,I^{(j)}_{M_{K}},\\
  I^{(j)}(M_{\Kc},\Pc)&=\frac{2}{3}x_{\pi K}^{-1/2}\,I^{(j)}_{M_{K}},\qquad j=2,4.
 \end{aligned}
\end{equation}
The integrals $I_D^{(j)}(M_{\Kc},\Pn)$ resp. $I_D^{(j)}(M_{\Kc},\Pc)$ can
be determined from the derivative of the chiral representation
$\partial_y\mathcal{F}_{\Kc}$ and read 
\begin{equation}\label{Eq:IDMKc}
 \begin{aligned}
  I_D^{(2)}(M_{\Kc},\Pn)&=I_D^{(2)}(M_{\Kc},\Pc)=0,\\
  I_D^{(4)}(M_{\Kc},\Pn)&=\frac{1}{2}\,I_D^{(4)}(M_{\Kc},\Pc)\\
                        &=x_{\pi K}B^2\bigg[-16N(4\Lr_2+\Lr_3)-\ell_{\pi}\,\frac{5x_{\pi K}}{1-x_{\pi K}}+\ell_\eta\,\frac{x_{\eta K}}{x_{\eta K}-1}\bigg.\\
                        &\qquad\quad\quad\ \ \,\bigg.        +\ell_K\bigg(\frac{5}{1-x_{\pi K}}-\frac{1}{x_{\eta K}-1}\bigg)\bigg]+S_D^{(4)}(M_{\Kc},\Pn).
 \end{aligned}
\end{equation}
Here, $N\!=\!(4\pi)^2$, $\ell_{P}\!=\!2\log(M_P/\mu)$, $x_{P Q}\!=\!M^2_P/M^2_Q$ for $P,Q\!=\!\pi,K,\eta$ and $\Lr_j$ are the renormalized LECs, see \cite{GaLe:85a}.
The expression of $S_D^{(4)}(M_{\Kc},\Pn)$ is given in \Eq\eqref{Eq:SD4MKc}.

The asymptotic formulae for the extra terms $\Delta\tSKc^\mu$ can be expressed in the form of \Eq\eqref{Eq:RtwXPs} replacing $\vartheta_{\mathscr{X}_{P}}=\vartheta_{\Sigma_{\Kc}}$ and $X_P=X_\pi=M_P=M_K$.
The integrals of $\vec{R}(\tSKc)$ can be evaluated from the chiral representation of $\mathcal{G}_{\Kc}$ and read
\begin{equation}\label{Eq:IMvtwKc}
 \begin{aligned}
  I^{(2)}(\tSKc)&=\pm\left\{-2\,x_{\pi K}^{1/2}\,B^2\right\}, \\
  I^{(4)}(\tSKc)&=\pm\Bigg\{x_{\pi K}^{1/2}\,B^2 \bigg[-16Nx_{\pi K}\Lr_5-\frac{\ell_\pi}{2}\,\frac{5x_{\pi K}^2}{1-x_{\pi K}}-\ell_\eta\frac{x_{\eta K}}{4}\bigg(3-\frac{1+x_{\pi K}}{x_{\eta K}-1}\bigg)\bigg.\Bigg.\\
                 &\qquad\qquad\qquad\:\:\:\bigg.\Bigg.   -\frac{\ell_K}{4}\bigg(7-\frac{10}{1-x_{\pi K}}+\frac{1+x_{\pi K}}{x_{\eta K}-1}\bigg)\bigg]+S^{(4)}(\tSKp)\Bigg\},
 \end{aligned}
\end{equation}
where $S^{(4)}(\tSKp)$ is given in \Eq\eqref{Eq:S4MvtwKp}. 
Note that the asymptotic formula for $\Delta\tSKn^\mu$ differs from that for $\Delta\tSKm^\mu$ only in terms of $\mathcal{G}_{\Kn}$, $\mathcal{G}_{\Km}$.
As the chiral representation of $\mathcal{G}_{\Kn}$ coincides with that of $\mathcal{G}_{\Km}$, the asymptotic formulae are equal in this case, and we can use $\vec{R}(\vartheta_{\Sigma_{\Km}})$ to estimate $\Delta\tSKn^\mu$.

\subsubsection{Decay constants}
      
The asymptotic formulae for the decay constants of charged kaons exhibit four contributions:
$R(F_{\Kc})=R(F_{\Kc},\Pn)+R(F_{\Kc},\Pc)+R_D(F_{\Kc},\Pn)+R_D(F_{\Kc},\Pc)$.
Their expressions at one loop are given by \Eq\eqref{Eq:RXPsPion} where $X_P=F_K$, $X_\pi=F_\pi$, $M_P=M_K$ and $D_P=D_{\Kc}$.
The integrals $I^{(j)}(F_{\Kc},\Pn)$ resp. $I^{(j)}(F_{\Kc},\Pc)$ can be evaluated from the chiral representation of $\mathcal{N}_{\Kc}$ and 
are related to the integrals $I^{(j)}_{F_K}$ of~\Ref\cite{CoDuHae:05} by virtue of\footnote{%
    In \Ref\cite{CoDuHae:05} there are two missprints: 
    in \Eq(57) the term $2\ell_\pi(x_{\pi\eta}-\frac{9}{4})$ should read $2(\ell_\eta-\ell_\pi\,\frac{9}{4})$ and 
    in \Eq(79) the factor of $S^{0,4}_{\eta K}$ should read $\frac{3}{32}(1+x_{\pi K})(5-2x_{\pi K}-3x_{\eta K})$ instead of $\frac{3}{32}(5-3x_{\eta K})(1-x^2_{\pi K})$. 
}
\begin{equation}\label{Eq:IFKc}
 \begin{aligned}
  I^{(j)}(F_{\Kc},\Pn)&=\frac{I^{(j)}_{F_K}}{3},\\
  I^{(j)}(F_{\Kc},\Pc)&=\frac{2}{3}\,I^{(j)}_{F_K},\qquad j=2,4.
 \end{aligned}
\end{equation}
The integrals $I_D^{(j)}(F_{\Kc},\Pn)$ resp. $I_D^{(j)}(F_{\Kc},\Pc)$ can
be determined from the derivative of the chiral representation
$\partial_y\mathcal{N}_{\Kc}$ and read 
\begin{equation}\label{Eq:IDFKc}
 \begin{aligned}
  I_D^{(2)}(F_{\Kc},\Pn)&=I_D^{(2)}(F_{\Kc},\Pc)=0,\\
  I_D^{(4)}(F_{\Kc},\Pn)&=\frac{1}{2}\,I_D^{(4)}(F_{\Kc},\Pc)\\
                        &=x_{\pi K}\,B^2\bigg[-16N(4\Lr_2+\Lr_3)-\ell_{\pi}\,\frac{5x_{\pi K}}{1-x_{\pi K}}+\ell_\eta\,\frac{x_{\eta K}}{x_{\eta K}-1}\bigg.\\
                        &\qquad\qquad\ \ \ \bigg.+\ell_K\bigg(\frac{5}{1-x_{\pi K}}-\frac{1}{x_{\eta K}-1}\bigg)\bigg]+S_D^{(4)}(F_{\Kc},\Pn),
 \end{aligned}
\end{equation}
where $S_D^{(4)}(F_{\Kc},\Pn)$ is given in \Eq\eqref{Eq:SD4FKc}. 

The asymptotic formulae for the extra terms $\Delta\tAKc^\mu$ can be expressed in the form of \Eq\eqref{Eq:RtwXPs} replacing $\vartheta_{\mathscr{X}_{P}}=\vartheta_{\mathscr{A}_{\Kc}}$, $X_P=F_K$, $X_\pi=F_\pi$ and $M_P=M_K$.
The integrals of $\vec{R}(\tAKc)$ can be evaluated from the chiral representation of $\mathcal{H}_{\Kc}$ and read
\begin{equation}\label{Eq:IFvtwKc}
 \begin{aligned}
  I^{(2)}(\tAKc)&=\pm\Big\{-2\,x_{\pi K}^{1/2}\,B^2\Big\}, \\
  I^{(4)}(\tAKc)&=\pm\Bigg\{x_{\pi K}^{1/2}\,B^2\bigg[-8Nx_{\pi K}\Lr_5-\frac{\ell_\pi}{4}\,\frac{5x_{\pi K}^2}{1-x_{\pi K}}-\ell_\eta\frac{x_{\eta K}}{8}\bigg(3-\frac{1+x_{\pi K}}{x_{\eta K}-1}\bigg)\bigg.\Bigg.\\
                &\qquad\qquad\qquad\:\ \bigg.\Bigg.  -\frac{\ell_K}{8}\bigg(7-\frac{10}{1-x_{\pi K}}+\frac{1+x_{\pi K}}{x_{\eta K}-1}\bigg)\bigg]+S^{(4)}(\tAKp)\Bigg\}, 
 \end{aligned}
\end{equation}
where $S^{(4)}(\tAKp)$ is given in \Eq\eqref{Eq:S4FvtwKp}.

\subsubsection{Pseudoscalar coupling constants}

The asymptotic formulae for the pseudoscalar coupling constants of charged kaons are given in \Eq\eqref{Eq:AFGPs}.
In section~\ref{subsubsec:ChWI} we have mentioned that $\mathcal{C}_{\Kc}$ is related to $\mathcal{F}_{\Kc}$, $\mathcal{N}_{\Kc}$ by virtue of chiral Ward identities. 
We use that relation to determine the chiral representation of $\mathcal{C}_{\Kc}$ at one loop.
Expanding the representation as in \Eq\eqref{Eq:XPs} the asymptotic formula exhibits four contributions: 
$R(G_{\Kc})=R(G_{\Kc},\Pn)+R(G_{\Kc},\Pc)+R_D(G_{\Kc},\Pn)+R_D(G_{\Kc},\Pc)$.
Their expressions at one loop are given by \Eq\eqref{Eq:RXPsPion} where $X_P=G_K$, $X_\pi=G_\pi$, $M_P=M_K$ and $D_P=D_{\Kc}$.
The integrals can be evaluated from those of~\Eqs(\ref{Eq:IMKc}, \ref{Eq:IFKc}) by means of
\begin{equation}
 \begin{aligned}
  I^{(j)}(G_{\Kc},\Pn)  &=\overset{\circ}{x}_{\pi K}x_{\pi K}^{-1}\bigg[I^{(j)}(F_{\Kc},\Pn)-\frac{F_K}{F_\pi}I^{(j)}(M_{\Kc},\Pn)\bigg],\\ 
  I^{(j)}(G_{\Kc},\Pc)  &=\overset{\circ}{x}_{\pi K}x_{\pi K}^{-1}\bigg[I^{(j)}(F_{\Kc},\Pc)-\frac{F_K}{F_\pi}I^{(j)}(M_{\Kc},\Pc)\bigg], 
 \end{aligned}
\end{equation}
where $j=2,4$ and $\overset{\circ}{x}_{\pi K}=\sMlopi/\sMloK$.
Here, similar relations hold also for $I_D^{(j)}(G_{\Kc},\Pn)$, $I_D^{(j)}(G_{\Kc},\Pc)$. 
In the evaluation, some attention must be paid: because of the prefactors (i.e. $\overset{\circ}{x}_{\pi K}x_{\pi K}^{-1}$ and $F_K/F_\pi$) the integrals with $j=2$ on the right-hand side generate terms that contribute to the integrals with $j=4$ on the left-hand side.

The asymptotic formulae for the extra terms $\Delta\tGKc^\mu$ are presented in \Eq\eqref{Eq:AFGDtPLPsc}. 
As mentioned in section~\ref{subsubsec:ChWI} the chiral representation of $\mathcal{K}_{\Kc}$ is related to that of $\mathcal{G}_{\Kc}$, $\mathcal{H}_{\Kc}$ by means of chiral Ward identities. 
We use that relation to determine the chiral representation of $\mathcal{K}_{\Kc}$ at one loop.
The asymptotic formulae for $\Delta\tGKc^\mu$ can be then expressed in the form of \Eq\eqref{Eq:RtwXPs} replacing $\vartheta_{\mathscr{X}_{P}}=\vartheta_{\mathscr{G}_{\Kc}}$, $X_P=G_K$, $X_\pi=G_\pi$ and $M_P=M_K$.
In this case, the equation must be divided by $\sMkaon$.
The integrals of $\vec{R}(\tGKc)$ can be evaluated from those of \Eqs(\ref{Eq:IMvtwKc}, \ref{Eq:IFvtwKc}) by means of 
\begin{equation}
  I^{(j)}(\tGKc)=\overset{\circ}{x}_{\pi K}x_{\pi K}^{-1}\bigg[I^{(j)}(\tAKc)-\frac{F_K}{F_\pi}I^{(j)}(\tSKc)\bigg],\qquad j=2,4. 
\end{equation}
Note that because of the prefactors (viz. $\overset{\circ}{x}_{\pi K}x_{\pi K}^{-1}$ and $F_K/F_\pi$) the integrals $I^{(2)}(\tSKc)$ and $I^{(2)}(\tAKc)$ generate terms that contribute to $I^{(4)}(\tGKc)$.

\subsection{Eta meson}\label{subsec:AFeta}

In the generic form of \Eq\eqref{Eq:RXPs} the asymptotic formula for the mass of the eta meson exhibits two contributions,
$R(M_\eta)=R(M_\eta,\Pn)+R(M_\eta,\Pc)$.
At one loop, $R(M_\eta,\Pn)$ and $R(M_\eta,\Pc)$ are given by the first two expressions of \Eq\eqref{Eq:RXPsPion} multiplied by $(-1/2)$ and replacing $X_P=X_\pi$ as well as $M_P=M_\eta$.
The integrals $I^{(j)}(M_\eta,\Pn)$ resp. $I^{(j)}(M_\eta,\Pc)$ can be evaluated from the chiral representation of $\mathcal{F}_{\eta}$ and are related to the integrals $I^{(j)}_{M_\eta}$ of \Ref\cite{CoDuHae:05} by virtue of 
\begin{equation}
 \begin{aligned}
  I^{(j)}(M_\eta,\Pn)&=\frac{x_{\pi\eta}^{-1/2}}{3}\,I^{(j)}_{M_\eta},\\
  I^{(j)}(M_\eta,\Pc)&=\frac{2}{3}x_{\pi\eta}^{-1/2}\,I^{(j)}_{M_\eta},\qquad j=2,4.
 \end{aligned}
\end{equation}

\section{Numerical results}\label{sec:NumericalResults}

\subsection{Numerical set-up}\label{subsec:SetUp}

\begin{table}[tbp]
 \centering
 \hspace{\stretch{2}}
 \subfloat[LECs of $\mathrm{SU}(2)$ ChPT.][$\mathrm{SU}(2)$ ChPT LECs.]{%
 \setlength{\fboxrule}{1.5pt}
 \framebox[\width]{ 
 \begin{tabular}{CRc}
  j & \elbPhys_j\phantom{-} & \Ref             \\
  \midrule
  1 & -0.36\pm0.59          & \cite{CoGaLe:01} \\
  2 &  4.31\pm0.11          & \cite{CoGaLe:01} \\
  3 &  3.05\pm0.99          & \cite{FLAG:14}   \\
  4 &  4.02\pm0.28          & \cite{FLAG:14}   \\
 \end{tabular}
 }
 }
 \hspace{\stretch{1}}
 \subfloat[LECs of $\mathrm{SU}(3)$ ChPT.][$\mathrm{SU}(3)$ ChPT LECs.]{%
 \setlength{\fboxrule}{1.5pt}
 \framebox[\width]{ 
 \begin{tabular}{CRc}
  j & \Lr_j(\mu)\cdot10^3\phantom{a} & \Ref            \\
  \midrule
  1 &  0.88\pm0.09                   & \cite{BiJe:11}  \\
  2 &  0.61\pm0.20                   & \cite{BiJe:11}  \\
  3 & -3.04\pm0.43                   & \cite{BiJe:11}  \\
  4 &  0.04\pm0.14                   & \cite{MILC:09A} \\
  5 &  0.84\pm0.38                   & \cite{MILC:09A} \\
  6 &  0.07\pm0.10                   & \cite{MILC:09A} \\
  7 & -0.16\pm0.15                   &                 \\
  8 &  0.36\pm0.09                   & \cite{MILC:09A} \\
 \end{tabular}
 }
 }
 \hspace{\stretch{2}}
  \caption[Values of LECs.]{Values of LECs used in the numerical analysis.  
                           Here, the renormalization scale is $\mu=\unit{0.770}{\GeV}$. 
                           The central value of $\Lr_7$ has been determined evaluating the expression of $M_\eta$ at the physical point (see text)
                           whereas its uncertainty 
                           is taken from \Tab5 (column ``All'') of \Ref\cite{BiJe:11}.}
 \label{Tab:LECnum}
\end{table}
\begin{figure}[bp]
   \centering
   \includegraphics[width=.6\columnwidth]{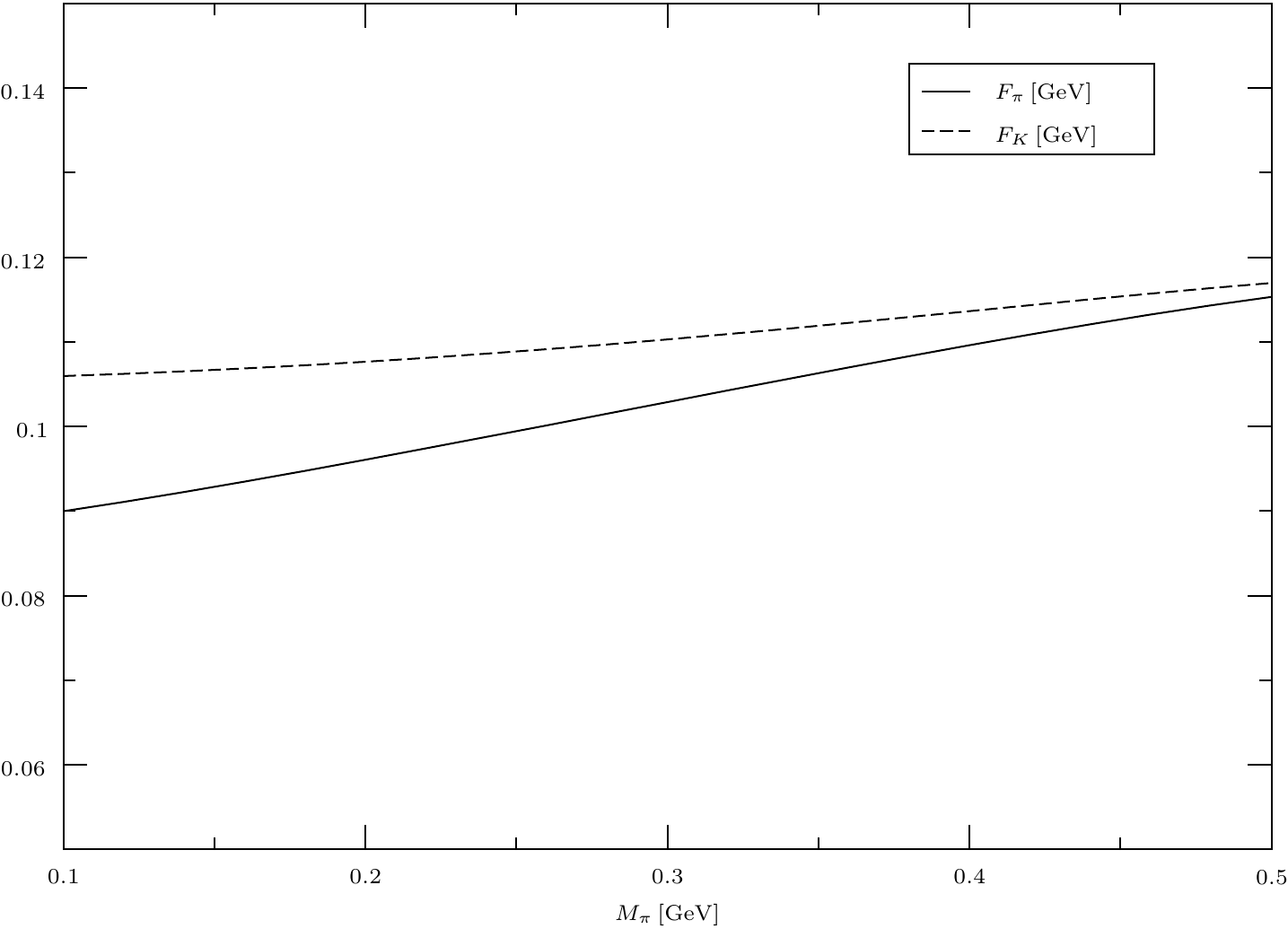}
   \caption[Pion mass dependence of decay constants in infinite volume.]{%
            Pion mass dependence of $F_\pi$, $F_K$ in infinite volume. }
  \label{Fig:Fp_Fk_vs_Mp}
\end{figure}

We adopt the numerical set-up of \Refs\cite{CoDu:04,CoDuHae:05} and express
the quantities in infinite volume appearing in the formulae (i.e. $F_\pi$,
$F_K$, $M_K$, $M_\eta$) as functions of $M_{\pi}$. 
For $F_\pi$ we use the expression at NNLO obtained with $\mathrm{SU}(2)$
ChPT 
while for $F_K$, $M_K$, $M_\eta$ we use expressions at NLO of
$\mathrm{SU}(3)$ ChPT. 
In any cases, the values of the relevant LECs are
summarized in \Tab\ref{Tab:LECnum}. 
If available, these values are taken from results of lattice simulations
with $N_f=2+1$ dynamical flavors. For LECs of $\mathrm{SU}(2)$ ChPT we take
the averages of the FLAG working group~\cite{FLAG:14} which are obtained
from \Refs\cite{Bor:12,RBCUKQCD:12,NPLQCD:11,MILC:10A}. 
For LECs of $\mathrm{SU}(3)$ ChPT we take the results of
\Ref\cite{MILC:09A} as recommended by FLAG~\cite{FLAG:14}. The remaining
values come from phenomenology~\cite{CoGaLe:01,BiJe:11} or have been
determined evaluating the expressions of $F_\pi$, $F_K$, $M_K$, $M_\eta$ at
the physical point (see later).

\subsubsection{Pion mass dependence in infinite volume}

In infinite volume, the expression of the pion decay constant at NNLO reads
\begin{equation}\label{Eq:FpionNNLO}
F_\pi=F\bigg\{1+\xi\elb_4+\xi^2\!\bigg[\frac{5}{4}\ell_\pi^2+\ell_\pi\bigg(\frac{7}{6}\elb_1+\frac{4}{3}\elb_2+\frac{\elb_3}{2}-\frac{\elb_4}{2}+2\!\bigg)
       +\frac{\elb_3}{2}\elb_4-\frac{\elb_1}{12}-\frac{\elb_2}{3}-\frac{13}{192}+r_F(\mu)\bigg]\bigg\}.
\end{equation}
This expression is obtained with $\mathrm{SU}(2)$ ChPT, see
\Ref\cite{BiCoTa:98}. 
Here, $\xi=M_\pi^2/(4\pi F)^2$ and $\ell_\pi=2\log(M_{\pi}/\mu)$.
The constants $\elb_j$ are given in \Eq\eqref{Eq:elbj} and the values of $\elbPhys_j$ are in \Tab\ref{Tab:LECnum}. 
The parameter $r_F(\mu)$ is a combination of LECs at NNLO.  
In \Ref\cite{CoDu:04} it was estimated as $r_F(\mu)=0\pm3$.

We can determine $F$ numerically, evaluating \Eq\eqref{Eq:FpionNNLO} at the physical point. 
Taking $M_{\pi}=\MPcPhys$ and inverting the expression, we find $F=\unit{(86.6\pm0.4)}{\MeV}$.
This value agrees with the result of~\Ref\cite{CoDu:04} and provides the ratio $F_\pi/F=(1.065\pm0.006)$ which in turn agrees with the FLAG average obtained from simulations with $N_f=2+1$ dynamical flavors, see \Refs\cite{FLAG:14,MILC:09A,NPLQCD:11,Bor:12}.

In \Fig\ref{Fig:Fp_Fk_vs_Mp} we represent the pion mass dependence of decay
constants. We observe that the dependence of $F_\pi$ is rather mild.
As in \Ref\cite{CoDu:04}, we conclude that this dependence is too mild to
violate the condition of \Eq\eqref{Eq:ChPTapp}. Thus, we can rely on ChPT
and apply the formulae of sections~\ref{sec:ChPTFV} and
\ref{sec:ApplicationAF} to estimate finite-volume corrections. 

At NLO the expressions of $M_K^2$, $M_\eta^2$, $F_K$
obtained with $\mathrm{SU}(3)$ ChPT read~\cite{CoDuHae:05}: 
\begin{subequations}\label{Eq:MkMeFkNLO}
 \begin{align}
  M_K^2  &=\sMloK\!+\!\frac{1}{F_\pi^2}\bigg\{M_\pi^4\bigg[\frac{1}{4N}\Big(\frac{\elllo_\eta}{3}-\ell_\pi\Big)-2k_1\bigg]
         \!\!+\!\!B_0 m_s (M_\pi^2+B_0m_s)\Big[8(k_1+2k_2)\!+\!\frac{4}{9N}\elllo_\eta\Big]\bigg\},\label{Eq:MkaonNLO}\\
 M_\eta^2&=\sMloeta+\frac{1}{F_\pi^2}\bigg\{M_\pi^4\bigg[\frac{16}{9}(-k_1+2k_3)-\frac{1}{3N}\Big(2\ell_\pi-\elllo_K\Big)\bigg]\bigg.\notag\\
         &\qquad\qquad\qquad\bigg.+M_\pi^2 B_0m_s\bigg[\frac{64}{9}(k_1+3k_2-2k_3)+\frac{4}{3N}\bigg(\elllo_K-\frac{2}{9}\elllo_\eta\bigg)\bigg]\bigg.\label{Eq:MetaNLO}\\
         &\qquad\qquad\qquad\bigg.+(B_0m_s)^2\bigg[\frac{128}{9}(k_1+3k_2/2+k_3)+\frac{4}{3N}\bigg(\elllo_K-\frac{8}{9}\elllo_\eta\bigg)\bigg]\bigg\},\notag\\
         \notag\\
     F_K &=F_\pi+\frac{1}{F_\pi^2}\bigg[4(M_K^2-M_\pi^2)\Lr_5+\frac{1}{N}\bigg(\frac{5}{8}M_\pi^2\ell_\pi-\frac{1}{4}M_K^2\ell_\eta-\frac{3}{8}M_\eta^2\ell_\eta\bigg)\bigg].\label{Eq:FkaonNLO}
 \end{align}
\end{subequations}
In these expressions, $k_1=2\Lr_8-\Lr_5$, $k_2=2\Lr_6-\Lr_4$, $k_3=3\Lr_7+\Lr_8$ and $\elllo_P=2\log(\MloPs/\mu)$ resp. $\ell_P=2\log(M_P/\mu)$ for $P=\pi,K,\eta$. 
The decay constant $F_\pi$ is expressed as in \Eq\eqref{Eq:FpionNNLO}.
Here, the circled masses are not just at leading order but are of hybrid nature,
\begin{equation}\label{Eq:sMloKeta}
 \sMloK=\frac{1}{2}(M_\pi^2+2B_0m_s) \qquad\text{and}\qquad \sMloeta=\frac{1}{3}(M_\pi^2+4B_0m_s).
\end{equation}
The part containing $M_\pi^2$ is at NLO while the part containing $B_0m_s$ is at leading order.
This is unavoidable if we want to study the dependence on the pion mass of
$M_K$, $M_\eta$, $F_K$ in ChPT.  
In practice, we use the first two expressions~(\ref{Eq:MkaonNLO},
\ref{Eq:MetaNLO}) to determine $B_0m_s$, $\Lr_7$ and the third
one~\eqref{Eq:FkaonNLO} to check the numerical results. 

Taking the LECs from \Tab\ref{Tab:LECnum} we evaluate \Eq\eqref{Eq:MkaonNLO} at $M_{\pi}=\MPcPhys$.  
Requiring $M_K=\MKcPhys$ we find $B_0m_s=\unit{0.241}{\GeV\squared}$.  The
uncertainty on this value is of order $\unit{\power{10}{-5}}{\GeV\squared}$
and may be neglected.  Then, we evaluate \Eq\eqref{Eq:MetaNLO} at $M_{\pi}=\MPcPhys$ and require
$M_\eta=\MetaPhys$.  We find $\Lr_7=-0.16\cdot\power{10}{-3}$.  This value
agrees with the result presented in \Tab5 (column ``All'') of
\Ref\cite{BiJe:11}. As uncertainty estimated we take the one given in
\cite{BiJe:11}, yielding $\Lr_7=(-0.16\pm0.15)\cdot\power{10}{-3}$. 
To check our numerical results on $B_0m_s$,
$\Lr_7$ we insert the expressions of $M_K^2$, $M_\eta^2$ in
\Eq\eqref{Eq:FkaonNLO} and evaluate $F_K$ at $M_{\pi}=\MPcPhys$. We find
$F_K=\unit{(0.106\pm0.004)}{\GeV}$ which agrees with the PDG result~\cite{PDG:14}
and with the FLAG average obtained from simulations with
$N_f=2+1$ dynamical flavors, see
\Refs\cite{FLAG:14,RBCUKQCD:12,MILC:10,HPQCDUKQCD:07}.

We stress that $F_\pi$ is expressed here with $\mathrm{SU}(2)$ ChPT even
in the $\mathrm{SU}(3)$ expressions of $M_K^2$, $M_\eta^2$, $F_K$.  
This choice was already made in \Ref\cite{CoDuHae:05} and for
$m_s=m_s^{\textup{phys}}$ it exactly reproduces what one would get in the
$\mathrm{SU}(3)$ framework. As lattice simulations are usually performed at
$m_s\approx m_s^{\textup{phys}}$ we expect that such choice remains a valid 
approximation also in our numerical analysis.

In \Fig\ref{Fig:MP_xiP_vs_Mp} we show the pion mass dependence of
$M_K$, $M_\eta$. We observe that the dependence on $M_{\pi}$ is mild.
The same holds for $F_K$ as one sees from \Fig\ref{Fig:Fp_Fk_vs_Mp}.
Note that for $M_{\pi}\approx\unit{0.500}{\GeV}$ we have
$M_K\approx\unit{0.610}{\GeV}$ and $M_\eta\approx\unit{0.640}{\GeV}$. 
In that case, the values of $M_{\pi}$, $M_K$, $M_\eta$ are all similar. 
If we consider $\xi_{P}=M_P^2/(4\pi F_\pi)^2$ for $P=\pi,K,\eta$ we expect
that the expansion parameter stays small for all pion masses in
$\mathopen{[}\unit{0.1}{\GeV}, \unit{0.5}{\GeV}\mathclose{]}$. 
This is confirmed by \Fig\ref{Fig:MP_xiP_vs_Mp} where the pion mass
dependence of $\xi_{P}$ is represented graphically.
In the numerical analysis we will use the expansion parameter as
in \Fig\ref{Fig:MP_xiP_vs_Mp} and consider $\xi_{P}$ as
exact, ignoring its uncertainty. 

{
\begin{figure}
   \centering
   \includegraphics[width=.48\columnwidth]{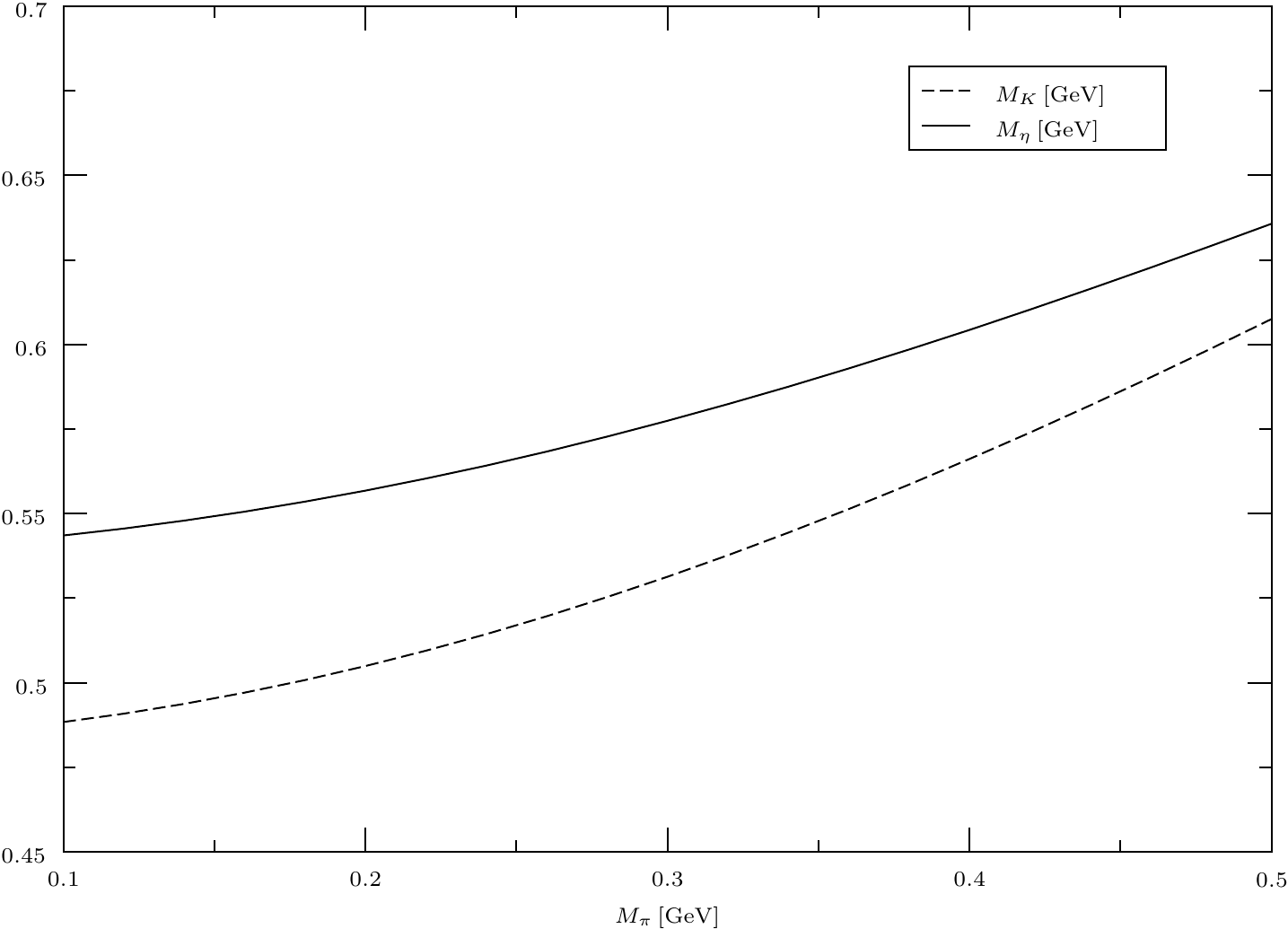}
   \includegraphics[width=.48\columnwidth]{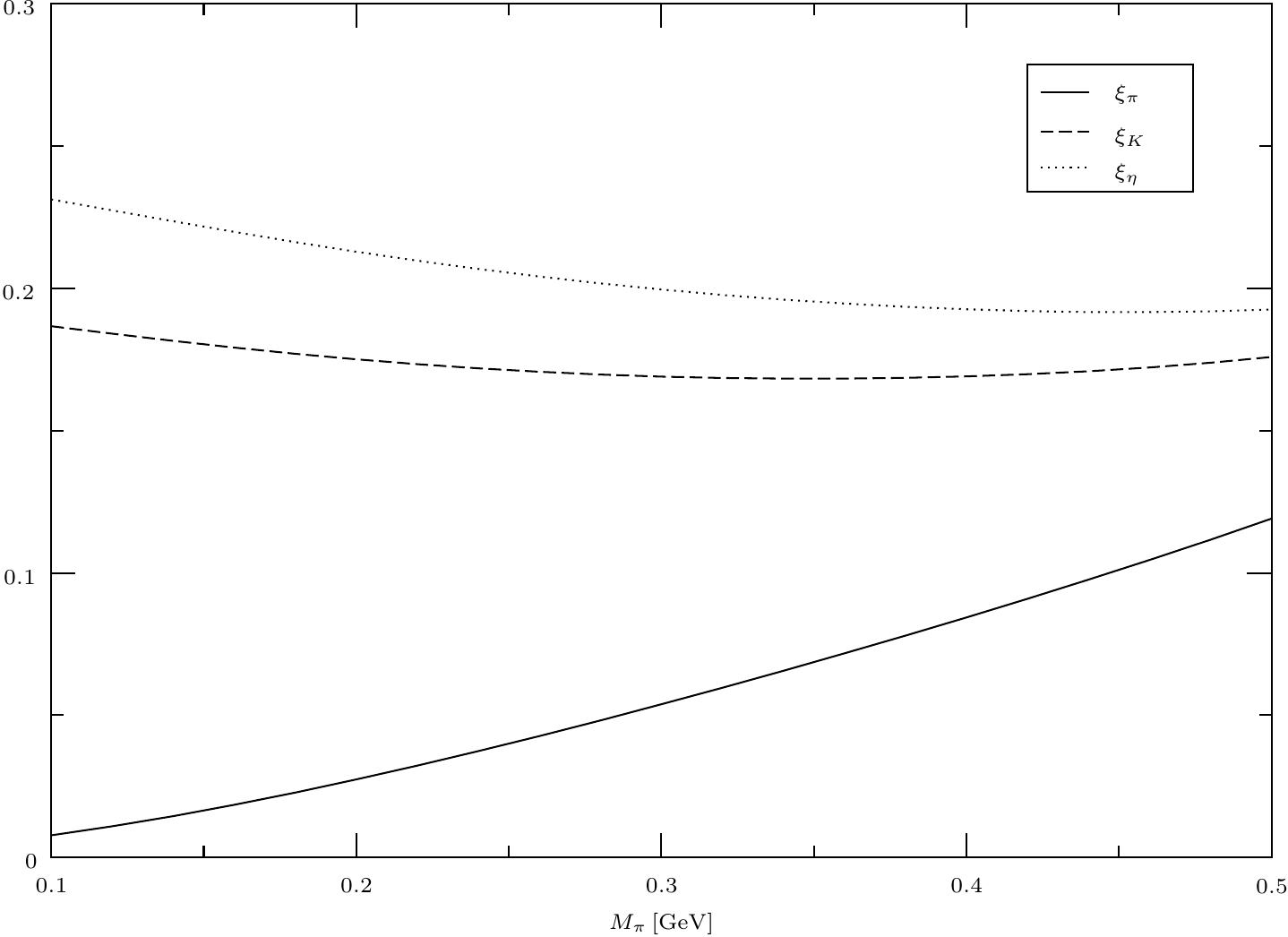}
   \caption[Pion mass dependence of expansion parameters.]{%
            Pion mass dependence of $M_K$, $M_\eta$ and of the expansion
            parameter $\xi_{P}$ in infinite volume.}
  \label{Fig:MP_xiP_vs_Mp}
\end{figure}
}

\subsubsection{Twisting angles and multiplicity}

To perform the numerical analysis we use the following configuration of twisting angles 
\begin{equation}\label{Eq:TW}
 \vec{\vartheta}_{\Pp}=\vec{\vartheta}_{K^{+}}=\frac{1}{L}\begin{pmatrix}\theta\\ 0\\ 0\end{pmatrix}\qquad\text{resp.}\qquad\vec{\vartheta}_{\Kn}=\vec{0}.
\end{equation}
In principle, the angle $\theta$ can be arbitrarily chosen.
Since we rely on ChPT, we require $\theta/L\ll4\pi F_\pi$.
If we consider $L\gg\unit{1}{\fm}$ --- as requested by
\Eq\eqref{Eq:ChPTapp} --- such condition is certainly satisfied for
$\theta\in\mathopen{[}0,2\pi\mathclose{]}$. 

The configuration~\eqref{Eq:TW} allows us to simplify a lot of
calculations. As twisting angles are aligned to one specific axis, we may
use the formula, 
\begin{equation}
  \sum_{%
	\begin{subarray}{c}
	  \vec{n}\in\Z{3}\\
	  \abs{\vec{n}}\neq0
	  \end{subarray}
	  }%
	  f(\,\abs{\vec{n}}\,)\ \e^{\i L\vec{n}\vtPp}=\sum_{n=1}^\infty f(\,\sqrt{n}\,)
	  \sum_{n_1=-{\scriptscriptstyle\myfloor{\sqrt{n}}}}^{{\scriptscriptstyle\myfloor{\sqrt{n}}}}m(n,n_1)\,\e^{\i n_1\theta},
\end{equation}
to rewrite the three sums over integer vectors as a nested sum over $n\in\N{}$. 
In general, this speeds up the numerical evaluation by a factor 15.
The notation $\myfloor{\,.\,}$ indicates the floor function 
and $m(n,n_1)$ is the multiplicity (i.e. the number of possibilities to
construct a vector $\vec{n}\in\Z{3}$ with $n=\abs{\vec{n}}^2$, having
previously fixed the value of the first component to $n_1\in\Z{}$). 
As an illustration, we list in \Tab\ref{Tab:mnn1} the values of the
multiplicity for $n\leq10$. 

In the following we present our numerical results. We plot the dependences
of the corrections on $M_{\pi}$ and $\theta$.  
The pion mass dependence is plotted for different values of $L$ and
$\theta$. Lines of different colors refer to different values of $L$
whereas lines of different hatchings refer to different values of $\theta$.
The dark (resp. light) yellow areas refer to the region $M_{\pi} L<2$ for
$\theta=0$ (resp. $\theta=\pi/3$). The dependence on the angle $\theta$ is plotted for
different values of $M_{\pi}$, at fixed $L=\unit{2}{\fm}$. Lines of
different hatchings refer to different values of $M_{\pi}$. For
$L=\unit{2}{\fm}$, the region $M_{\pi} L<2$ begins for masses smaller than 
$M_{\pi}=\unit{0.197}{\GeV}$. We remind the reader that in that region
formulae obtained in the $p$-regime are not reliable any more. 
\begin{table}[tbp]
 \centering
 \setlength{\fboxrule}{1.5pt}
 \framebox[\width]{
 \begin{tabular}{RR|RRRR} 
  m(n,n_1)   & n_1  & 0   & \pm1 & \pm2 & \pm3 \\
  n          &      &     &      &      &      \\
 \hline
  1          &      & 4   & 1    & 0    & 0    \\
  2          &      & 4   & 4    & 0    & 0    \\
  3          &      & 0   & 4    & 0    & 0    \\
  4          &      & 4   & 0    & 1    & 0    \\
  5          &      & 8   & 4    & 4    & 0    \\
  6          &      & 0   & 8    & 4    & 0    \\
  7          &      & 0   & 0    & 0    & 0    \\
  8          &      & 4   & 0    & 4    & 0    \\
  9          &      & 4   & 4    & 8    & 1    \\
 10          &      & 8   & 4    & 0    & 4    \\
 \end{tabular}
 }
 \caption{Multiplicity of vectors $\vec{n}\in\Z{3}{}$ with $n:=\abs{\vec{n}}^2\leq10$.}
 \label{Tab:mnn1}
\end{table}

\subsection{Finite-volume corrections at NLO}\label{subsec:FVCatNLO}

The corrections of masses, decay constants and vector form factors were
numerically evaluated at NLO in \Refs\cite{JiTi:07,BiRe:14}.  
As we have reproduced their plots we refrain from presenting results for these
quantities at this order. We just focus on pseudoscalar coupling constants
and scalar form factors for which no numerical analysis was published,
yet. 

\begin{figure}[tbp]
\centering
  \subfloat[][Pion mass dependence of $-\delta G_{\Pn}$.]
  {\label{Fig:m_deltaGPn_vs_Mp}{
   \includegraphics[width=.95\columnwidth]{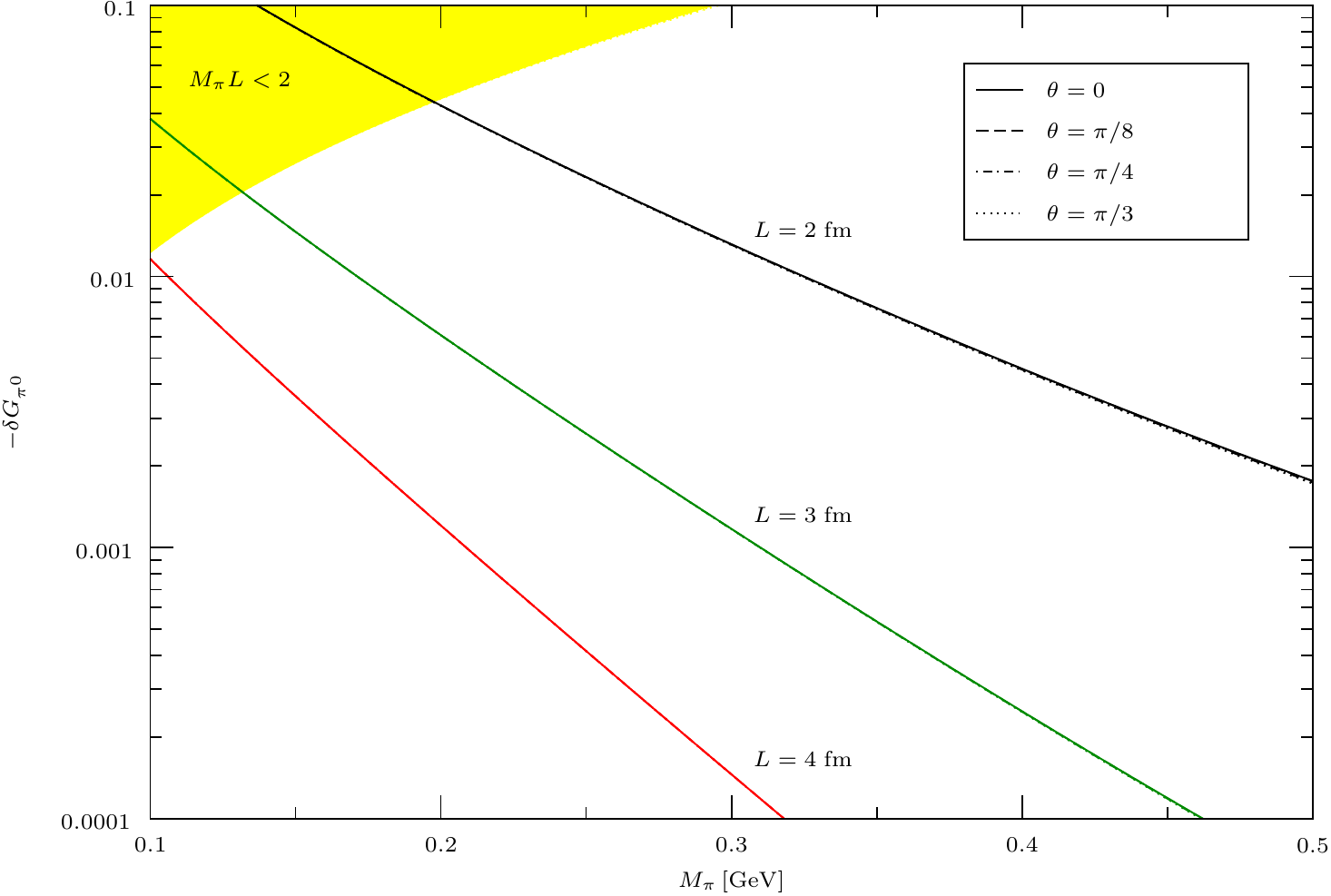}
  }}\\
  \subfloat[][Pion mass dependence of $-\delta G_\eta$.]
  {\label{Fig:m_deltaGeta_vs_Mp}{
   \includegraphics[width=.95\columnwidth]{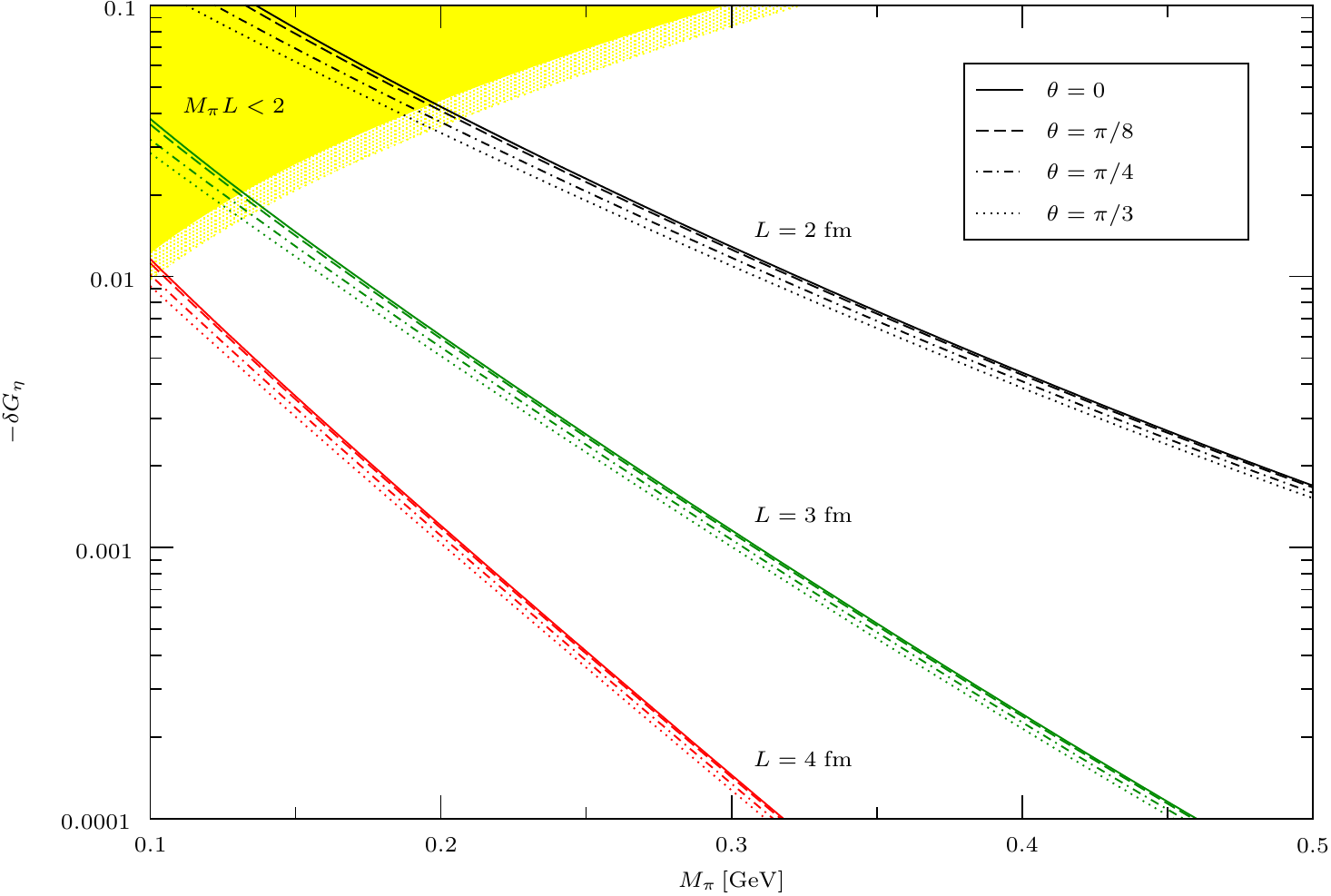}
  }}
 \caption[Corrections of pseudoscalar coupling constants at NLO.]{%
          Corrections of pseudoscalar coupling constants at NLO.
         }
\end{figure}

We start with the pseudoscalar coupling constants. 
At NLO the corrections of the pseudoscalar coupling constants exhibit all a dependence on twisting angles. 
In section~\ref{subsec:MassesDecayCouplConst} we have seen that the corrections of the pseudoscalar coupling constants are given by the sum of the corrections of masses and decay constants, see \Eq\eqref{Eq:dGPndGPc}.
As $\delta M_{\Pc}$, $\delta M_{K^{\pm}}$ and $\delta F_{\Pc}$, $\delta F_{K^{\pm}}$ were numerically evaluated in \Ref\cite{BiRe:14} one can use those results to determine $\delta G_{\Pc}$, $\delta G_{K^{\pm}}$.
Moreover, $\delta G_{\Kn}$ can be determined from $\delta G_{K^{\pm}}$ substituting $\tKn\leftrightarrow\tKp$, cfr. \Eq\eqref{Eq:CouplConst}.
Thus, we just study the dependence of $\delta G_{\Pn}$, $\delta G_\eta$ on the pion mass. 

In \Fig\ref{Fig:m_deltaGPn_vs_Mp} (resp. \ref{Fig:m_deltaGeta_vs_Mp}) we represent the pion mass dependence of $-\delta G_{\Pn}$ (resp. $-\delta G_\eta$).
The logarithmic graphs illustrate the exponential decay $\Ord{\e^{-\lambda_{\pi}}}$ of the corrections.
The line slopes depend on $L$ while the $y$-intercepts on $\theta$.
In \Fig\ref{Fig:m_deltaGPn_vs_Mp} the lines are so close that they overlap in the graph: $\delta G_{\Pn}$ is practically insensitive to $\theta$.
On the contrary, 
$\delta G_\eta$ is noticeably sensitive to $\theta$. 
In general, the corrections are negative and for $\theta\in\set{0,\,\pi/8,\,\pi/4,\,\pi/3}$ their absolute values decrease with the angle.
Note that 
$\delta G_{\Pn}$, $\delta G_\eta$ 
reach the percentage level before entering yellow areas and are thus comparable with the statistical precision of lattice simulations.

\begin{figure}[tbp]
\centering
  \subfloat[][Pion mass dependence of $\delta\GSPn$ at $q^2=0$.]
  {\label{Fig:dGSPn0_vs_Mp}{
   \includegraphics[width=.95\columnwidth]{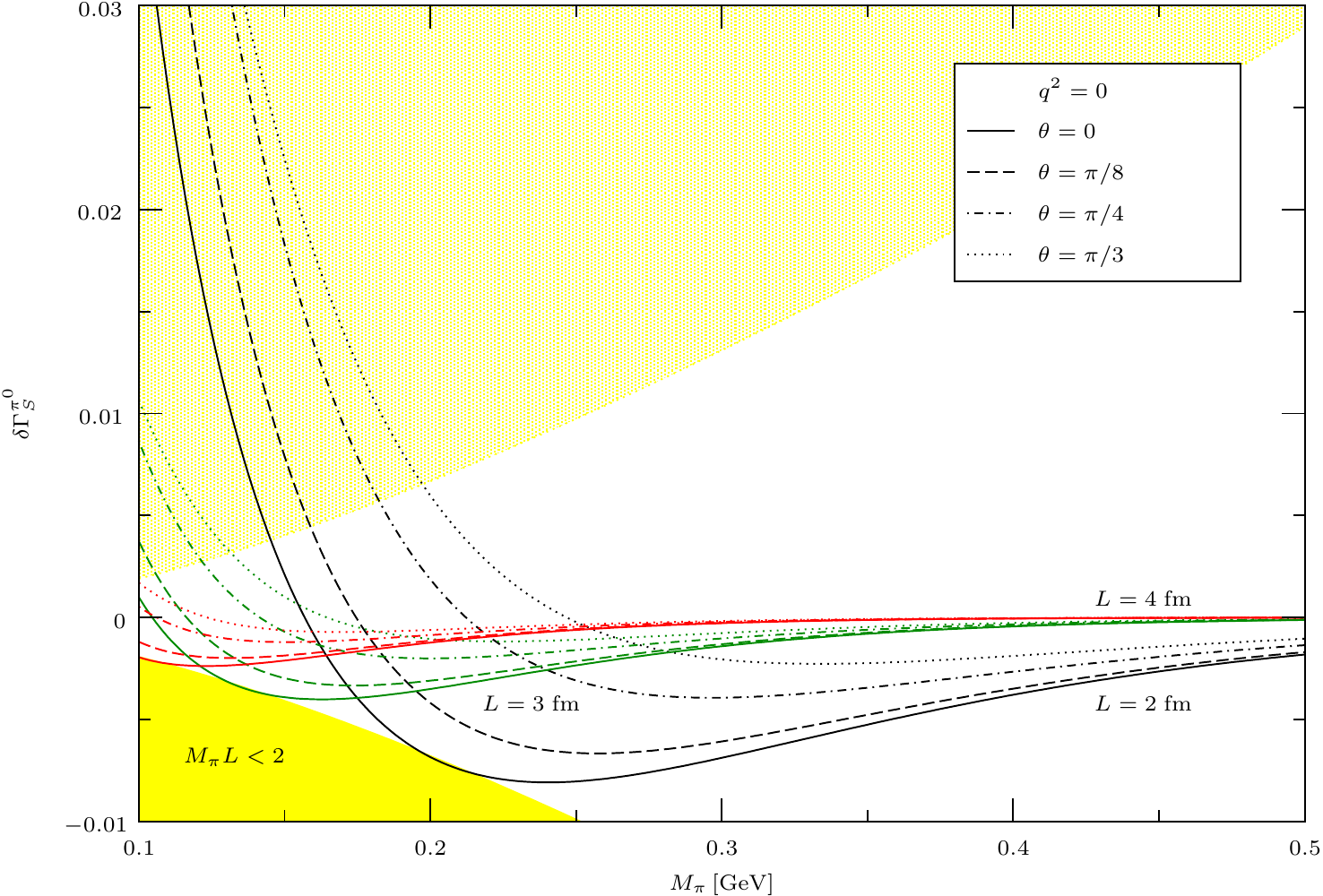}
  }}\\
  \subfloat[][Pion mass dependence of $\delta\GSPp$ at $q^2=0$.]
  {\label{Fig:dGSPp0_vs_Mp}{
   \includegraphics[width=.95\columnwidth]{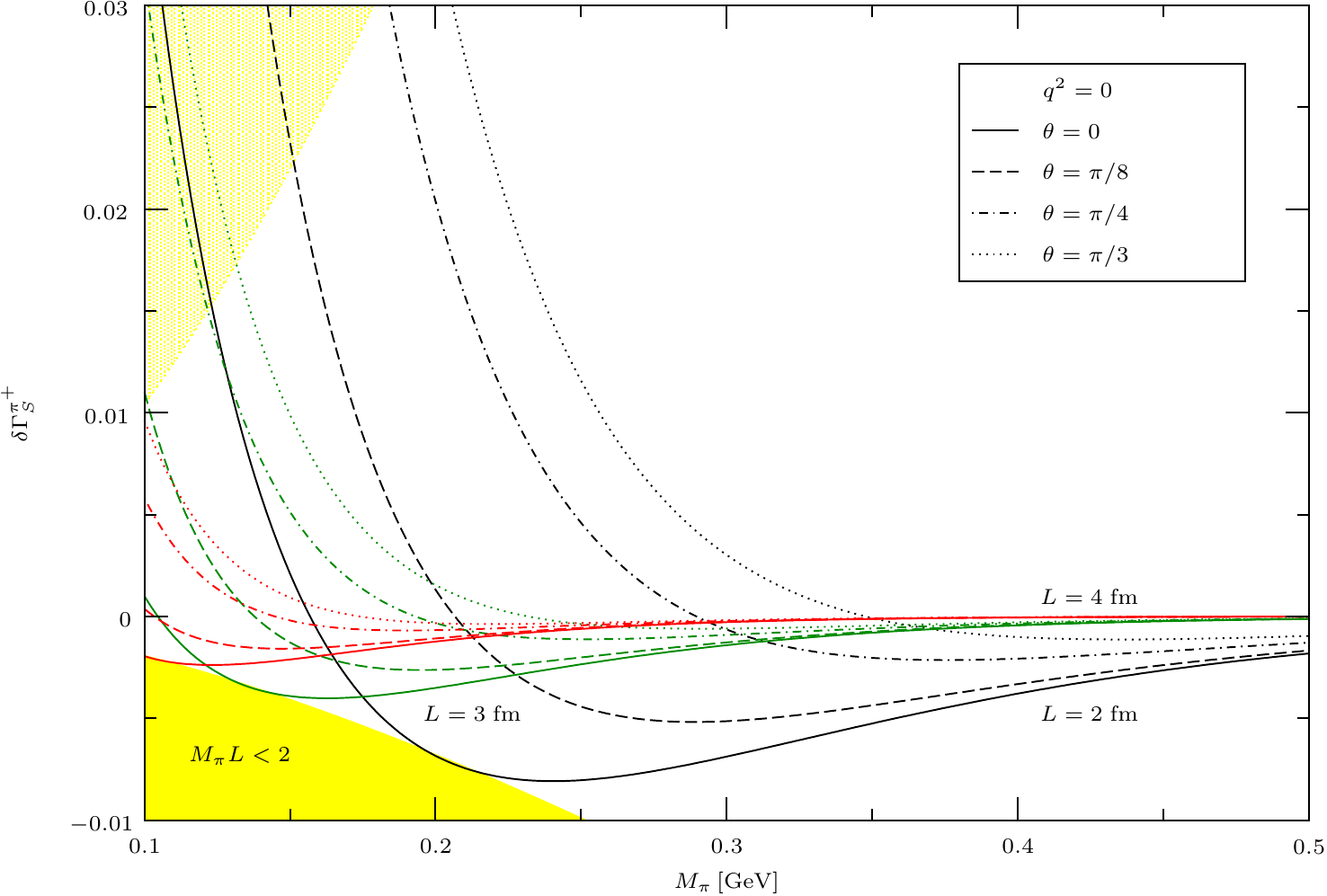}
  }}
 \caption[Corrections of the matrix elements of the scalar form factor at NLO.]{%
          Corrections of the matrix elements of the scalar form factor at NLO.
         }
 \label{Fig:dGSPn0}
\end{figure}

In \Fig\ref{Fig:dGSPn0_vs_Mp} and \ref{Fig:dGSPp0_vs_Mp} we represent the
corrections of the matrix elements of the pion scalar form factor at
vanishing momentum transfer. The pion mass dependence is represented in
linear graphs where yellow areas refer to the region $M_{\pi} L<2$. The
solid lines ($\theta=0$) reach that region when --- starting from the
right-hand side of the figure --- they first touch the dark yellow area. 
The dotted lines ($\theta=\pi/3$) reach this region when they enter in the
light yellow area. 
We observe that the corrections decay exponentially as
$\Ord{\e^{-\lambda_{\pi}}}$ and are mainly negative. 
They may turn positive depending on the pion mass. 
In the region where the $p$-regime is guaranteed, the corrections are
less than the percentage level and thus negligible. 

\begin{figure}[tbp]
\centering
  \subfloat[][Pion mass dependence of $-\delta\GSPn$ at $q^2=q^2_{\textup{min}}$.]
  {\label{Fig:m_dGSPn_w_qqPnL_vs_Mp}{
   \includegraphics[width=.95\columnwidth]{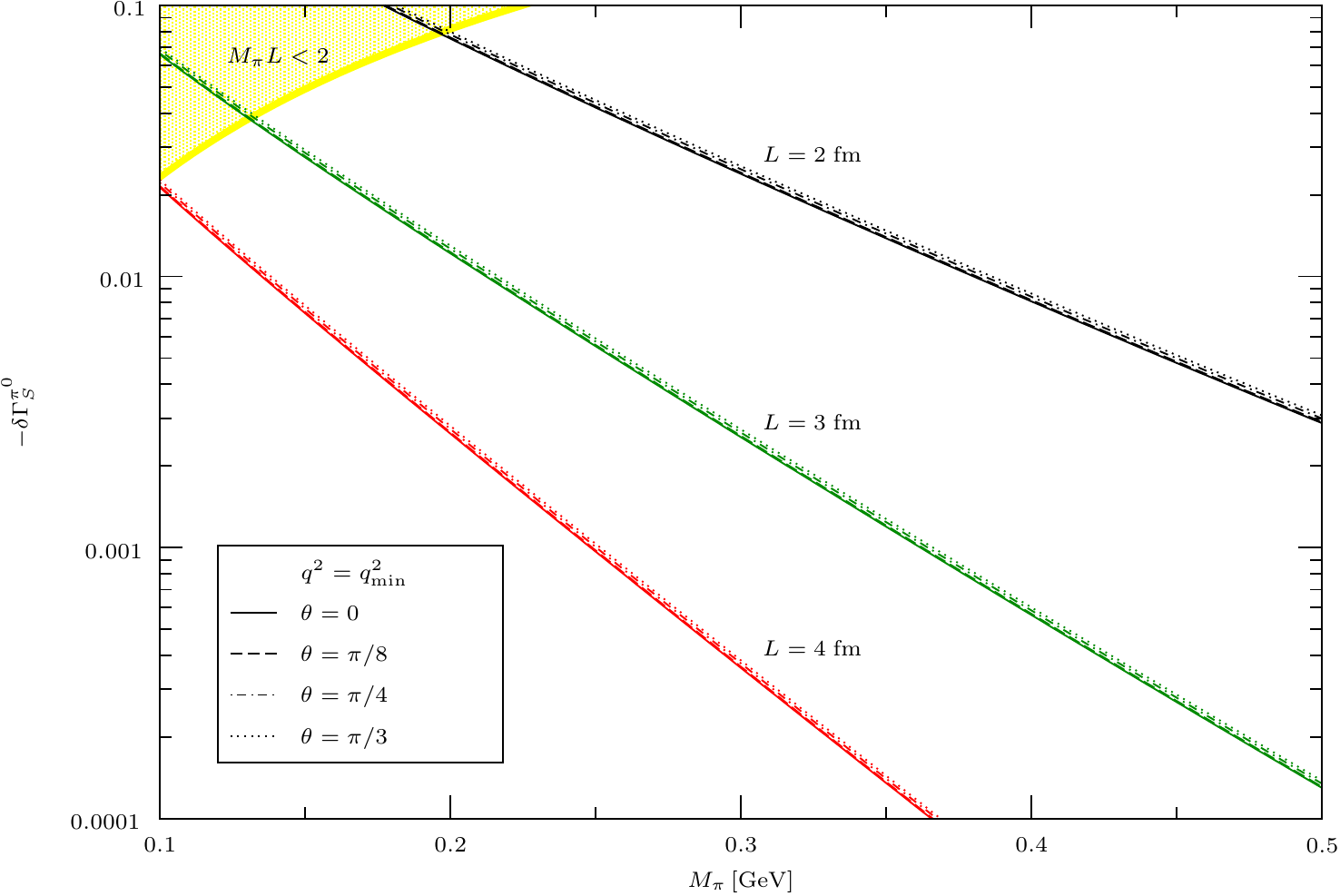}
  }}\\
   \subfloat[][Pion mass dependence of $-\delta\GSPp$ at $q^2=q^2_{\textup{min}}$.]
  {\label{Fig:m_dGSPp_w_qqPpL_vs_Mp}{
   \includegraphics[width=.95\columnwidth]{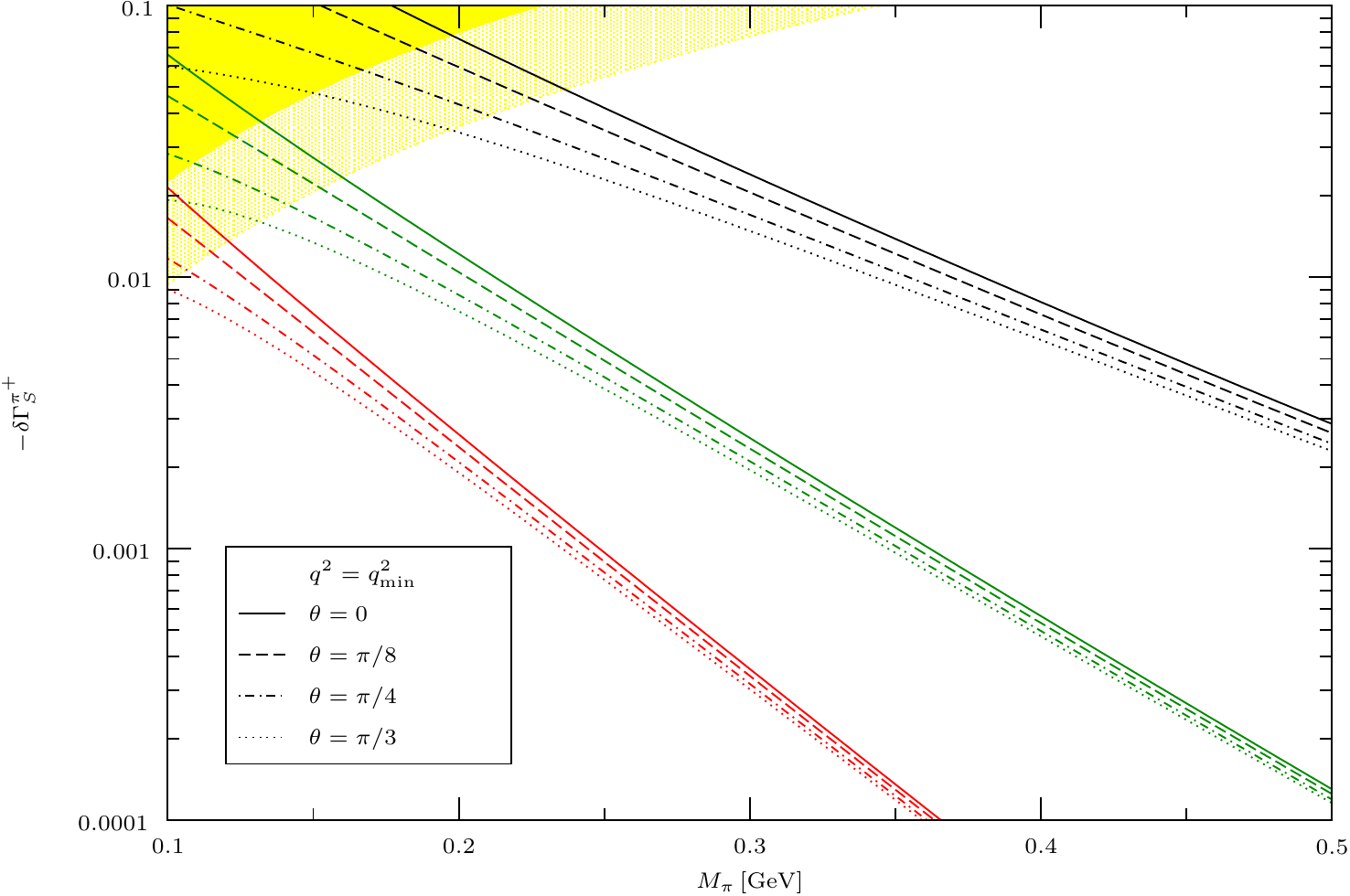}
  }}
 \caption[Corrections of the matrix element of the scalar form factor at NLO.]{%
          Corrections of the matrix element of the scalar form factor at NLO.
         }
 \label{Fig:dGSPn_w_qqPnL}
\end{figure}

To estimate the corrections at a non-zero momentum transfer we consider the incoming pion at rest (i.e. $\vec{p}=\vec{0}$) and the outgoing pion moving along the first axis carrying the first non-zero momentum (i.e. $\abs{\vpp}=2\pi/L$).
This kinematics provides the momentum transfer, 
\begin{equation}
 \qmin^\mu=\begin{pmatrix}q^0\\\vec{q}\end{pmatrix}\qquad\text{with}\qquad\vec{q}=\frac{2\pi}{L}\begin{pmatrix}1\\0\\0\end{pmatrix}.
\end{equation}
The zeroth component corresponds to the energy transfer among external pions,
\begin{equation}
\begin{aligned}
  q^0&=E'_{\Pn}(L)-M_{\Pn}(L)& &\text{for external}\ \Pn,\\
  q^0&=E'_{\Pp}(L)-E_{\Pp}(L)& &\text{for external}\ \Pp.
\end{aligned}
\end{equation}
For the configuration~\eqref{Eq:TW} the pion energies take the following forms,
\begin{equation}
\begin{aligned}
  E'_{\Pn}(L)&=\sqrt{M_{\Pn}^2(L)+(2\pi)^2/L^2},\\
   E_{\Pp}(L)&=\sqrt{M_{\Pc}^2(L)+(\theta/L)^2+2\,\theta\Delta\vartheta/L+\Ord{\xi^2_{\pi}}},\\
  E'_{\Pp}(L)&=\sqrt{M_{\Pc}^2(L)+(2\pi+\theta)^2/L^2+2\,(2\pi+\theta)\Delta\vartheta/L+\Ord{\xi^2_{\pi}}}.
\end{aligned}
\end{equation}
Here, $M_{\Pn}(L)$, $M_{\Pc}(L)$ are the pion masses in finite volume and
for $\mathrm{SU}(2)$ ChPT, their corrections are given from \Eq\eqref{Eq:Masses}
discarding the contributions of kaons and the eta meson.  
The quantity,  
\begin{equation}
 \Delta\vartheta=\xi_\pi\sum_{n=1}^\infty \frac{4\i}{Ln}\KB_2(\lambda_{\pi}\sqrt{n})
	         \sum_{n_1=-{\scriptscriptstyle\myfloor{\sqrt{n}}}}^{{\scriptscriptstyle\myfloor{\sqrt{n}}}}m(n,n_1)\,n_1\e^{\i n_1\theta},
\end{equation}
corresponds to the first component of the extra term~\eqref{Eq:DtL} in the
configuration~\eqref{Eq:TW}. We refrain from presenting numerical
results of the square radius as according to \Ref\cite{Hall:12} it is more effective to correct the
matrix elements of form factors and from those, extract the square radii. 

In \Fig\ref{Fig:m_dGSPn_w_qqPnL_vs_Mp}
(resp. \ref{Fig:m_dGSPp_w_qqPpL_vs_Mp}) we represent the pion mass
dependence of $-\delta\GSPn$ (resp. $-\delta\GSPp$) at
$q^2=q_{\textup{min}}^2$. The logarithmic graphs illustrate the exponential
decay $\Ord{\e^{-\lambda_{\pi}}}$ of the corrections. The corrections are
mainly negative. For $\theta\in\set{0,\,\pi/8,\,\pi/4,\,\pi/3}$ the
absolute value of $\delta\GSPn$ increases (resp. that of $\delta\GSPp$
decreases) with the angle. Note that the corrections reach the percentage
level before entering yellow areas and they should be subtracted when the
scalar form factor is extracted from lattice data.

\subsection{Finite-volume corrections beyond NLO}\label{subsec:FVCbeyondNLO}

\subsubsection{Masses and extra terms of self-energies}

\begin{figure}[tbp]
\centering
  \subfloat[][Pion mass dependence of $R(M_{\Pn})$.]
  {\label{Fig:RMPn_vs_Mp}{
   \includegraphics[width=.95\columnwidth]{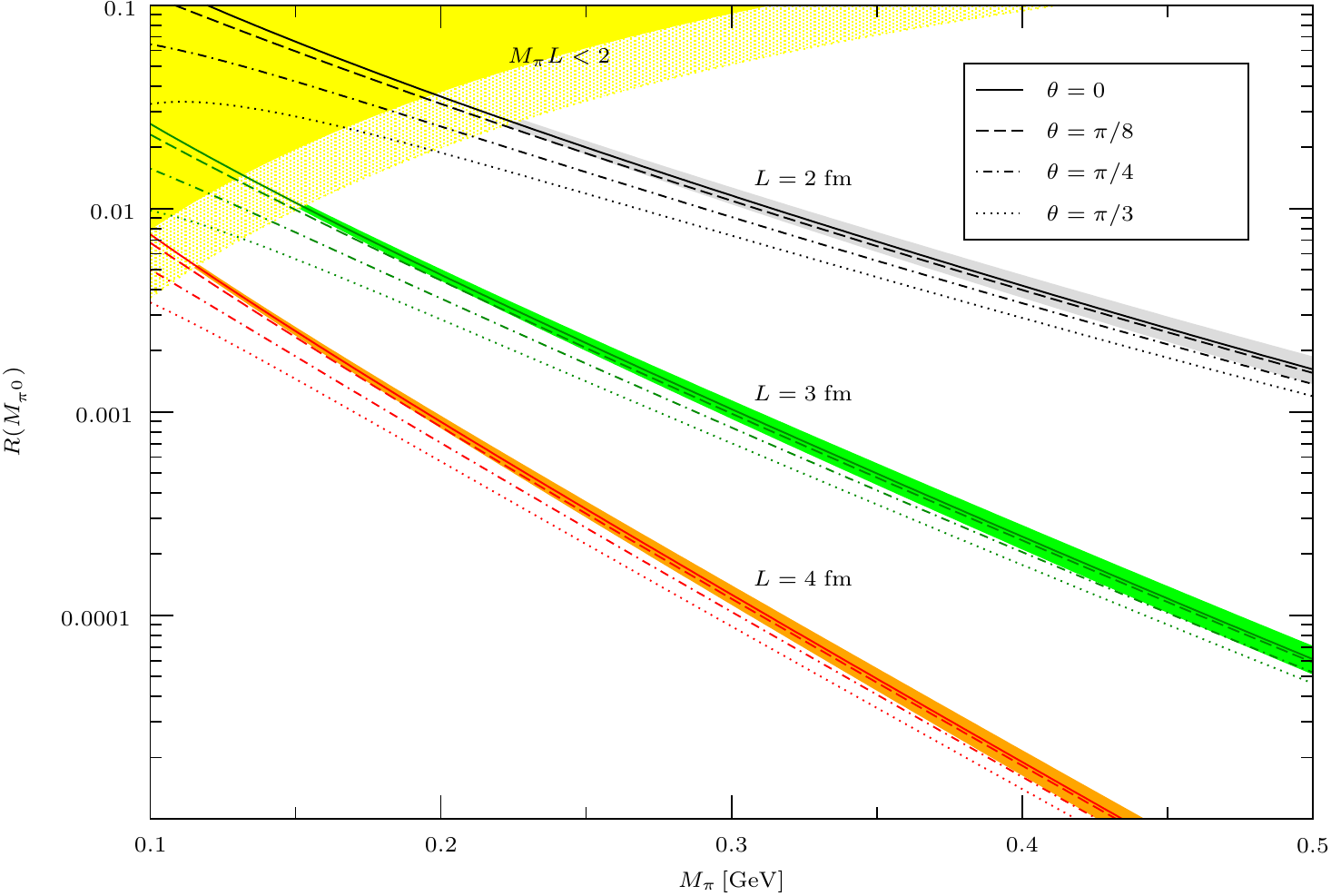}
  }}\\
  \subfloat[][Pion mass dependence of $R(M_{\Pc})$.]
  {\label{Fig:RMPc_vs_Mp}{
   \includegraphics[width=.95\columnwidth]{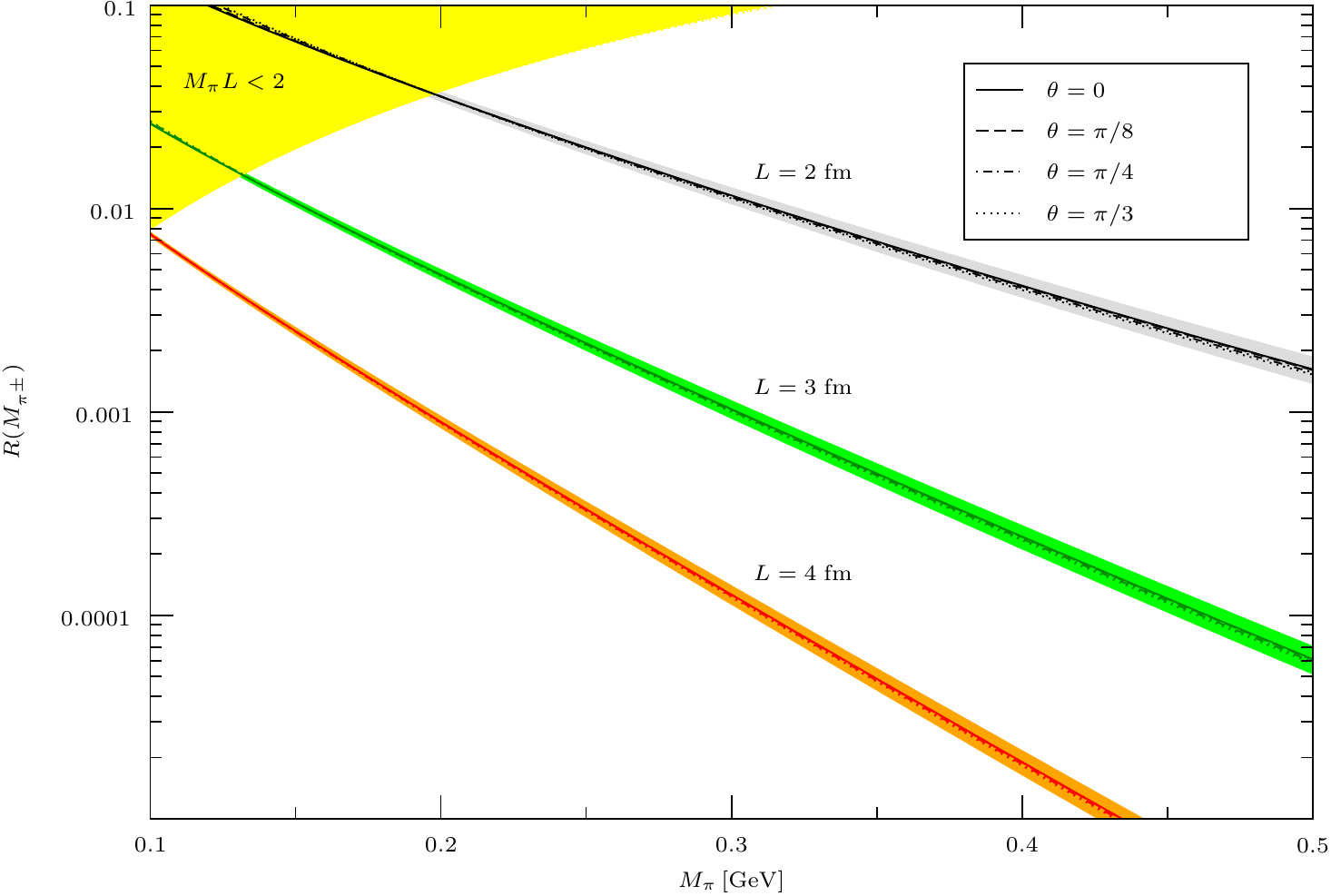}
  }}
 \caption[Mass corrections of pions beyond NLO.]{%
          Mass corrections of pions beyond NLO.        
         }
 \label{Fig:RMPn}
\end{figure}

\begin{table}[tbp]
 \centering
 \hspace{\stretch{2}}
 \subfloat[Mass corrections for $\theta=0$.][Mass corrections for $\theta=0$.]{%
 \setlength{\fboxrule}{1.5pt}
 \framebox[\width]{
 \begin{tabular}{CCC}
  M_{\pi}\,[\GeV] &  \delta M_{\Pn}     & R(M_{\Pn}) \\
  \midrule
  0.100           &  0.0240             & 0.0334(5)  \\
  0.120           &  0.0163             & 0.0234(5)  \\
  0.140           &  0.0113             & 0.0167(5)  \\                                            
  0.160           &  0.0079             & 0.0121(4)  \\                                            
  0.180           &  0.0057             & 0.0088(4)  \\
  0.200           &  0.0040             & 0.0065(3)  \\
  0.220           &  0.0029             & 0.0048(3)  \\
  0.240           &  0.0021             & 0.0038(2)  \\
  0.260           &  0.0015             & 0.0027(2)  \\
  0.280           &  0.0011             & 0.0020(2)  \\
  0.300           &  0.0008             & 0.0015(1)  \\
  0.320           &  0.0006             & 0.0012(1)  \\
  0.340           &  0.0004             & 0.0009(1)  \\
  0.360           &  0.0003             & 0.0007(1)  \\
  0.380           &  0.0002             & 0.0005(1)  \\
  0.400           &  0.0002             & 0.0004(0)  \\
  0.420           &  0.0001             & 0.0003(0)  \\
  0.440           &  0.0001             & 0.0002(0)  \\
  0.460           &  0.0001             & 0.0002(0)  \\
  0.480           &  0.0001             & 0.0001(0)  \\
  0.500           &  0.0000             & 0.0001(0)  \\
  \end{tabular}
 }
 }
 \hspace{\stretch{1}}
 \subfloat[Mass corrections for $\theta=\pi/3$.][Mass corrections for $\theta=\pi/3$.]{%
 \setlength{\fboxrule}{1.5pt}
 \framebox[\width]{
 \begin{tabular}{CCC}
  M_{\pi}\,[\GeV] & \delta M_{\Pn} & R(M_{\Pn})  \\
  \midrule
  0.100           & 0.0050           & 0.0119(4) \\
  0.120           & 0.0050           & 0.0102(4) \\
  0.140           & 0.0043           & 0.0082(4) \\
  0.160           & 0.0034           & 0.0065(4) \\
  0.180           & 0.0027           & 0.0050(3) \\
  0.200           & 0.0020           & 0.0039(3) \\
  0.220           & 0.0015           & 0.0030(3) \\
  0.240           & 0.0012           & 0.0023(2) \\
  0.260           & 0.0009           & 0.0017(2) \\
  0.280           & 0.0006           & 0.0013(2) \\
  0.300           & 0.0005           & 0.0010(1) \\
  0.320           & 0.0004           & 0.0008(1) \\
  0.340           & 0.0003           & 0.0006(1) \\
  0.360           & 0.0002           & 0.0005(1) \\
  0.380           & 0.0001           & 0.0004(1) \\
  0.400           & 0.0001           & 0.0003(0) \\
  0.420           & 0.0001           & 0.0002(0) \\
  0.440           & 0.0001           & 0.0002(0) \\
  0.460           & 0.0000           & 0.0001(0) \\
  0.480           & 0.0000           & 0.0001(0) \\
  0.500           & 0.0000           & 0.0001(0) \\
 \end{tabular}
 }
 }
 \hspace{\stretch{2}}
 \caption[Comparison of corrections evaluated at NLO with ChPT and estimated beyond NLO with asymptotic formulae.]{
          Comparison of corrections evaluated at NLO with ChPT and estimated beyond NLO with asymptotic formulae.
          Here, we use $L=\unit{2.83}{\fm}$ so that $M_{\pi} L=2$ for $M_{\pi}=\MPcPhys$.
          The uncertainties of $R(M_{\Pn})$ originate from the errors of LECs contained in the integrals $I^{(4)}(M_{\Pn},\Pn)$, $I^{(4)}(M_{\Pn},\Pc)$.
          Note that results with $M_{\pi}<\unit{0.140}{\GeV}$ should be taken with a grain of salt as they are in the region where the $p$-regime is no more guaranteed. 
          }
 \label{Tab:Mass_NLO_AF}
\end{table}

\begin{figure}[tbp]
\centering
  \subfloat[][Angle dependence of $R(M_{\Pn})$.]
  {\label{Fig:RMPn_vs_TWx_L2fm}{
   \includegraphics[width=.925\columnwidth]{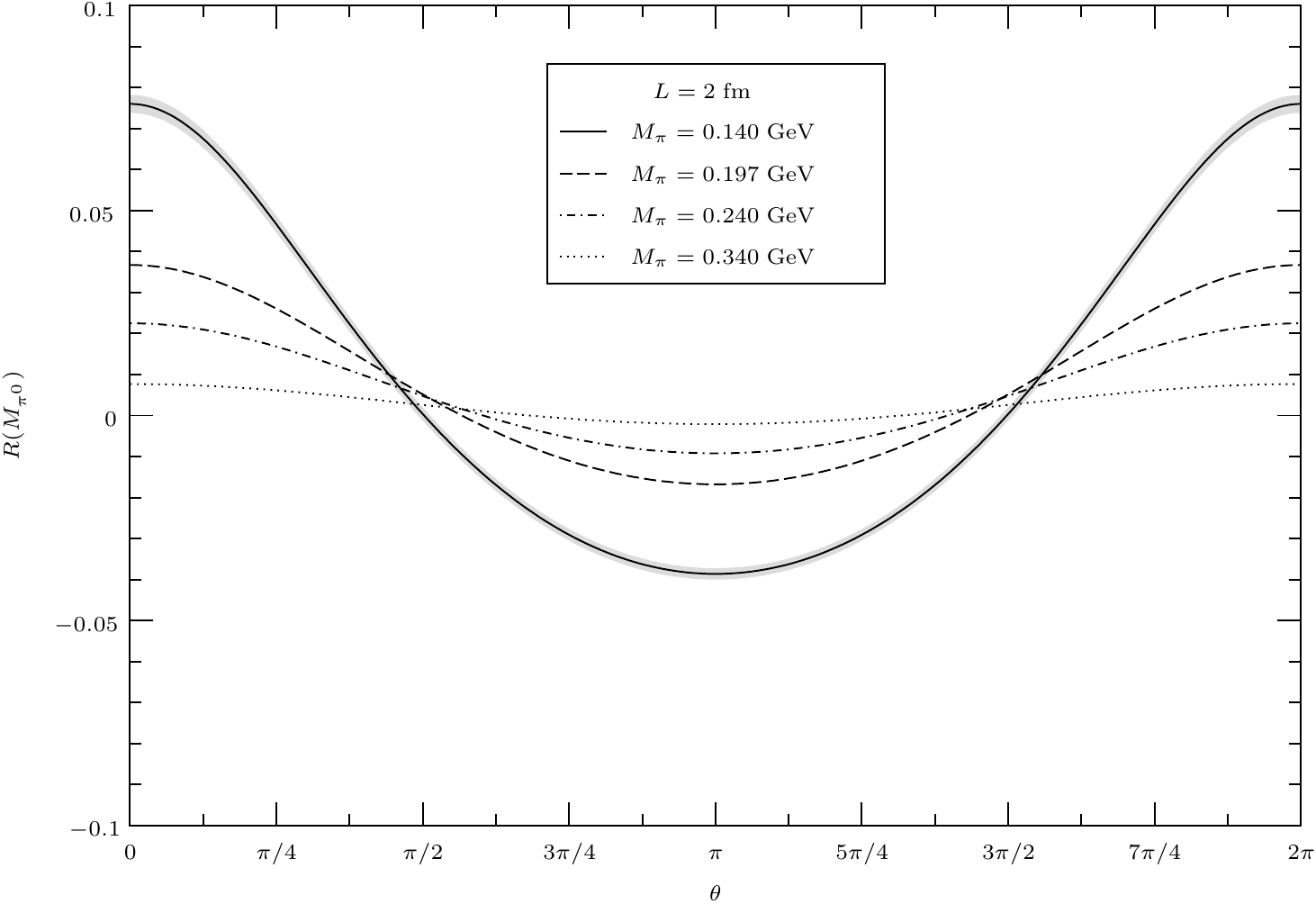}
  }
  }\\
  \subfloat[][Angle dependence of $R(M_{\Pc})$.]
  {\label{Fig:RMPc_vs_TWx_L2fm}{
   \includegraphics[width=.925\columnwidth]{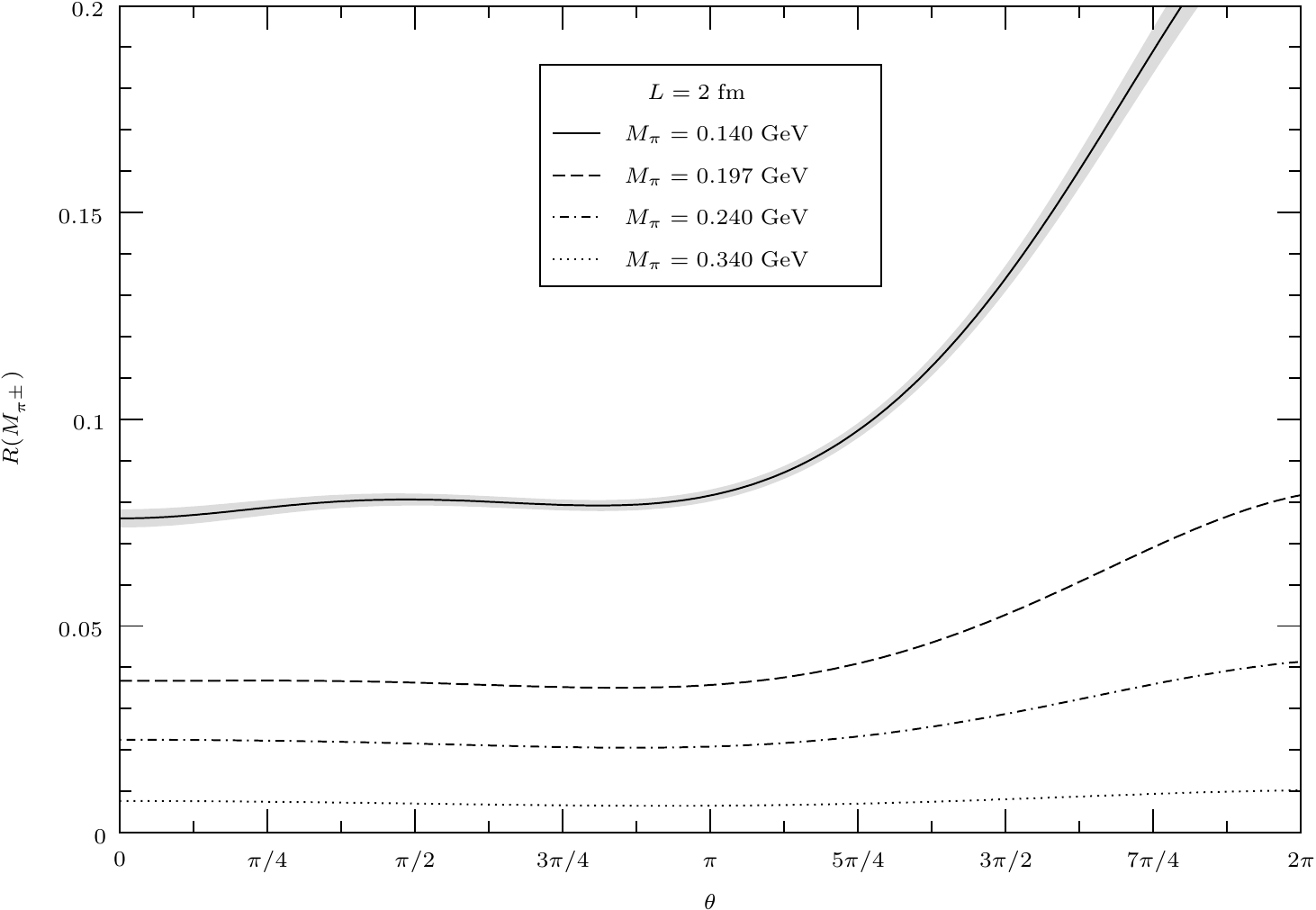}
  }}
 \caption[Mass corrections of pions beyond NLO.]{%
          Mass corrections of pions beyond NLO.        
         }
 \label{Fig:RMPc}
\end{figure}

In \Fig\ref{Fig:RMPn} and \ref{Fig:RMPc} we represent the mass corrections
estimated with $R(M_{\Pn})$, $R(M_{\Pc})$. The pion mass dependences of
$R(M_{\Pn})$, $R(M_{\Pc})$ are represented in logarithmic plots whereas
the angle dependences in linear ones. In these graphs the bands of the
uncertainty are displayed for solid lines (i.e. for $\theta=0$ in
\Fig\ref{Fig:RMPn} resp. for $M_{\pi}=\unit{0.140}{\GeV}$ in
\Fig\ref{Fig:RMPc}). Bands of different colors refer to different values of
$L$. The uncertainty bands are calculated with the usual formula of the
error propagation. As unique source of error, we have taken the
uncertainties on the LECs contained in the integrals
$I^{(4)}(M_{\Pn},\Pn),\dots,I^{(4)}_D(M_{\Pc},\Pc)$.  

The logarithmic plots neatly illustrate the exponential decay
$\Ord{\e^{-\lambda_{\pi}}}$ of the corrections.  
In \Fig\ref{Fig:RMPn_vs_Mp} the lines are almost straight and can be
distinguished for different angles. In \Fig\ref{Fig:RMPc_vs_Mp} the lines
are exactly straight and are so close that they overlap in the graph. This
indicates that for small angles, $R(M_{\Pn})$ is more sensitive to
$\theta$. The corrections are significantly bigger than those at
NLO. \label{Pag:SubleadingFSE} 
As illustration, we list in \Tab\ref{Tab:Mass_NLO_AF}a
(resp. \Tab\ref{Tab:Mass_NLO_AF}b) numerical values for $\delta M_{\Pn}$,
$R(M_{\Pn})$ at $L=\unit{2.83}{\fm}$ for $\theta=0$ (resp. for
$\theta=\pi/3$). In some cases, $R(M_{\Pn})$ is of order $50\%$ with
respect to $\delta M_{\Pn}$. Such significant subleading effects were
already observed in finite volume with PBC,
see~\Ref\cite{CoDuSo:02}. However, the comparison of numerical results
obtained from asymptotic formulae with amplitudes at two
loops~\cite{CoDu:04,CoDuHae:05} has showed that subsubleading effects are
small and that the expansion does have a good converging behaviour for
$M_{\pi} L>2$. We are confident that this is also true for TBC and that
our numerical estimates are reliable, at least for small angles
(i.e. $\theta<\pi$). 

In \Fig\ref{Fig:RMPc} we display the angle dependences of
$R(M_{\Pn})$, $R(M_{\Pc})$. We observe that $R(M_{\Pn})$ depends on
$\theta$ as a cosine function. The corrections oscillate with a period of
$2\pi$ and have maxima (resp. minima) at even (resp. odd) integer multiples
of $\pi$. If we consider $M_{\pi}=\unit{0.197}{\GeV}$, the difference among
maxima and minima is $5.4\%$ at $L=\unit{2}{\fm}$. This is a sizable effect
and should be taken into account when physical observables are extrapolated
from lattice data. In \Fig\ref{Fig:RMPc_vs_TWx_L2fm} we observe that
$R(M_{\Pc})$ depends on $\theta$ as $(a+\cos\theta)\sqrt{1+\theta^2}$ with
$a>0$. This dependence originates from contributions~$\Ord{\xi^2_\pi}$ and
provides large corrections at large angles. However, one should here retain
only results in the interval $\theta\in\mathopen{[}0,\pi\mathclose{]}$. The
reason is, $R(M_{\Pc})$ is derived by means of an expansion which is valid
for small external twisting angles. Considering
$M_{\pi}=\unit{0.197}{\GeV}$, the difference among the local maximum and
the local minimum in $\theta\in\mathopen{[}0,\pi\mathclose{]}$ is $0.2\%$
at $L=\unit{2}{\fm}$. This is a negligible effect. Note that the corrections
estimated with $R(M_{K^{\pm}})$, $R(M_\eta)$ are less than a percent for
small angles and can be neglected. 

\begin{figure}
\centering
  \subfloat[][Pion mass dependence of $-R^{\mu}(\tSPp)$ for $\mu=1$.]
  {\label{Fig:m_RMTWPp_vs_Mp}{
   \includegraphics[width=.925\columnwidth]{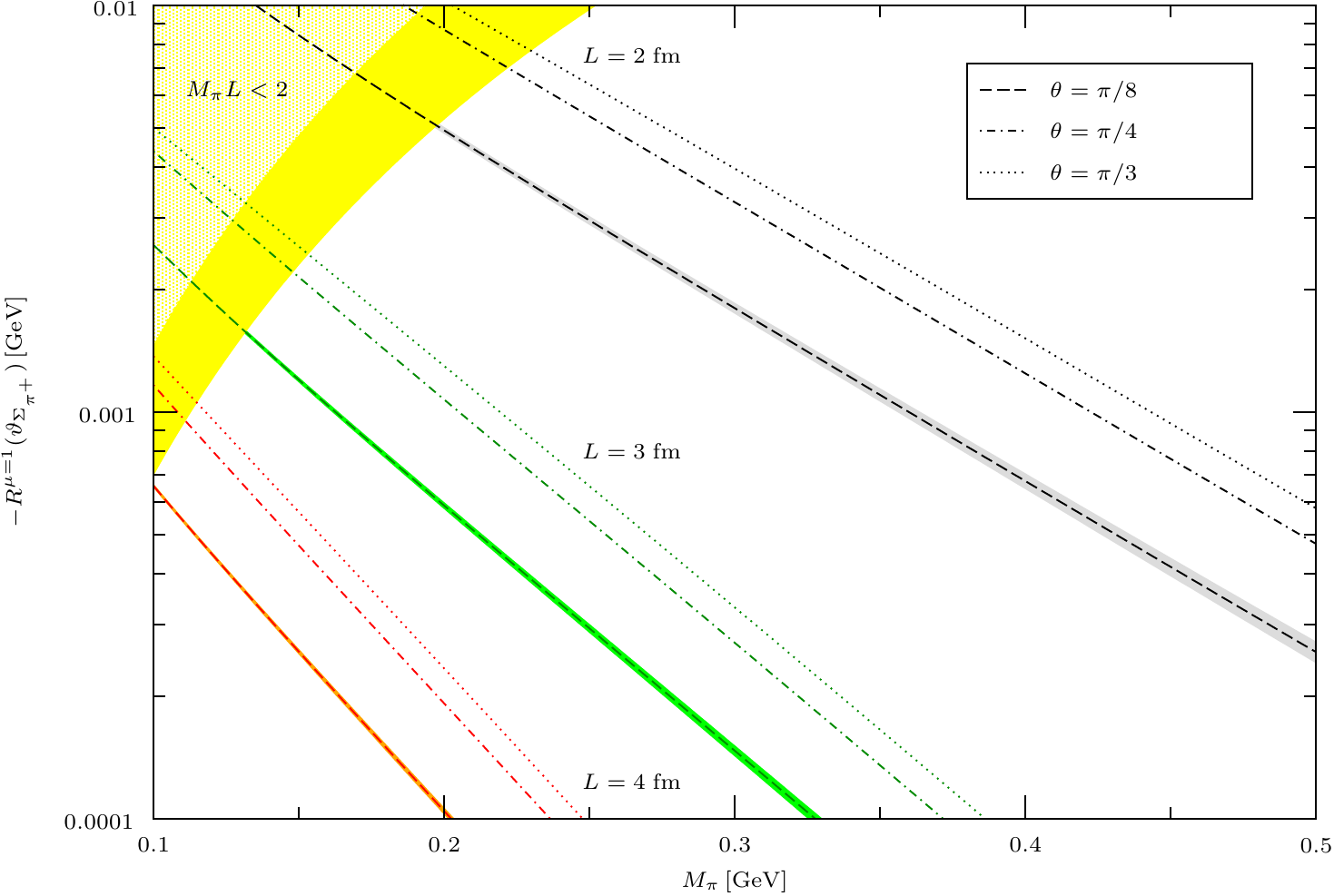}
  }}\\
  \subfloat[][Angle dependence of $R^{\mu}(\tSPp)$ for $\mu=1$.]
  {\label{Fig:RMTWPp_x_vs_TWx_L2fm}{
   \includegraphics[width=.925\columnwidth]{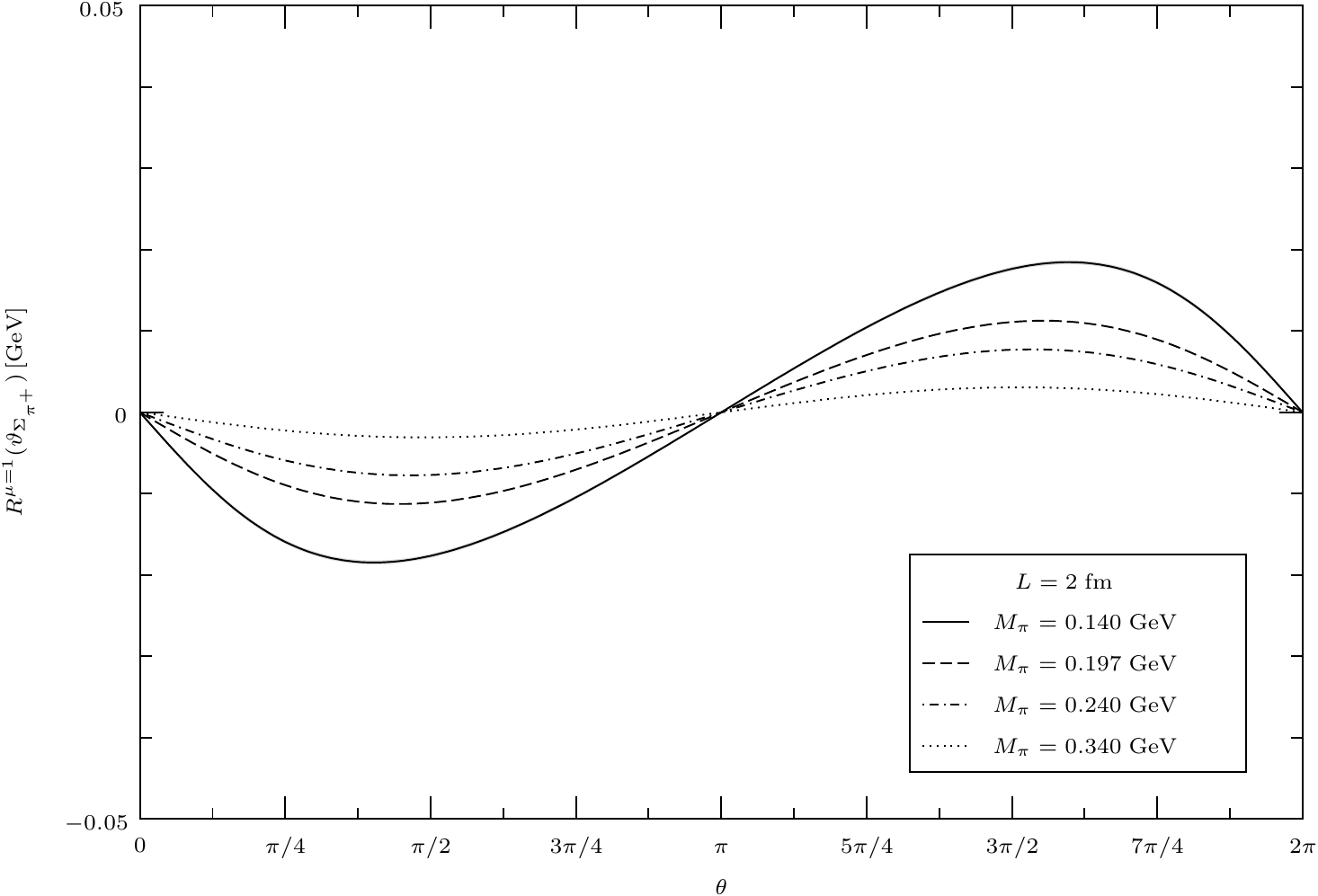}
  }}
 \caption[Extra term of positive pion beyond NLO.]{%
          Extra term of positive pion beyond NLO.
          Note that in \textrm{(a)} the dark yellow area refers to the region $M_{\pi} L<2$ for $\theta=\pi/8$.
         }
 \label{Fig:RMTWPp}
\end{figure}

In \Fig\ref{Fig:RMTWPp} we plot the extra term
$\Delta\vartheta^\mu_{\Sigma_{\Pp}}$ estimated by means of
$R^{\mu}(\tSPp)$. We represent the first spatial component as it is the
only one which is non-zero for configuration~\eqref{Eq:TW}. The values on
the $y$-axis are in $\GeV$ since $R^{\mu}(\tSPp)$ is a dimensionful
quantity. Uncertainty bands are displayed for $\theta=\pi/8$ in
\Fig\ref{Fig:m_RMTWPp_vs_Mp} and for $M_{\pi}=\unit{0.140}{\GeV}$ in
\Fig\ref{Fig:RMTWPp_x_vs_TWx_L2fm}. From the logarithmic graph of
\Fig\ref{Fig:m_RMTWPp_vs_Mp} we observe that the extra term decays
exponentially as $\Ord{\e^{-\lambda_{\pi}}}$. For
$\theta\in\set{\pi/8,\,\pi/4,\,\pi/3}$ its absolute value increases with
the angle. This can also be seen in \Fig\ref{Fig:RMTWPp_x_vs_TWx_L2fm}
where $R^{\mu}(\tSPp)$ is represented as a function of $\theta$. We observe
that the extra term depends on $\theta$ almost exactly as a (negative) sine
function. The zeros correspond to integer multiples of $\pi$ and extrema
are close to half-integer multiples of $\pi$. In this graph, one should
retain only results for $\theta\in\mathopen{[}0,\pi\mathclose{]}$ as by
derivation, $R^{\mu}(\tSPp)$ is valid for small external angles.  Note that
$R^{\mu}(\tSKp)$ has similar dependences on $M_{\pi}$ resp. $\theta$ and
its absolute value is in general smaller than that of $R^{\mu}(\tSPp)$.

\subsubsection{Decay constants and extra terms in axialvector matrix elements}

\begin{figure}[tbp]
\centering
  \subfloat[][Pion mass dependence of $-R(F_{\Pc})$.]
  {\label{Fig:m_RFPc_vs_Mp}{
   \includegraphics[width=.95\columnwidth]{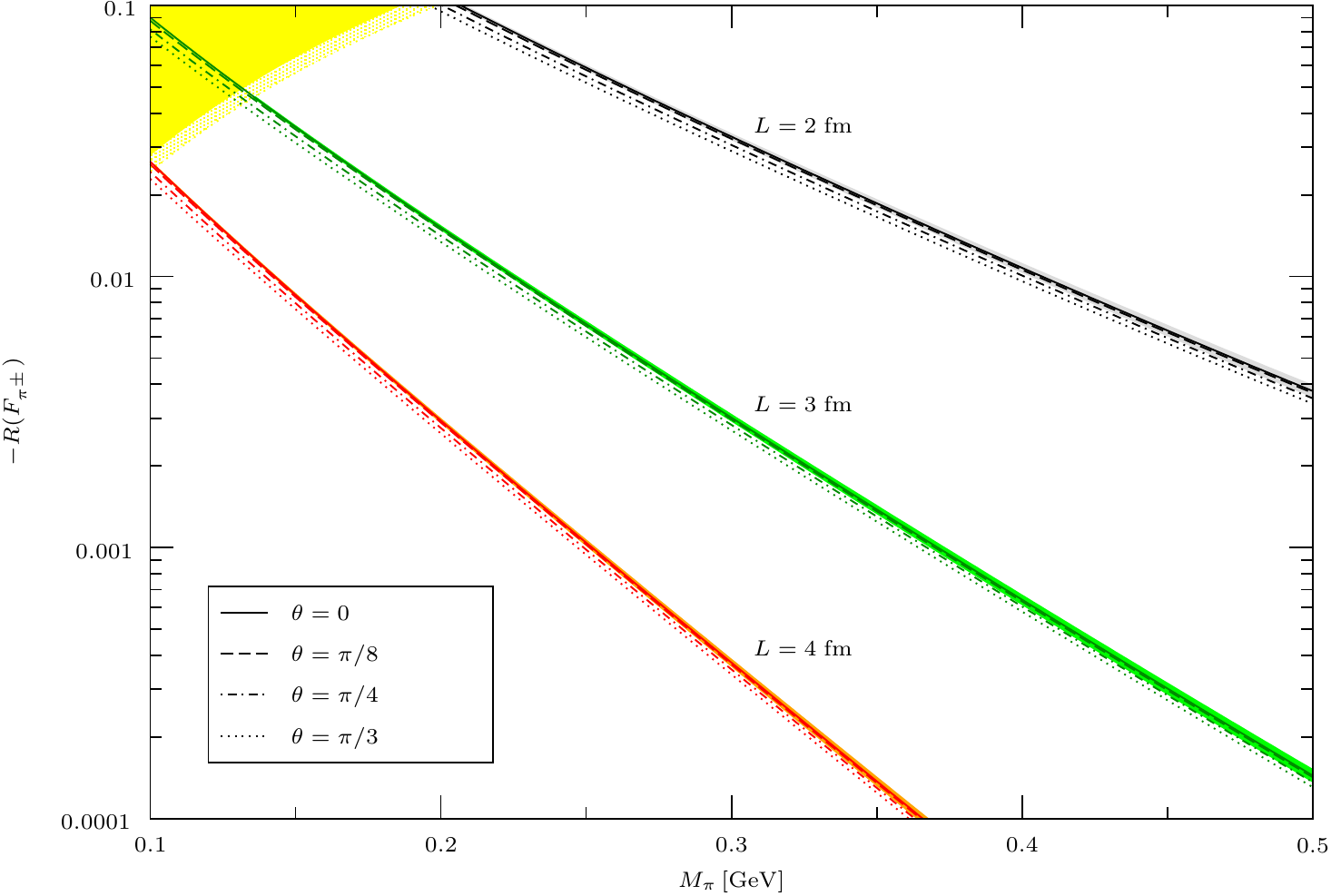}
  }}\\
 \subfloat[][Pion mass dependence of $-R(F_{K^{\pm}})$.]
  {\label{Fig:m_RFKc_vs_Mp}{
   \includegraphics[width=.95\columnwidth]{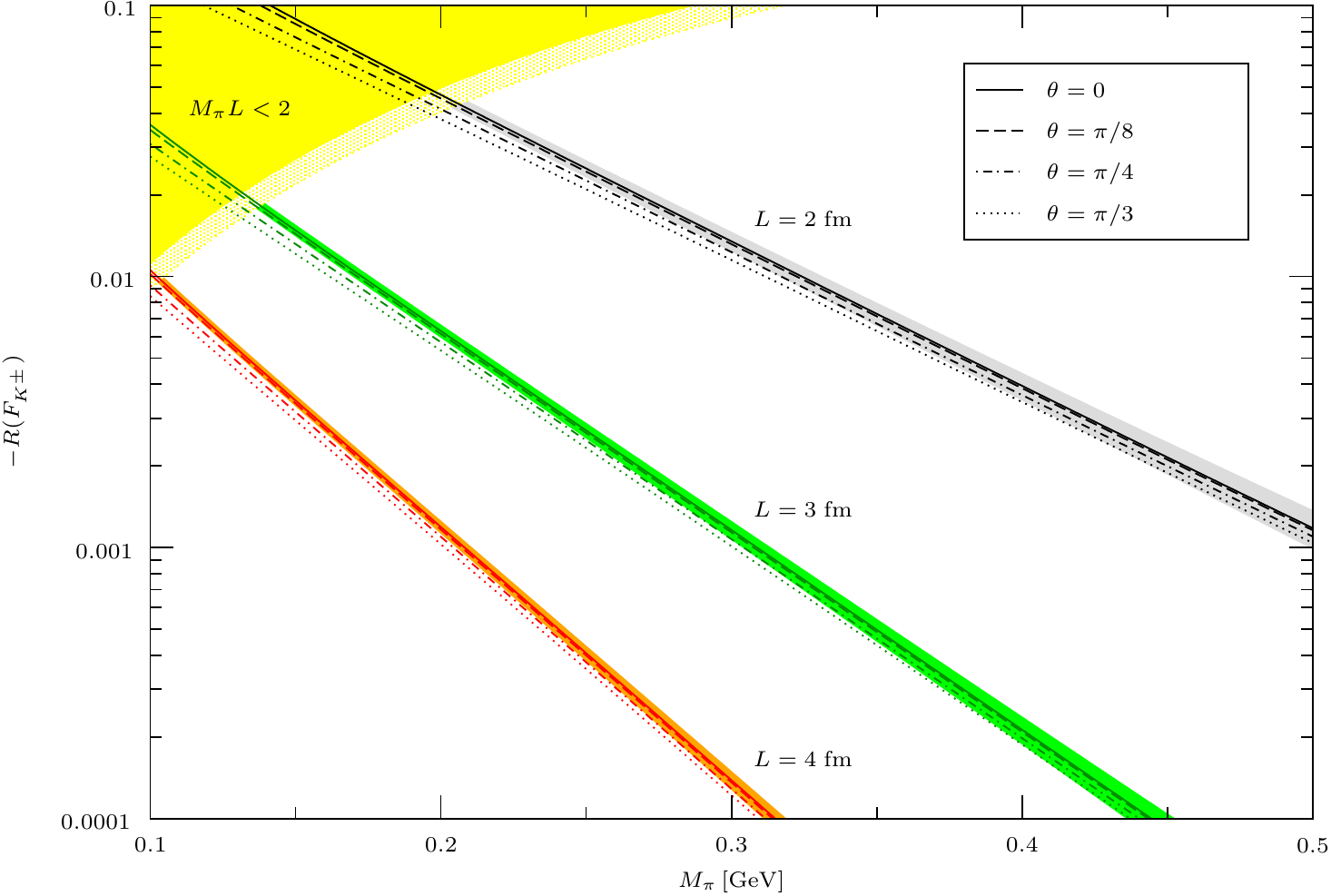}
  }}
 \caption[Corrections of decay constants beyond NLO.]{%
          Corrections of decay constants beyond NLO.
         }
 \label{Fig:RFPc}
\end{figure}

In \Fig\ref{Fig:RFPc} we represent the corrections of decay constants
estimated with $R(F_{\Pc})$, $R(F_{K^{\pm}})$.  The logarithmic graphs
illustrate the exponential decay $\Ord{\e^{-\lambda_{\pi}}}$ of the
corrections.  The corrections are negative and for
$\theta\in\set{0,\,\pi/8,\,\pi/4,\,\pi/3}$ their absolute values decrease
with the angle. Note that the corrections may reach more than $10\%$
before entering yellow areas and they should be subtracted before decay
constants are extracted from lattice data.

In \Fig\ref{Fig:m_RFTWPp_vs_Mp} we represent the extra term
$\Delta\vartheta^\mu_{\mathscr{A}_{\Pp}}$ estimated with $R^{\mu}(\tAPp)$.
We represent the first component as it is the only one which is non-zero
for configuration~\eqref{Eq:TW}. Also this extra term decays exponentially
with $\lambda_\pi$. Comparing \Fig\ref{Fig:m_RFTWPp_vs_Mp} with
\Fig\ref{Fig:m_RMTWPp_vs_Mp} we observe that for a fixed angle
$-R^{\mu}(\tAPp)$ is smaller than $-R^{\mu}(\tSPp)$.  This difference
is~$\Ord{\xi^2_\pi}$ and is proportional to the extra term $R^{\mu}(\tGPp)$. 
A similar observation can be made for charged
kaons where the difference among $R^{\mu}(\tAKp)$ and $R^{\mu}(\tSKp)$
is~$\Ord{\xi^2_\pi}$ and is proportional to $R^{\mu}(\tGKp)$.

\subsubsection{Pseudoscalar coupling constants}

In \Fig\ref{Fig:m_RGPn_vs_Mp} we represent the pion mass dependence of
$-R(G_{\Pn})$ which is as well exponential. In general, the corrections are
negative and for $\theta\in\set{0,\,\pi/8,\,\pi/4,\,\pi/3}$ their absolute
value increases with the angle. In this case, the corrections are smaller
than those at NLO. This can be explained if we look at the
contributions~$\Ord{\xi^2_\pi}$, see~\Eq\eqref{Eq:IGPn}. For
$\theta\in\set{0,\,\pi/8,\,\pi/4,\,\pi/3}$ the contribution originating
from integral $I^{(4)}(G_{\Pn},\Pn)$ is negative but that
from~$I^{(4)}(G_{\Pn},\Pc)$ is positive. As the negative contribution is
smaller than the positive one, the corrections estimated with $-R(G_{\Pn})$
are smaller than those evaluated with $-\delta G_{\Pn}$ at NLO. 
{
\begin{figure}
  \centering
  \includegraphics[width=.925\columnwidth]{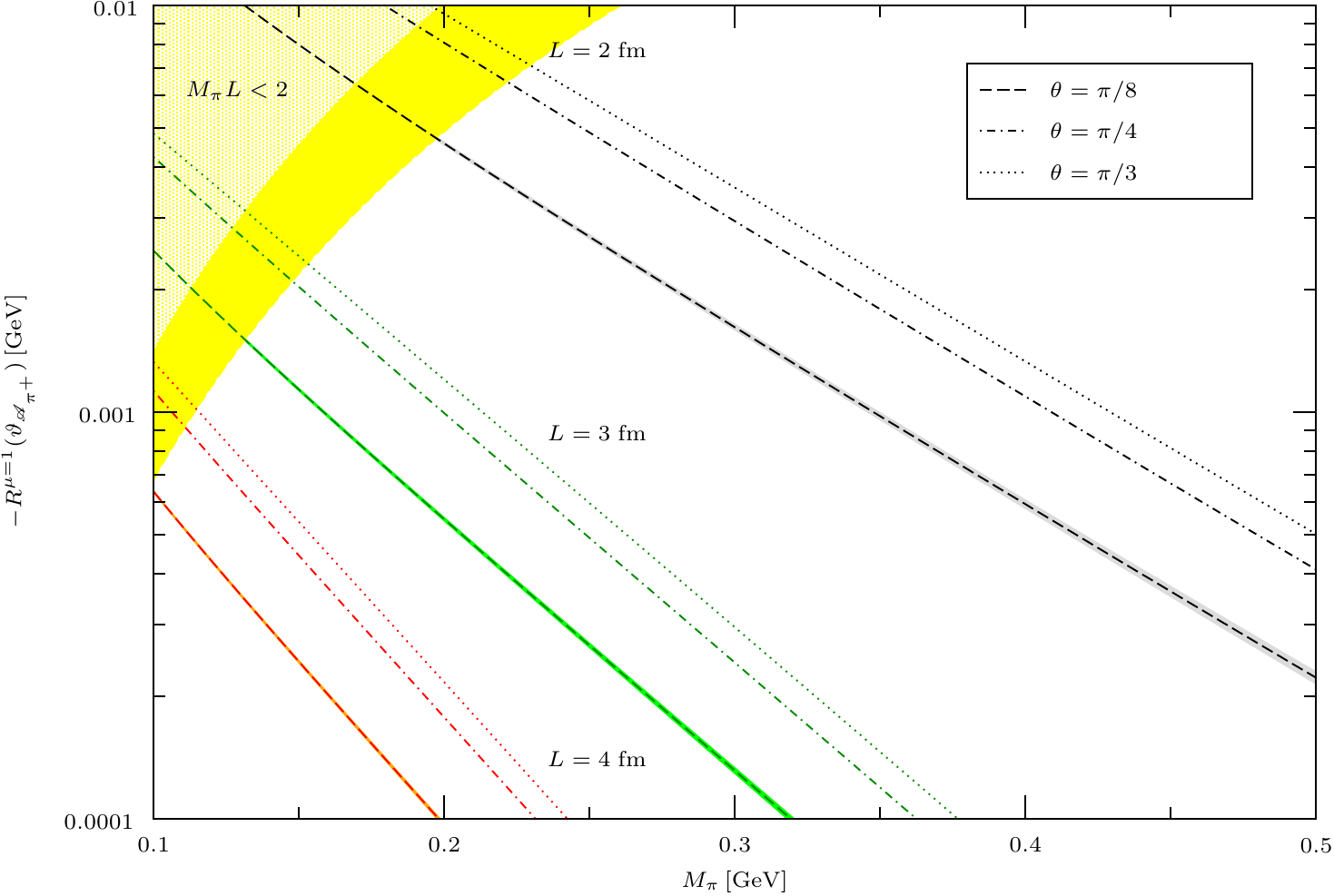}
  \caption[Pion mass dependence of $-R^{\mu}(\tAPp)$ for $\mu=1$.]{%
            Pion mass dependence of $-R^{\mu}(\tAPp)$ for $\mu=1$.
            Note that the dark yellow area refers to the region $M_{\pi} L<2$ for $\theta=\pi/8$.
            }
  \label{Fig:m_RFTWPp_vs_Mp}
\end{figure}
\begin{figure}
  \centering
  \includegraphics[width=.925\columnwidth]{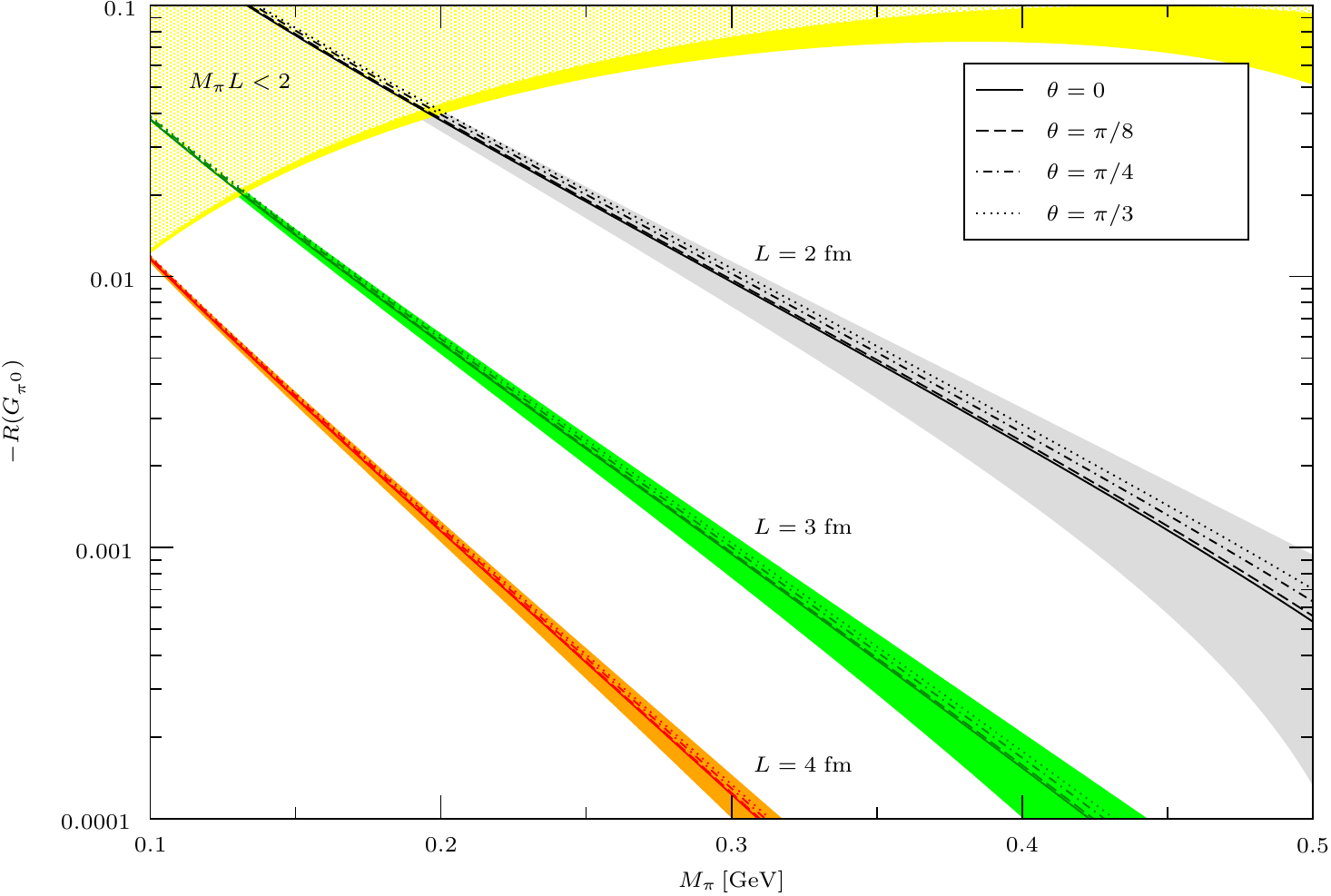}
  \caption[Pion mass dependence of $-R(G_{\Pn})$.]{%
           Pion mass dependence of $-R(G_{\Pn})$.
          }
\label{Fig:m_RGPn_vs_Mp}
\end{figure}
}

\begin{figure}[tbp]
\centering
  \subfloat[][Pion mass dependence of $R(\GSPn)$ at $q^2=0$.]
  {\label{Fig:RGSPn0_vs_Mp}{
   \includegraphics[width=.95\columnwidth]{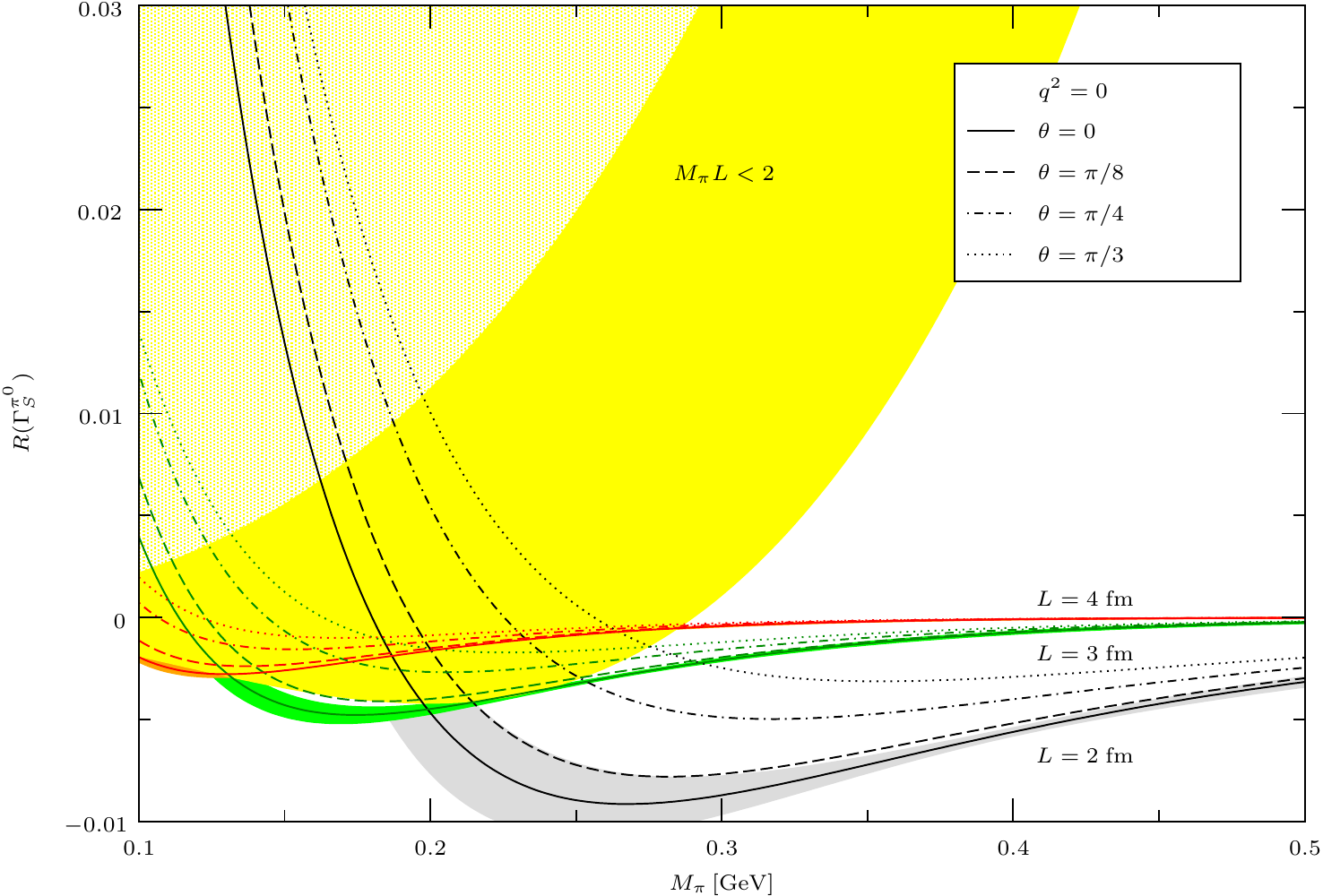}
  }}\\
  \subfloat[][Pion mass dependence of $R(\GSPp)$ at $q^2=0$.]
  {\label{Fig:RGSPp0_vs_Mp}{
   \includegraphics[width=.95\columnwidth]{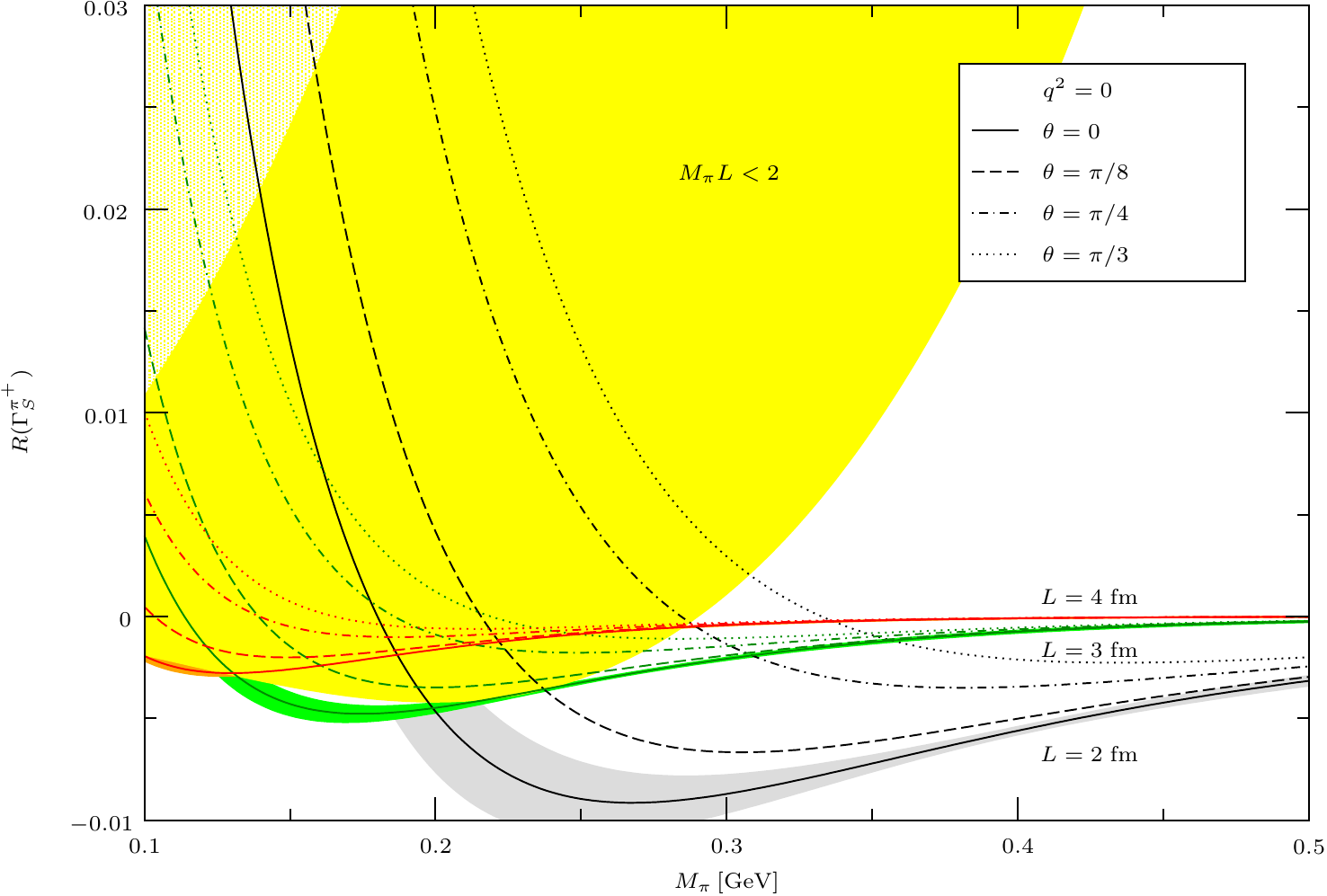}
  }}
 \caption[Corrections of matrix elements of the scalar form factor beyond NLO.]{%
          Corrections of matrix elements of the scalar form factor beyond NLO.
         }
 \label{Fig:RGSPn0}
\end{figure}

\subsubsection{Pion form factors}

We consider pions at rest and estimate the corrections of the matrix
elements of the scalar form factor with $R(\GSPn)$, $R(\GSPp)$.  
In \Fig\ref{Fig:RGSPn0_vs_Mp} [resp. \ref{Fig:RGSPp0_vs_Mp}] we represent
the pion mass dependence of $R(\GSPn)$ [resp. $R(\GSPp)$] at $q^2=0$. 
From the graphs we observe that the corrections decay exponentially as
$\Ord{\e^{-\lambda_{\pi}}}$. In general, they are negative but they may
turn positive depending on the pion mass. Note that the corrections reach
percentage level before entering yellow areas and they should be
subtracted when the scalar form factor is extracted from lattice data.

\section{Conclusions}

In this work we studied the effects of a finite cubic volume with twisted
boundary conditions on pseudoscalar mesons. We applied chiral perturbation
theory (ChPT) in the $p$-regime and introduced twisting angles by means of
a constant vector field, see~\Ref\cite{SaVi:05}. The corrections for
masses, decay constants, pseudoscalar coupling constants were recalculated
at next-to-leading order (NLO) and new results for pion form factors were
presented.  In the calculations we adopted the mass definition of
\Ref\cite{SaVi:05,JiTi:07} which treats new extra terms as renormalization
terms of twisting angles, and argued in some detail about the reasons
behind this choice. These extra terms originate from the breaking of
the cubic invariance and can be reabsorbed in the on-shell conditions
modifying the mass definition in finite volume.  We found that the
Feynman-Hellmann theorem~\cite{Hell:37,Fey:39} as well as the Ward-Takahashi
identity~\cite{Green:53,Taka:57a,Taka:57b} are satisfied.  To prove the
Ward-Takahashi identity we constructed an effective field theory for
charged pions invariant under gauge transformations which reproduces
results obtained with ChPT.

We generalized the derivation of L{\"u}scher~\cite{Lu:85} and derived
asymptotic formulae for twisted boundary conditions. We showed that the
asymptotic formulae for masses, decay constants, pseudoscalar coupling
constants are related by means of chiral Ward identities where extra terms
satisfy such relations in an independent way. Applying asymptotic formulae
in combination with ChPT, we estimated corrections beyond NLO and found
that, as in the case of PBC, NNLO corrections can be very significant,
indeed almost as large as NLO corrections. This underlines the importance
of using asymptotic formulae combined with NLO chiral calculations of the
relevant infinite-volume amplitude to reliably estimate finite-volume
corrections. From our numerical analysis we see that the corrections can be
comparable (or even larger) than the statistical precision reached in
simulations of lattice QCD and hence, should be taken into account. 

\acknowledgments

We thank Christoph H{\"a}feli, Giovanni Villadoro, Johan Bijnens, Fu-Jiun Jiang and Brian C. Tiburzi for private communications at different stages of the work.
This work was funded by the Swiss National Science Foundation.

\appendix
\section{Sums in finite volume}\label{app:Sums}

We list some results which are useful in the evaluation of loop diagrams in finite volume.
For convenience, we define
\begin{equation}
\sumintp g(k):=\frac{1}{L^3}\sum_{\begin{subarray}{c}
				\vec{k}=\frac{2\pi}{L}\vec{m}\\
				\vec{m}\in\Z{3}
				\end{subarray}
				}%
				\int_{\R{}}\frac{\d{k_{0}}{}}{2\pi}g(k)-\int_{\R{4}}\frac{\d{k}{4}}{(2\pi)^4}g(k),
\end{equation}
where $g$ is a generic function in momentum space and $L$ is the side length of the finite cubic box. 
The right-hand side of the equation represents the difference among
contributions in finite and infinite volume. 
For loop diagrams encountered in this work, this difference is finite and
can be calculated by means of the Poisson resummation
formula~\cite{HaLe:89}:  
\begin{equation}\label{Eq:Poisson}
\sumintp g(k)=\sum_{
		  \begin{subarray}{c}
		  \vec{n}\in\Z{3}\\
		  \abs{\vec{n}}\neq0
		  \end{subarray}}\int_{\R{4}}\frac{\d{k}{4}}{(2\pi)^4}\, g(k)\,\e^{\i L\vec{n}\vec{k}}.
\end{equation}

The first group of results is
\begin{equation}\label{Eq:SumFVtadpole}
\begin{aligned}
  \sumintp\frac{\Gamma(r)}{\i\big[M_P^2-(k+\vartheta)^2\big]^r}                                    &=\phantom{+}\frac{M_P^2}{(4\pi)^2}\g_r(\lambda_{P},\vartheta), \\
  \sumintp\frac{\Gamma(r)\,(k+\vartheta)^\mu}{\i\big[M_P^2-(k+\vartheta)^2\big]^r}                 &=-\frac{M_P^2}{(4\pi)^2}\f_r^{\mu}(\lambda_{P},\vartheta), \\
  \sumintp\frac{\Gamma(r)\,(k+\vartheta)^\mu(k+\vartheta)^\nu}{\i\big[M_P^2-(k+\vartheta)^2\big]^r}&=-\frac{M_P^2}{(4\pi)^2}\bigg[\frac{g^{\mu\nu}}{2}\g_{r-1}(\lambda_{P},\vartheta)+\h_r^{\mu\nu}(\lambda_{P},\vartheta)\bigg],
\end{aligned} 
\end{equation}
for $r\in\N{}$.
Here, $\Gamma(r)$ is the gamma function, $g^{\mu\nu}$ is the metric of Minkowski space-time, $\vartheta^\mu=\big(\begin{smallmatrix}0\\ \vec{\vartheta}\end{smallmatrix}\big)$ is a twisting angle and $\lambda_{P}=M_P L$.
The functions on the right-hand side can be expressed in terms of modified Bessel functions of the second kind, $\KB_{r}(x)$.
They read
\begin{equation}\label{Eq:gRfRhRTransferZero}
\begin{aligned}
 \g_r(\lambda_{P},\vartheta)&=\sum_{%
		\begin{subarray}{c}
		\vec{n}\in\Z{3}\\
		\abs{\vec{n}}\neq0
		\end{subarray}
		}%
		\frac{2}{M_P^2}\bigg[\frac{L^2\abs{\vec{n}}}{2 \lambda_{P}}\bigg]^{r-2}\KB_{r-2}(\lambda_{P}\abs{\vec{n}})\ \e^{\i L\vec{n}\vt},\\ 
 \f_r^{\mu}(\lambda_{P},\vartheta)&=\sum_{%
		\begin{subarray}{c}
		\vec{n}\in\Z{3}\\
		\abs{\vec{n}}\neq0
		\end{subarray}
		}%
		\frac{\i L}{M_P^2}\,n^{\mu}\bigg[\frac{L^2\abs{\vec{n}}}{2 \lambda_{P}}\bigg]^{r-3}\KB_{r-3}(\lambda_{P}\abs{\vec{n}})\ \e^{\i L\vec{n}\vt}, \\ 
 \h_r^{\mu\nu}(\lambda_{P},\vartheta)&=\sum_{%
		\begin{subarray}{c}
		\vec{n}\in\Z{3}\\
		\abs{\vec{n}}\neq0
		\end{subarray}
		}%
		\frac{L^2}{2M_P^2}\,n^{\mu}n^{\nu}\bigg[\frac{L^2\abs{\vec{n}}}{2 \lambda_{P}}\bigg]^{r-4}\KB_{r-4}(\lambda_{P}\abs{\vec{n}})\ \e^{\i L\vec{n}\vt}, 
\end{aligned} 
\end{equation}
with $n^{\mu}=\big(\begin{smallmatrix}0 \\ \vec{n}\end{smallmatrix}\big)$.
Corrections of masses and decay constants were calculated using the results~\eqref{Eq:SumFVtadpole} for $r=1$.

To evaluate loop diagrams with two different propagators one can use the Feynman parametrization,
\begin{equation}\label{Eq:FeynmanPara}
  \frac{1}{AB}=\int_{0}^{1}\frac{\d{z}{} }{\big[zA+(1-z)B\big]^2}.
\end{equation} 
Here, we consider $A=M_P^2-(k+\vartheta)^2$ and $B=M_P^2-(k+\vartheta+q)^2$
where $q^\mu$ is an external momentum.
The second group of results we present, is 
\begin{equation}\label{Eq:FishgfhR}
\begin{aligned}
  \sumintp\frac{\Gamma(r)}{\i\left[AB\right]^{\frac{r}{2}}}                                    &=\phantom{+}\frac{M_P^2}{(4\pi)^2}\intz\g_r(\lambda_{z},q,\vartheta),\\ 
  \sumintp\frac{\Gamma(r)\,(k+\vartheta)^\mu}{\i\left[AB\right]^{\frac{r}{2}}}                 &=-\frac{M_P^2}{(4\pi)^2}\intz\left[\f_r^{\mu}(\lambda_{z},q,\vartheta)+(1-z)q^\mu\g_r(\lambda_{z},q,\vartheta)\right],\\ 
  \sumintp\frac{\Gamma(r)\,(k+\vartheta)^\mu(k+\vartheta)^\nu}{\i\left[AB\right]^{\frac{r}{2}}}&=\phantom{+}\frac{M_P^2}{(4\pi)^2}\intz\left\{(1-z)\left[\f_r^{\mu}(\lambda_{z},q,\vartheta)q^\nu+\f_r^{\nu}(\lambda_{z},q,\vartheta)q^\mu\right]\right\}\\
										   &\phantom{+}-\!\frac{M_P^2}{(4\pi)^2}\intz\left[\frac{g^{\mu\nu}}{2}\g_{r-1}(\lambda_{z},q,\vartheta)+\h_r^{\mu\nu}(\lambda_{z},q,\vartheta)\right]\\
										   &\phantom{+}+\!\frac{M_P^2}{(4\pi)^2}\intz\left[(1-z)^2q^\mu q^\nu\g_r(\lambda_{z},q,\vartheta)\right],
\end{aligned} 
\end{equation}
with $\lambda_{z}=M_P L\sqrt{1+z(z-1)q^2/M_P^2}$. 
The functions on the right-hand side can be expressed as 
\begin{equation}\label{Eq:FishSums}
\begin{aligned}
 \g_r(\lambda_{z},q,\vartheta)&=\sum_{%
		\begin{subarray}{c}
		\vec{n}\in\Z{3}\\
		\abs{\vec{n}}\neq0
		\end{subarray}
		}%
		\frac{2}{M_P^2}\bigg[\frac{L^2\abs{\vec{n}}}{2 \lambda_{z}}\bigg]^{r-2}\KB_{r-2}(\lambda_{z}\abs{\vec{n}})\ \e^{\i L\vec{n}\big[\vt+\vec{q}(1-z)\big]},\\
 \f_r^{\mu}(\lambda_{z},q,\vartheta)&=\sum_{%
		\begin{subarray}{c}
		\vec{n}\in\Z{3}\\
		\abs{\vec{n}}\neq0
		\end{subarray}
		}%
		\frac{\i L}{M_P^2}\,n^{\mu}\bigg[\frac{L^2\abs{\vec{n}}}{2 \lambda_{z}}\bigg]^{r-3}\KB_{r-3}(\lambda_{z}\abs{\vec{n}})\ \e^{\i L\vec{n}\big[\vt+\vec{q}(1-z)\big]},\\
 \h_r^{\mu\nu}(\lambda_{z},q,\vartheta)&=\sum_{%
		\begin{subarray}{c}
		\vec{n}\in\Z{3}\\
		\abs{\vec{n}}\neq0
		\end{subarray}
		}%
		\frac{L^2}{2M_P^2}\,n^{\mu}n^{\nu}\bigg[\frac{L^2\abs{\vec{n}}}{2 \lambda_{z}}\bigg]^{r-4}\KB_{r-4}(\lambda_{z}\abs{\vec{n}})\ \e^{\i L\vec{n}\big[\vt+\vec{q}(1-z)\big]}.
\end{aligned} 
\end{equation}
For $q^2=0$ the functions $\g_r(\lambda_{z},q,\vartheta),\f_r^{\mu}(\lambda_{z},q,\vartheta),\h_r^{\mu\nu}(\lambda_{z},q,\vartheta)$ reduce to the expressions of \Eq\eqref{Eq:gRfRhRTransferZero}. 
Note that if $\vec{q}=\frac{2\pi}{L}\vec{l}$ with $\vec{l}\in\Z{3}$, the
results~\eqref{Eq:FishgfhR} can be simplified by means of substitutions
$z\mapsto(1-z)$ and $\vec{n}\mapsto-\vec{n}$. 
This leads to the results of section~\ref{subsec:FormFactors} and
appendix~\ref{app:WTI}.

\section{Gauge symmetry in finite volume}\label{app:GITBC}

To explain the results of section~\ref{subsec:FormFactors} we construct an
an effective theory for charged pions which is invariant under
electromagnetic gauge transformations.
The theory reproduces the expression obtained at vanishing momentum
transfer and indicates that the gauge symmetry is preserved in this case. 
Relying on this observation, we show that the Ward-Takahashi
identity~\cite{Green:53,Taka:57a,Taka:57b} holds in finite volume as long
as the momentum transfer is discrete. 
Only the differential form of the identity --- the Ward
identity~\cite{Ward:50} --- is violated due to the discretization of the
spatial components. These considerations were presented for PBC in
\Ref\cite{HuJiTi:07} and are here generalized to TBC.

\subsection{Construction of a gauge invariant effective theory}\label{app:GIEFT}

We consider a finite cubic box of side length~$L$ on which we impose TBC.
In presence of two light flavors, we can introduce the electromagnetic
gauge field through the external vector field, 
\begin{equation}
  v^\mu=-eA^\mu(x)\,\mathcal{Q}. 
\end{equation}
Here, $e$ is the elementary electric charge of the positron and $\mathcal{Q}=\diag(2/3,-1/3)$.
As long as $\mathcal{Q}$ is diagonal we may redefine the fields so that they are periodic and introduce the twisting angles by means of a constant vector field. 

Since $A^\mu(x)$ as well as other fields are periodic, we can proceed in a
similar way as in \Ref\cite{HuJiTi:07} to construct the effective theory.
The only difference is that the effective Lagrangian contains additional
terms due to the constant vector field proportional to the twisting angles. 
At low energies, the relevant degrees of freedom are pions and for
simplicity, we just consider the charged ones in the following. 
In absence of the electromagnetic interaction the Lagrangian of the
effective theory reads 
\begin{equation}
 \mathcal{L}=\frac{1}{4}\braket{\hat{D}_{\mu}\Phi[\hat{D}^{\mu}\Phi]^{\dagger}-M_{\Pc}^2(L)\,\Phi^\dagger\Phi},\qquad\text{where}\qquad
        \Phi=\begin{pmatrix}
                0             & \sqrt{2}\ \Pp\\
                \sqrt{2}\ \Pm & 0                     
             \end{pmatrix},
\end{equation}
and $M_{\Pc}(L)$ is the mass of charged pions in finite volume.
The kinetic term contains the derivative
$\hat{D}^{\mu}\Phi=\partial^{\mu}\Phi-\i\left[w_{\vartheta}^{\mu},\Phi\right]$
with 
\begin{equation}
 w_{\vartheta}^{\mu}=\left(\tPp^\mu+\Delta\vartheta^\mu_{\Gamma_{\Pp}}\right)\frac{\tau_3}{2}.
\end{equation}
The constant vector field $w_{\vartheta}^{\mu}$ is proportional to the
third Pauli matrix, $\tau_3=\diag(1,-1)$ and introduces the twisting angle
$\tPp^\mu$ as well as the extra term $\Delta\vartheta^\mu_{\Gamma_{\Pp}}$. 
Here, $\tPp^\mu$, $\Delta\vartheta^\mu_{\Gamma_{\Pp}}$ break Lorentz invariance. 
For $\tPp^\mu\to0$ the field $w_{\vartheta}^{\mu}$ disappears and the cubic
invariance is restored: in this case the theory respects PBC. 
Note that $M_{\Pc}(L)$, $\Delta\vartheta^\mu_{\Gamma_{\Pp}}$ implicitly
depend on parameters of the effective theory (like the LECs).

To add the electromagnetic interaction we must include all possible
operators which are invariant under gauge transformations. 
This can be achieved using Wilson loops, see \Ref\cite{HuJiTi:07}.
We limit ourselves to include operators containing the zero mode of the
gauge field $A^\mu(x)$ as these allow us to study the electromagnetic form
factor at vanishing momentum transfer. Proceeding in a similar way as
\Ref\cite{HuJiTi:07} we obtain the following effective Lagrangian in
presence of the electromagnetic interaction,  
\begin{equation}\label{Eq:LaGIEFT}
 \mathcal{L}=\frac{1}{4}\braket{\hat{D}_{\mu}\Phi[\hat{D}^{\mu}\Phi]^{\dagger}-M_{\Pc}^2(L)\,\Phi^\dagger\Phi}-\frac{\i}{2}Q(L)_{\mu\nu}W_{-}^\mu\braket{\mathcal{Q}[(\hat{D}^{\nu}\Phi)^\dagger\Phi-\Phi^\dagger(\hat{D}^{\nu}\Phi)]}+\dots, 
\end{equation}
where $\hat{D}^{\mu}\Phi=\partial^{\mu}\Phi+\i
eA^\mu(x)[\mathcal{Q},\Phi]-\i\left[w_{\vartheta}^{\mu},\Phi\right]$. 
The operator $W^\mu_-=(\begin{smallmatrix}0\\\vec{W}_-\end{smallmatrix})$
  is constructed from Wilson loops, see \Ref\cite{HuJiTi:07}. 
The expression~\eqref{Eq:LaGIEFT} needs some explanations.
The dots at the end indicate that we have just written down the relevant
terms of the effective Lagrangian. The most general effective Lagrangian
contains terms with arbitrary many insertions of $W^\mu_-$, which we are
not writing explicitly. The expansion of $W^\mu_-$ starts with a term
linear in the zero mode which allows us to study the electromagnetic form
factor at vanishing momentum transfer. The tensor~$Q(L)^{\mu\nu}$ breaks
the Lorentz as well as the cubic invariances and must be determined by
matching. 
For $\tPp^\mu=0$ we expect that~$Q(L)^{\mu\nu}$ reproduces the result for
PBC~\cite{HuJiTi:07} and that it disappears for $L\to\infty$. 

We match~$Q(L)^{\mu\nu}$ with the results~\eqref{Eq:DGVPc0} of
section~\ref{subsec:FormFactors}. From the Lagrangian~\eqref{Eq:LaGIEFT} we
take the terms linear in the zero mode and evaluate them at the first order
in $e$. We obtain
\begin{equation}
 \braket{\Pc|J^\mu|\Pc}=2e\,Q_e\left[(p+\tPc+\Delta\vartheta_{\Gamma_{\Pc}})^\mu+\frac{L}{3}\,(p+\tPc)_\nu Q(L)^{\mu\nu}\right]+\Ord{e^2},
\end{equation}
where $Q_e=\pm1$ is the electric charge of $\Pc$ in elementary units.
Matching this expression with \Eq\eqref{Eq:DGVPc0} we find
\begin{equation}\label{Eq:QLmunu}
 Q(L)^{\mu\nu}=\frac{6}{L}\,\xi_\pi\,\h_2^{\mu\nu}(\lambda_{\pi},\tPp),
\end{equation}
where $\h_2^{\mu\nu}(\lambda_{\pi},\tPp)$ is defined in
\Eq\eqref{Eq:gRfRhRTransferZero}. For $\tPc^\mu=0$ the
tensor~\eqref{Eq:QLmunu} coincides with the result of~\Eq(33) of
\Ref\cite{HuJiTi:07}. 

The effective theory of \Eq\eqref{Eq:LaGIEFT} reproduces the expression of
the vector form factor 
at vanishing momentum transfer. 
The presence of Wilson loops ensures that the theory is invariant under
gauge transformations. As long as $A^\mu(x)$ is periodic this invariance is
preserved.  Starting from this observation, we show in
appendix~\ref{app:WTI} that the Ward-Takahashi identity holds for TBC  
and that the corrections to the vector form factor are related to
inverse propagators.

\subsection{Ward-Takahashi identity}\label{app:WTI}

In infinite volume gauge symmetry implies that the electromagnetic
vertex function~$\Gamma^\mu$ satisfies the Ward-Takahashi
identity~\cite{Green:53,Taka:57a,Taka:57b}: 
\begin{equation}\label{Eq:WTI}
 -\i q_\mu\Gamma^\mu=\i Q_{e}\left[\Delta^{-1}(\pp)-\Delta^{-1}(p)\right].
\end{equation}
Here, $q^\mu=(\pp-p)^\mu$ is the momentum transfer, $\Delta(\pp)$
resp. $\Delta(p)$ are the propagators of outgoing and incoming particles
and $Q_{e}=Q/e$ is the electric charge of external particles in units of
the positron charge. 
In the limit $q^\mu\to0$, the identity tends to a differential form, known
as Ward identity~\cite{Ward:50}: 
\begin{equation}\label{Eq:WI}
 -\i \Gamma^\mu=\i Q_{e}\,\frac{\partial}{\partial p_\mu}\Delta^{-1}(p).
\end{equation}
For external charged pions, we can calculate the electromagnetic vertex
function from the matrix elements,
$\i\Gamma^\mu=\bra{\Pc(\pp)}V_{3}^{\mu}\ket{\Pc(p)}$. 
In \Ref\cite{GaLe:84} these matrix elements are evaluated in ChPT at NLO
and amount to  
\begin{equation}\label{Eq:VertexFct}
 \begin{aligned}
  \Gamma^\mu&=Q_e\left\{(\pp+p)^\mu\left[1+f(q^2)\right]-\frac{q^\mu}{q^2}(\pp^2-p^2)\,f(q^2)\right\},\\ 
  f(q^2)    &=\frac{1}{6F_\pi^2}\left[\left(q^2-4M_\pi^2\right)\bar{J}(q^2)+\frac{q^2}{(4\pi)^2}\left(\elb_6-\frac{1}{3}\right)\right]+\Ord{q^4},
 \end{aligned}
\end{equation}
where $\bar{J}(q^2)$ is the finite part of the loop integral~\eqref{Eq:Jq2}. 
Here, we display all terms, even those that disappear as external momenta
are on-shell. For on-shell momenta only the term proportional to
$(\pp+p)^\mu$ contributes and provides the vector form factor, see
\Ref\cite{GaLe:84}. One can show that the vertex
function~\eqref{Eq:VertexFct} satisfies the Ward-Takahashi identity by
contracting with $q_\mu$ and arranging the surviving terms in inverse
propagators. Taking the limit $q^\mu\to0$, the same vertex function
satisfies the Ward identity, indicating that the electromagnetic current as
well as the electric charge are conserved. 

In finite volume the vertex function receives additional corrections:
$\Gamma^\mu(L)=\Gamma^\mu+\Delta\Gamma^\mu$. 
The first term corresponds to \Eq\eqref{Eq:VertexFct} with momenta shifted
by $\vartheta^\mu_{\Pc}$,  
\begin{equation}\label{Eq:VertexfctFV1}
 \Gamma^\mu=Q_e\left\{P^\mu[1+f(q^2)]-\frac{q^\mu}{q^2}\left(P_\nu q^\nu\right)f(q^2)\right\},
\end{equation}
where $q^\mu,P^\mu$ are defined in \Eqs(\ref{Eq:Transfer} \ref{Eq:PDTPc}).
The second term includes corrections arising from loop diagrams,
\begin{equation}\label{Eq:VertexfctFV2}
 \begin{aligned}
   \Delta\Gamma^\mu&=Q_e\left\{P^\mu \G_1+2\HH_2^{\mu\nu}P_\nu-q^\mu\F_2^\nu P_\nu
                            -\frac{P_\nu q^\nu}{q^2}\left[q^\mu \G_1+2\HH_2^{\mu\rho}q_\rho-q^\mu\F_2^\rho q_\rho\right]\right\}\\
                   &+Q_e\left\{2 \Delta\tGammaPc^\mu + \left[2M_\pi^2-\left(\pp+\tPc\right)^2-\left(p+\tPc\right)^2-q^2\right]\Delta\TPc^\mu\right\}.
 \end{aligned}
\end{equation}
The Lorentz vectors $\Delta\tGammaPc^\mu$, $\Delta\TPc^\mu$ are given in
\Eqs(\ref{Eq:DtL}, \ref{Eq:PDTPc}) and the new functions are defined as 
\begin{equation}
\begin{aligned}
 \G_1          &=\xi_\pi\bigg[\intz\g_1(\lambda_{z},q,\tPp)-\g_1(\lambda_{},\tPp)\bigg],\\
 \F_2^\mu      &=\xi_\pi\intz\left(1-2z\right)\f_2^{\mu}(\lambda_{z},q,\tPp),\\ 
 \HH_2^{\mu\nu}&=\xi_\pi\intz\h_2^{\mu\nu}(\lambda_{z},q,\tPp).
\end{aligned}
\end{equation}
In the case of on-shell momenta, the second term $\Delta\Gamma^\mu$ reduces
to the corrections~\eqref{Eq:DGVPc}. 
We note that
\begin{equation}\label{Eq:PI}
 \begin{aligned}
     \F_2^\mu    &=\frac{2}{q^2}\, q_\nu\HH_2^{\mu\nu},\\ 
  \Delta\TPc^\mu &=\pm\xi_{\pi}\intz\left[\f_2^{\mu}(\lambda_{z},q,\tPp)+ q^{\mu}\left(1/2-z\right)\g_2(\lambda_{z},q,\tPp)\right]\\
                 &=\pm\xi_{\pi}\intz\left\{\f_2^{\mu}(\lambda_{z},q,\tPp)-\frac{q^{\mu}}{q^2}\left[q_\nu\f_2^{\nu}(\lambda_{z},q,\tPp)\right]\right\}, 
 \end{aligned}
\end{equation}
if $q^\mu$ is non-vanishing and if $\vec{q}=\frac{2\pi}{L}\vec{l}$ with
$\vec{l}\in\Z{3}\setminus\set{\vec{0}}$. 
These relations can be obtained by partial integration and by using
properties of the derivatives of the modified Bessel functions of second
kind. 

We now show that in this case, the Ward-Takahashi identity holds in finite volume.
We contract the vertex function with $q_\mu$ and use the relations~(\ref{Eq:PI}).
The term $q_\mu\Delta\TPc^\mu$ disappears and many others mutually cancel.
The surviving terms can be arranged to form inverse propagators, 
\begin{align}\label{Eq:WTIFVC}
  -\i q_\mu\Gamma^\mu(L)&=-\i Q_{e}\Big[q_\mu P^\mu+2 q_\mu\Delta\tGammaPc^\mu\Big]=\notag\\
                        &=\i Q_{e}\Big[(p+\tPc)^2+2(p+\tPc)_\mu\Delta\tGammaPc^\mu-\big(\pp+\tPc\big)^2-2\big(\pp+\tPc\big)_\mu\Delta\tGammaPc^\mu\Big]\notag\\
                        &=\i Q_{e}\Big[\Delta_{\Pc,L}^{-1}(\pp)-\Delta_{\Pc,L}^{-1}(p)\Big].
\end{align}
In the last step of \Eq\eqref{Eq:WTIFVC} we added terms canceling each other
and used the fact that at NLO the extra terms $\Delta\tGammaPc^\mu$
coincide with those of self-energies,
$\Delta\vartheta^\mu_{\Sigma_{\Pc}}$. 
This allows us to form the propagators with self-energies $\Delta\Sigma_{\Pc}$ at NLO,
\begin{equation}
 \Delta_{\Pc,L}(p)=\frac{1}{M_\pi^2-(p+\tPc)^2-\Delta\Sigma_{\Pc}}.
\end{equation}

Eq.~\eqref{Eq:WTIFVC} shows that the Ward-Takahashi identity holds even for
TBC. Necessary conditions are: the discretization of $q^\mu$ and that 
$\Delta\tGammaPc^\mu$ coincide with $\Delta\vartheta^\mu_{\Sigma_{\Pc}}$. 
Note that the limit $q^\mu\to0$ can not be taken due to the discretization
of $q^\mu$. This invalidates the differential form of the identity,
i.e. the Ward identity~\eqref{Eq:WI}. In this case, the Ward identity is
violated for the spatial components but it remains valid for the zeroth
component.

\section{\boldmath Terms \texorpdfstring{$S^{(4)}$}{S}}\label{app:IntegralsS4}

We list some explicit expressions of the terms $S^{(4)}$ introduced in
section~\ref{sec:ApplicationAF} indicating the equation where they appear.
Other terms $S^{(4)}$ can be found in appendix~A of \Ref\cite{CoDuHae:05}.

\subsection{Pions}

We begin with the terms $S^{(4)}$ appearing in the asymptotic formulae for pions.
Note that the functions $R_{0}^{k}$, $(R_{0}^{k})'$, $(R_{0}^{k})''$, $Q_{0}^{k}$, $(Q_{0}^{k})'$ are defined in \Eq\eqref{Eq:R0kQ0k}.

\noindent
The terms appearing in \Eq\eqref{Eq:IMPn} are
\begin{equation}\label{Eq:S4MPn}
  S^{(4)}(M_{\Pn},\Pn)=3R_0^0-8R_0^1-8R_0^2\quad\text{and}\quad
  S^{(4)}(M_{\Pn},\Pc)=\frac{4}{3}\left(R_0^0+2R_0^1-4R_0^2\right).
\end{equation} 
The terms appearing in \Eq\eqref{Eq:IDMPc} are
\begin{equation}\label{Eq:SD4MPc}
 \begin{aligned}
  S_D^{(4)}(M_{\Pc},\Pn)&=\frac{4}{3}\left(R_0^1-4R_0^2\right)-\frac{4}{3}\left[(R_0^1)'-2(R_0^2)'-4(R_0^3)'\right],\\
  S_D^{(4)}(M_{\Pc},\Pc)&=-\frac{4}{3}\left(5 R_0^1+16 R_0^2\right)-\frac{2}{3}\left[11(R_0^1)'+20(R_0^2)'-32(R_0^3)'\right].
 \end{aligned}
\end{equation} 
The term appearing in \Eq\eqref{Eq:IMvtwPc} is
\begin{equation}\label{Eq:SM4vtwPc}
 S^{(4)}(\tSPp)=-\frac{1}{3}\left(11R_0^1+20R_0^2-8R_0^3\right).
\end{equation}
The terms appearing in \Eq\eqref{Eq:IDFPc} are
\begin{equation}\label{Eq:SD4FPc}
 \begin{aligned}
  S^{(4)}_D(F_{\Pc},\Pn)&=+\frac{4}{3}\left(R_0^1-4R_0^2\right)-\frac{2}{3}\left[3(R_0^1)'-8(R_0^2)'-8(R_0^3)'\right]\\
                        &\phantom{=}\
                          +\!\frac{2}{3}\left[(R_0^1)''-2(R_0^2)''-4(R_0^3)''\right],\\
  S^{(4)}_D(F_{\Pc},\Pc)&=-\frac{8}{3}\left(R_0^1+8 R_0^2\right)+\frac{2}{3}\left[3(R_0^1)'+8(R_0^2)'+32(R_0^3)'\right]\\
                        &\phantom{=}\
                          +\!\frac{1}{3}\left[11(R_0^1)''+20(R_0^2)''-32(R_0^3)''\right].
 \end{aligned}
\end{equation}
The term appearing in \Eq\eqref{Eq:IFvtwPc} is
\begin{equation}\label{Eq:SF4vtwPp}
 S^{(4)}(\tAPp)=-\frac{2}{3}\left(R_0^1+4R_0^2-4R_0^3\right)+\frac{1}{6}\left[11(R_0^1)'+20(R_0^2)'-8(R_0^3)'\right].
\end{equation}
The terms appearing in \Eq\eqref{Eq:IGPn} are
\begin{equation}\label{Eq:S4GPn}
 \begin{aligned}
  S^{(4)}(G_{\Pn},\Pn)&=-3R_0^0+4R_0^1-\frac{1}{2}\left[3(R_0^0)'-8(R_0^1)'-8(R_0^2)'\right]\\
  S^{(4)}(G_{\Pn},\Pc)&=-\frac{2}{3}\left[(R_0^0)'+2(R_0^1)'-4(R_0^2)'\right].
 \end{aligned}
\end{equation} 
The terms appearing in \Eq\eqref{Eq:IGSPn} are
\begin{equation}\label{Eq:S4GSPn}
 \begin{aligned}
  S^{(4)}(\GSPn,\Pn)&=5\left(3R_0^0-8R_0^1-8R_0^2\right)-(3Q_0^0-8Q_0^1-8Q_0^2),\\
  S^{(4)}(\GSPn,\Pc)&=\frac{20}{3}\left(R_0^0+2R_0^1-4R_0^2\right)-\frac{4}{3}\left(Q_0^0+2Q_0^1-4Q_0^2\right).
 \end{aligned}
\end{equation}
The terms appearing in \Eqs(\ref{Eq:IGSPc}, \ref{Eq:IDGSPc}, \ref{Eq:IGSvTwPc}) are
\begin{equation}\label{Eq:S4GSPcSD4GSPcSGS4vTwPc} 
 \begin{aligned}
    S^{(4)}(\GSPc,\Pn)&=\frac{10}{3}\left(R_0^0+2R_0^1-4R_0^2\right)-\frac{2}{3}\left(Q_0^0+2Q_0^1-4Q_0^2\right),\\
    S^{(4)}(\GSPc,\Pc)&=\frac{5}{3}\left(11R_0^0-20R_0^1-32R_0^2\right)-\frac{1}{3}\left(11Q_0^0-20Q_0^1-32Q_0^2\right),\\
  S_D^{(4)}(\GSPc,\Pn)&=+\frac{4}{3}[4-C_{\Pc}]\left[R_0^1-4R_0^2-(R_0^1)'+2(R_0^2)'+4(R_0^3)'\right]\\
                      &\phantom{=}\ 
                        -\!\frac{4}{3}\left[Q_0^1-4Q_0^2-(Q_0^1)'+2(Q_0^2)'+4(Q_0^3)'\right],\\
  S_D^{(4)}(\GSPc,\Pc)&=-\frac{2}{3}[4-C_{\Pc}]\left[10 R_0^1+32 R_0^2+11(R_0^1)'+20(R_0^2)'-32(R_0^3)'\right]\\
                      &\phantom{=}\ 
                        +\!\frac{2}{3}\left[10 Q_0^1+32 Q_0^2+11(Q_0^1)'+20(Q_0^2)'-32(Q_0^3)'\right],\\
   S^{(4)}(\TPp)&=-\frac{5}{3}\left(11R_0^1+20R_0^2-8R_0^3\right)+\frac{1}{3}\left(11Q_0^1+20Q_0^2-8Q_0^3\right).
 \end{aligned}
\end{equation}
Here, $C_{\Pc}=M_\pi/(M_\pi+D_{\Pc})$ and $D_{\Pc}=\sqrt{M_\pi^2+\abs{\vtPc}^2}-M_\pi$.\\ 
\noindent
The functions $R_{0}^{k}$, $(R_{0}^{k})'$, $(R_{0}^{k})''$, $Q_{0}^{k}$, $(Q_{0}^{k})'$ entering the above expressions are defined as 
\begin{align}\label{Eq:R0kQ0k}
  (R_{0}^{k})^{(\ ,\prime,\prime\prime)}
  &=(R_{0}^{k})^{(\ ,\prime,\prime\prime)}(\lambda_{\pi}\abs{\vec{n}})\notag\\
  &=\begin{cases}\Re\\\Im\end{cases}\!\!\int_{\R{}}\d{y}{}y^k\e^{-\lambda_{\pi}\abs{\vec{n}}\sqrt{1+y^2}}g^{(\ ,\prime,\prime\prime)}(2+2\i y)\quad\text{for}\ \begin{cases}k\ \text{even}\\k\ \text{odd}\end{cases}\!\!,\notag\\
     (Q_{0}^{k})^{(\ ,\prime)}&=(Q_{0}^{k})^{(\ ,\prime)}(\lambda_{\pi}\abs{\vec{n}})\\
              &=\begin{cases}\Re\\\Im\end{cases}\!\!\int_{\R{}}\d{y}{}y^k\lambda_{\pi}\abs{\vec{n}}\sqrt{1+y^2}\e^{-\lambda_{\pi}\abs{\vec{n}}\sqrt{1+y^2}}g^{(\ ,\prime)}(2+2\i y)\quad\text{for}\ \begin{cases}k\ \text{even}\\k\ \text{odd}\end{cases}\!\!,\notag
\end{align}
where 
\begin{equation}
  g(x)=\sigma\log\left(\frac{\sigma-1}{\sigma+1}\right)+2 \qquad\text{with}\qquad\sigma=\sqrt{1-4/x},
\end{equation}
and $g'(x)$, $g''(x)$ are the first and second derivative of $g(x)$ with respect to $x$.
Note that $g(x)=(4\pi)^2\,\bar{J}(xM_\pi^2)$ with $\bar{J}(q^2)=J(q^2)-J(0)$ the loop-integral function evaluated in $d=4$ dimensions,
\begin{equation}\label{Eq:Jq2}
     J(q^2)=\int\frac{\d{k}{d}}{(2\pi)^d}\,\frac{1}{\i\left[M_\pi^2-(k+q)^2\right]\left[M_\pi^2-k^2\right]}.
\end{equation}

\subsection{Kaons}

We list the terms $S^{(4)}$ appearing in the asymptotic formulae for kaons.
Note that the functions $S^{kl}_{P Q}$ are defined in \Eq\eqref{Eq:STPsQskl}. 
In the next expressions, we denote the ratio of mass squares as $x_{P Q}=M_P^2/M_Q^2$ for $P,Q=\pi,K,\eta$. 

\noindent
The term appearing in \Eq\eqref{Eq:IDMKc} is
 \begin{align}\label{Eq:SD4MKc}
 S_D^{(4)}(M_{\Kc},\Pn)=\bigg\{&\!\!-\frac{5}{8}(1+x_{\pi K})S^{1,1}_{K\pi}-\frac{3}{16}(1+x_{\pi K})^2 S^{1,2}_{K\pi}+\frac{13}{8}(1-x_{\pi K}) S^{1,3}_{K\pi} \bigg.\notag\\
                               &\!\!+\frac{3}{8}(1-x_{\pi K}^2)S^{1,4}_{K\pi}-\frac{19}{4}\big(S^{2,1}_{K\pi}-S^{3,2}_{K\pi}\big)-\frac{5}{4}(1+x_{\pi K}) S^{2,2}_{K\pi}\notag\\
                               &\!\!+\frac{13}{4}(1-x_{\pi K}) S^{2,4}_{K\pi}+3\big(x_{\pi K} S^{1,6}_{K\pi}-S^{2,5}_{K\pi}+S^{3,6}_{K\pi}\big)-\frac{1}{8}(1+x_{\pi K}) S^{1,1}_{\eta K}\notag\\ 
                               &\!\!-\frac{1}{48}(1+x_{\pi K})^2 S^{1,2}_{\eta K}+\frac{3}{8}(1-2x_{\pi K}+x_{\eta K}) S^{1,3}_{\eta K}-\frac{1}{4}(1+x_{\pi K}) S^{2,2}_{\eta K}\notag\\ 
                               &\!\!+\frac{1}{8}(1+x_{\pi K})(5-2x_{\pi K}-3x_{\eta K})S^{1,4}_{\eta K}-\frac{3}{4}\big(S^{2,1}_{\eta K}-S^{3,2}_{\eta K}\big)\notag\\ 
                               &\!\!+\frac{3}{4}(1-2x_{\pi K}+x_{\eta K}) S^{2,4}_{\eta K}\bigg.+3\big(x_{\pi K}S^{1,6}_{\eta K}-S^{2,5}_{\eta K}+S^{3,6}_{\eta K}\big)\bigg\}x_{\pi K}^{-1/2}.
\end{align}
\noindent
The term appearing in \Eq\eqref{Eq:IMvtwKc} is
\begin{align}\label{Eq:S4MvtwKp}
  S^{(4)}(\tSKp)=\bigg\{&\!\!-\frac{3}{16}(1+x_{\pi K})^2S^{1,1}_{K\pi}+\frac{3}{8}(1-x_{\pi K}^2)S^{1,3}_{K\pi}-\frac{5}{4}(1+x_{\pi K}) S^{2,1}_{K\pi}\bigg.\notag\\
                        &\!\!+\frac{13}{4}(1-x_{\pi K}) S^{2,3}_{K\pi}+\frac{3}{4}S^{3,1}_{K\pi}+3\big(x_{\pi K}S^{1,5}_{K\pi}+S^{3,5}_{K\pi}\big)-\frac{1}{48}(1+x_{\pi K})^2S^{1,1}_{\eta K}\notag\\
                        &\!\!+\frac{1}{8}(1+x_{\pi K})\big(5-2x_{\pi K}-3x_{\eta K}\big)S^{1,3}_{\eta K}-\frac{1}{4}(1+x_{\pi K})S^{2,1}_{\eta K}\notag\\ 
                        &\!\!\bigg.+\frac{3}{4}(1-2x_{\pi K}+x_{\eta K}) S^{2,3}_{\eta K}+\frac{3}{4}S^{3,1}_{\eta K}+3\big(x_{\pi K}S^{1,5}_{\eta K}+S^{3,5}_{\eta K}\big)\bigg\}x_{\pi K}^{-1}.
 \end{align}
The term appearing in \Eq\eqref{Eq:IDFKc} is
\begin{align}\label{Eq:SD4FKc}
 S_D^{(4)}(F_{\Kc},\Pn)=\bigg\{&\!\!-\frac{5}{16}(1+x_{\pi K})(S^{1,1}_{K\pi}-S^{1,2}_{K\pi})+\frac{1}{8}(16-5x_{\pi K})S^{1,3}_{K\pi}-\frac{1}{4}(4-7x_{\pi K})S^{1,4}_{K\pi}\bigg.\notag\\
                               &\!\!+\frac{3}{32}(1+x_{\pi K})^2S^{1,7}_{K\pi}-\frac{3}{16}(1-x_{\pi K}^2)S^{1,8}_{K\pi}+\frac{1}{8}(14-5x_{\pi K})S^{2,2}_{K\pi}\notag\\
                               &\!\!+\frac{1}{4}(16-5x_{\pi K})S^{2,4}_{K\pi}+\frac{5}{8}(1+x_{\pi K})S^{2,7}_{K\pi}-\frac{13}{8}(1-x_{\pi K})S^{2,8}_{K\pi}\notag\\
                               &\!\!+\frac{3}{2}\big(x_{\pi K}S^{1,6}_{K\pi}-x_{\pi K}S^{1,9}_{K\pi}+S^{2,6}_{K\pi}-S^{3,9}_{K\pi}\big)-\frac{19}{8}\big(2\,S^{2,1}_{K\pi}-2\,S^{3,2}_{K\pi}+S^{3,7}_{K\pi}\big)\notag\\
                               &\!\!+\frac{5}{2}\big(S^{2,3}_{K\pi}-S^{3,4}_{K\pi}\big)-3\big(S^{2,5}_{K\pi}-S^{3,6}_{K\pi}\big)-\frac{1}{16}(1+x_{\pi K})(S^{1,1}_{\eta K}-S^{1,2}_{\eta K})\notag\\
                               &\!\!+\frac{1}{8}(4-2x_{\pi K}+3x_{\eta K})S^{1,3}_{\eta K}+\frac{1}{4}(2+2x_{\pi K}-3x_{\eta K})S^{1,4}_{\eta K}\notag\\
                               &\!\!-\frac{1}{16}(1+x_{\pi K})\big(5-2x_{\pi K}-3x_{\eta K}\big)S^{1,8}_{\eta K}+\frac{1}{96}(1+x_{\pi K})^2S^{1,7}_{\eta K}\notag\\
                               &\!\!+\frac{1}{8}(2-x_{\pi K})S^{2,2}_{\eta K}+\frac{1}{4}(4-2x_{\pi K}+3x_{\eta K})S^{2,4}_{\eta K}-\frac{3}{8}(1-2x_{\pi K}+x_{\eta K})S^{2,8}_{\eta K}\notag\\
                               &\!\!+\frac{3}{2}\big(S^{2,3}_{\eta K}-S^{3,4}_{\eta K}\big)-\frac{3}{8}\big(2\,S^{2,1}_{\eta K}-2\,S^{3,2}_{\eta K}+S^{3,7}_{\eta K}\big)+\frac{1}{8}(1+x_{\pi K})S^{2,7}_{\eta K}\notag\\
                               &\!\!+\frac{3}{2}\big(x_{\pi K}S^{1,6}_{\eta K}-x_{\pi K}S^{1,9}_{\eta K}+S^{2,6}_{\eta K}-S^{3,9}_{\eta K}\big)-3\big(S^{2,5}_{\eta K}-S^{3,6}_{\eta K}\big)\bigg.\bigg\}x_{\pi K}^{-1/2}.
\end{align}
\noindent
The term appearing in \Eq\eqref{Eq:IFvtwKc} is
\begin{align}\label{Eq:S4FvtwKp}
  S^{(4)}(\tAKp)=\bigg\{&\!\!+\frac{3}{32}(1+x_{\pi K})^2S^{1,2}_{K\pi}-\frac{3}{16}(1-5x_{\pi K})S^{1,3}_{K\pi}\bigg.-\frac{3}{16}(1-x_{\pi K}^2)S^{1,4}_{K\pi}\notag\\
                        &\!\!-\frac{5}{8}(1+x_{\pi K})\big(S^{2,1}_{K\pi}-S^{2,2}_{K\pi}\big)+\frac{1}{4}(16-5x_{\pi K}) S^{2,3}_{K\pi}-\frac{13}{8}(1-x_{\pi K})S^{2,4}_{K\pi}\notag\\
                        &\!\!+\frac{3}{4}S^{3,1}_{K\pi}-\frac{3}{8}S^{3,2}_{K\pi}-\frac{5}{2}S^{3,3}_{K\pi}+3\,S^{3,5}_{K\pi}+\frac{3}{2}\big(x_{\pi K}S^{1,5}_{K\pi}-x_{\pi K}S^{1,6}_{K\pi}-S^{3,6}_{K\pi}\big)\notag\\
                        &\!\!+\frac{1}{96}(1+x_{\pi K})^2S^{1,2}_{\eta K}+\frac{1}{16}(11+2x_{\pi K}-9x_{\eta K})S^{1,3}_{\eta K}\notag\\ 
                        &\!\!-\frac{1}{16}(1+x_{\pi K})(5-2x_{\pi K}-3x_{\eta K})S^{1,4}_{\eta K}-\frac{1}{8}(1+x_{\pi K})\big(S^{2,1}_{\eta K}-S^{2,2}_{\eta K}\big)\notag\\ 
                        &\!\!+\frac{1}{4}(4-2x_{\pi K}+3x_{\eta K}) S^{2,3}_{\eta K}-\frac{3}{8}(1-2x_{\pi K}+x_{\eta K})S^{2,4}_{\eta K}+\frac{3}{4}S^{3,1}_{\eta K}-\frac{3}{8}S^{3,2}_{\eta K}\notag\\
                        &\!\!-\frac{3}{2}S^{3,3}_{\eta K}+3\,S^{3,5}_{\eta K}\bigg.+\frac{3}{2}\big(x_{\pi K}S^{1,5}_{\eta K}-x_{\pi K}S^{1,6}_{\eta K}-S^{3,6}_{\eta K}\big)\bigg\}x_{\pi K}^{-1}.
\end{align}
\noindent
The functions $S^{k,l}_{P Q}$ entering the above expressions are defined as
\begin{align}\label{Eq:STPsQskl}
  S^{k,l}_{P Q}&=S^{k,l}_{P Q}(\lambda_{\pi}\abs{\vec{n}})\\
               &=\begin{cases}\Re\\\Im\end{cases}\!\!\!\!
                                Nx_{\pi K}^{(k+1)/2}\!\int_{\R{}}\!\!\d{y}{}y^k\e^{-\lambda_{\pi}\abs{\vec{n}}\sqrt{1+y^2}}g^{(l)}_{P Q}(M_K^2\!+\!M_\pi^2\!+\!2\i M_K M_\pi y),\quad\text{for}\begin{cases}k\ \text{even}\\k\ \text{odd}\end{cases}\!\!\!.\notag
\end{align}
Here, $N=(4\pi)^2$ and 
\begin{equation}
 \begin{aligned}
  g^{(1)}_{PQ}(x)&=\bar{J}_{PQ}(x), &&&&& g^{(2)}_{PQ}(x)&=M_K^2\,\bar{J}'_{PQ}(x), &&&&& g^{(7)}_{PQ}(x)&=M_K^4\,\bar{J}''_{PQ}(x),\\
  g^{(3)}_{PQ}(x)&=K_{PQ}(x),       &&&&& g^{(4)}_{PQ}(x)&=M_K^2\,K'_{PQ}(x),       &&&&& g^{(8)}_{PQ}(x)&=M_K^4\,K''_{PQ}(x),\\
  g^{(5)}_{PQ}(x)&=\bar{M}_{PQ}(x), &&&&& g^{(6)}_{PQ}(x)&=M_K^2\,\bar{M}'_{PQ}(x), &&&&& g^{(9)}_{PQ}(x)&=M_K^4\,\bar{M}''_{PQ}(x).
 \end{aligned}
\end{equation}
The explicit forms of $g^{(l)}_{PQ}(x)$ were presented in \Ref\cite{BiCoGa:94}. 
They can be expressed in terms of the loop-integral function $\bar{J}_{PQ}(q^2)=J_{PQ}(q^2)-J_{PQ}(0)$ evaluated in $d=4$ dimensions,
\begin{equation}
     J_{PQ}(q^2)=\int\frac{\d{k}{d}}{(2\pi)^d}\,\frac{1}{\i\big[M_P^2-(k+q)^2\big]\big[M_Q^2-k^2\big]}.
\end{equation}
Using the abbreviations ($t=q^2$, $M=M_P$, $m=M_Q$), 
\begin{equation*}
\begin{aligned}
 \bar{J}(t)&=\bar{J}_{PQ}(t), &&& K(t)&= K_{PQ}(t) &&&\bar{M}(t)&=\bar{M}_{PQ}(t) \\
  \Delta&=M^2-m^2,   &&&   \Sigma&=M^2+m^2, &&& \rho&=(t+\Delta)^2-4tM^2, 
\end{aligned}
\end{equation*}
the above functions take the forms
\begin{equation}\label{Eq:JKM}
 \begin{aligned}
  \bar{J}(t)&=\frac{1}{2N}\bigg[2+\frac{\Delta}{t}\ln\frac{m^2}{M^2}-\frac{\Sigma}{\Delta}\ln\frac{m^2}{M^2}-\frac{\sqrt{\rho}}{t}\ln\frac{(t+\sqrt{\rho})^2-\Delta^2}{(t-\sqrt{\rho})^2-\Delta^2}\bigg],  \\      K(t)&=\frac{\Delta}{2t}\,\bar{J}(t),\\
  \bar{M}(t)&=\frac{1}{12t}[t-2\Sigma]\bar{J}(t)+\frac{\Delta^2}{3t^2}\,\bar{J}(t)+\frac{1}{18N}-\frac{1}{6Nt}\bigg[\Sigma+\frac{2M^2m^2}{\Delta}\ln\frac{m^2}{M^2}\bigg].
 \end{aligned}
\end{equation}

We conclude with a remark on the use of the loop-integral functions in the
asymptotic formulae. The loop functions \eqref{Eq:JKM} need to be evaluated
for complex values of their arguments. For $\bar{J}_{PQ}(M_P^2+M_Q^2+2\i
M_P M_Q y)$ there is an ambiguity due to the negative value of
$\rho=-4M_P^2M_Q^2(1+y^2)$, which \eqref{Eq:JKM} does not resolve explicitly.
An explicit analytic continuation was provided in \Ref\cite{CoDuHae:05} but
unfortunately was not correct. The correct prescription is as follows: take the positive
value of the square root $\sqrt{\rho}=2\i M_PM_Q\,\sqrt{1+y^2}$ for which
the logarithm in~\eqref{Eq:JKM} becomes ($t=M_P^2+M_Q^2+2\i M_PM_Q\,y$)
\begin{equation}
 \ln\frac{(t+\sqrt{\rho})^2-\Delta^2}{(t-\sqrt{\rho})^2-\Delta^2}=\ln\frac{\left(1+y^2\right)^{\frac{1}{2}}+y}{\left(1+y^2\right)^{\frac{1}{2}}-y}+\i\pi,\qquad\text{for all $ y\in\R{}$} \, .
\end{equation}

\bibliographystyle{JHEP}
\bibliography{biblio_preprint} 

\providecommand{\href}[2]{#2}\begingroup\raggedright\begin{thebibliography}{10}

\bibitem{deDiPeTa:04}
G.~de~Divitiis, R.~Petronzio and N.~Tantalo, \emph{{On the discretization of
  physical momenta in lattice QCD}},
  \href{http://dx.doi.org/10.1016/j.physletb.2004.06.035}{\emph{Phys. Lett.}
  {\bf B595} (2004) 408--413},
  [\href{http://arxiv.org/abs/hep-lat/0405002}{{\tt hep-lat/0405002}}].

\bibitem{deDiTa:04}
G.~M. de~Divitiis and N.~Tantalo, \emph{{Non-leptonic two-body decay amplitudes
  from finite volume calculations}},
  \href{http://arxiv.org/abs/hep-lat/0409154}{{\tt hep-lat/0409154}}.

\bibitem{GuMeSi:05}
D.~Guadagnoli, F.~Mescia and S.~Simula, \emph{{Lattice study of semileptonic
  form factors with twisted boundary conditions}},
  \href{http://dx.doi.org/10.1103/PhysRevD.73.114504}{\emph{Phys. Rev.} {\bf
  D73} (2006) 114504}, [\href{http://arxiv.org/abs/hep-lat/0512020}{{\tt
  hep-lat/0512020}}].

\bibitem{SaVi:05}
C.~T. Sachrajda and G.~Villadoro, \emph{{Twisted boundary conditions in lattice
  simulations}},
  \href{http://dx.doi.org/10.1016/j.physletb.2005.01.033}{\emph{Phys. Lett.}
  {\bf B609} (2005) 73--85}, [\href{http://arxiv.org/abs/hep-lat/0411033}{{\tt
  hep-lat/0411033}}].

\bibitem{Ti:05}
B.~C. Tiburzi, \emph{{Flavor twisted boundary conditions and the nucleon axial
  current}},
  \href{http://dx.doi.org/10.1016/j.physletb.2005.05.006}{\emph{Phys. Lett.}
  {\bf B617} (2005) 40--48}, [\href{http://arxiv.org/abs/hep-lat/0504002}{{\tt
  hep-lat/0504002}}].

\bibitem{Ti:06}
B.~C. Tiburzi, \emph{{Flavor twisted boundary conditions and isovector form
  factors}},
  \href{http://dx.doi.org/10.1016/j.physletb.2006.08.059}{\emph{Phys. Lett.}
  {\bf B641} (2006) 342--349},
  [\href{http://arxiv.org/abs/hep-lat/0607019}{{\tt hep-lat/0607019}}].

\bibitem{JiTi:07}
F.-J. Jiang and B.~C. Tiburzi, \emph{{Flavor twisted boundary conditions, pion
  momentum, and the pion electromagnetic form factor}},
  \href{http://dx.doi.org/10.1016/j.physletb.2006.12.041}{\emph{Phys. Lett.}
  {\bf B645} (2007) 314--321},
  [\href{http://arxiv.org/abs/hep-lat/0610103}{{\tt hep-lat/0610103}}].

\bibitem{JiTi:08}
F.-J. Jiang and B.~C. Tiburzi, \emph{{Flavor twisted boundary conditions in the
  breit frame}},
  \href{http://dx.doi.org/10.1103/PhysRevD.78.037501}{\emph{Phys. Rev.} {\bf
  D78} (2008) 037501}, [\href{http://arxiv.org/abs/0806.4371}{{\tt
  0806.4371}}].

\bibitem{BriDaLuuSav:13}
R.~A. Briceno, Z.~Davoudi, T.~C. Luu and M.~J. Savage, \emph{{Two-baryon
  systems with twisted boundary conditions}},
  \href{http://dx.doi.org/10.1103/PhysRevD.89.074509}{\emph{Phys. Rev.} {\bf
  D89} (2014) 074509}, [\href{http://arxiv.org/abs/1311.7686}{{\tt
  1311.7686}}].

\bibitem{BiRe:14}
J.~Bijnens and J.~Relefors, \emph{{Masses, decay constants and electromagnetic
  form factors with twisted boundary conditions}},
  \href{http://arxiv.org/abs/1402.1385}{{\tt 1402.1385}}.

\bibitem{Lu:85}
M.~L{\"u}scher, \emph{{Volume dependence of the energy spectrum in massive
  quantum field theories. 1. Stable particle states}},
  \href{http://dx.doi.org/10.1007/BF01211589}{\emph{Commun. Math. Phys.} {\bf
  104} (1986) 177}.

\bibitem{CoHae:04}
G.~Colangelo and C.~H{\"a}feli, \emph{{An asymptotic formula for the pion decay
  constant in a large volume}},
  \href{http://dx.doi.org/10.1016/j.physletb.2004.03.080}{\emph{Phys. Lett.}
  {\bf B590} (2004) 258--264},
  [\href{http://arxiv.org/abs/hep-lat/0403025}{{\tt hep-lat/0403025}}].

\bibitem{CoDu:04}
G.~Colangelo and S.~D{\"u}rr, \emph{{The pion mass in finite volume}},
  \href{http://dx.doi.org/10.1140/epjc/s2004-01593-y}{\emph{Eur. Phys. J.} {\bf
  C33} (2004) 543--553}, [\href{http://arxiv.org/abs/hep-lat/0311023}{{\tt
  hep-lat/0311023}}].

\bibitem{CoDuHae:05}
G.~Colangelo, S.~D{\"u}rr and C.~H{\"a}feli, \emph{{Finite-volume effects for
  meson masses and decay constants}},
  \href{http://dx.doi.org/10.1016/j.nuclphysb.2005.05.015}{\emph{Nucl. Phys.}
  {\bf B721} (2005) 136--174},
  [\href{http://arxiv.org/abs/hep-lat/0503014}{{\tt hep-lat/0503014}}].

\bibitem{CoFuLa:10}
G.~Colangelo, A.~Fuhrer and S.~Lanz, \emph{{Finite-volume effects for nucleon
  and heavy meson masses}},
  \href{http://dx.doi.org/10.1103/PhysRevD.82.034506}{\emph{Phys. Rev.} {\bf
  D82} (2010) 034506}, [\href{http://arxiv.org/abs/1005.1485}{{\tt
  1005.1485}}].

\bibitem{CoHae:06}
G.~Colangelo and C.~H{\"a}feli, \emph{{Finite-volume effects for the pion mass
  at two loops}},
  \href{http://dx.doi.org/10.1016/j.nuclphysb.2006.03.010}{\emph{Nucl. Phys.}
  {\bf B744} (2006) 14--33}, [\href{http://arxiv.org/abs/hep-lat/0602017}{{\tt
  hep-lat/0602017}}].

\bibitem{Hae:08}
C.~H{\"a}feli, \emph{{Private communications and notes}},  2008.

\bibitem{Hell:37}
H.~Hellmann, \emph{{Einf{\"u}hrung in die Quantenchemie}}, p.~285.
\newblock Leipzig: Franz Deuticke, 1937.

\bibitem{Fey:39}
R.~P. Feynman, \emph{{Forces in molecules}},
  \href{http://dx.doi.org/10.1103/PhysRev.56.340}{\emph{Phys. Rev.} {\bf 56}
  (1939) 340--343}.

\bibitem{Green:53}
H.~Green, \emph{{A prerenormalized quantum electrodynamics}},
  \href{http://dx.doi.org/10.1088/0370-1298/66/10/303}{\emph{Proc. Phys. Soc.}
  {\bf A66} (1953) 873--880}.

\bibitem{Taka:57a}
Y.~Takahashi{\emph{, Nuovo Cim.} {\bf Ser 10} (1957) 370}.

\bibitem{Taka:57b}
Y.~Takahashi, \emph{{On the generalized Ward identity}},
  \href{http://dx.doi.org/10.1007/BF02832514}{\emph{Nuovo Cim.} {\bf 6} (1957)
  371}.

\bibitem{FritGellLe:73}
H.~Fritzsch, M.~Gell-Mann and H.~Leutwyler, \emph{{Advantages of the color
  octet gluon picture}},
  \href{http://dx.doi.org/10.1016/0370-2693(73)90625-4}{\emph{Phys. Lett.} {\bf
  B47} (1973) 365--368}.

\bibitem{Wei:73}
S.~Weinberg, \emph{{Non-abelian gauge theories of the strong interactions}},
  \href{http://dx.doi.org/10.1103/PhysRevLett.31.494}{\emph{Phys. Rev. Lett.}
  {\bf 31} (1973) 494--497}.

\bibitem{GaLe:85a}
J.~Gasser and H.~Leutwyler, \emph{{Chiral perturbation theory: expansions in
  the mass of the strange quark}},
  \href{http://dx.doi.org/10.1016/0550-3213(85)90492-4}{\emph{Nucl. Phys.} {\bf
  B250} (1985) 465}.

\bibitem{Co:04}
G.~Colangelo, \emph{{Finite-volume effects in chiral perturbation theory}},
  \href{http://dx.doi.org/10.1016/j.nuclphysbps.2004.11.195}{\emph{Nucl. Phys.
  Proc. Suppl.} {\bf 140} (2005) 120--126},
  [\href{http://arxiv.org/abs/hep-lat/0409111}{{\tt hep-lat/0409111}}].

\bibitem{GaLe:86}
J.~Gasser and H.~Leutwyler, \emph{{Light quarks at low temperatures}},
  \href{http://dx.doi.org/10.1016/0370-2693(87)90492-8}{\emph{Phys. Lett.} {\bf
  B184} (1987) 83}.

\bibitem{GaLe:87:PL}
J.~Gasser and H.~Leutwyler, \emph{{Thermodynamics of chiral symmetry}},
  \href{http://dx.doi.org/10.1016/0370-2693(87)91652-2}{\emph{Phys. Lett.} {\bf
  B188} (1987) 477}.

\bibitem{GaLe:87:NP}
J.~Gasser and H.~Leutwyler, \emph{{Spontaneously broken symmetries: effective
  Lagrangians at finite volume}},
  \href{http://dx.doi.org/10.1016/0550-3213(88)90107-1}{\emph{Nucl. Phys.} {\bf
  B307} (1988) 763}.

\bibitem{HaLe:89}
P.~Hasenfratz and H.~Leutwyler, \emph{{Goldstone boson related finite-size
  effects in field theory and critical phenomena with $\mathrm{O}(N)$
  symmetry}}, \href{http://dx.doi.org/10.1016/0550-3213(90)90603-B}{\emph{Nucl.
  Phys.} {\bf B343} (1990) 241--284}.

\bibitem{Han:90}
F.~Hansen, \emph{{Finite-size effects in spontaneously broken
  $\mathrm{SU}(N)\times\mathrm{SU}(N)$ theories}},
  \href{http://dx.doi.org/10.1016/0550-3213(90)90405-3}{\emph{Nucl. Phys.} {\bf
  B345} (1990) 685--708}.

\bibitem{HanLe:90}
F.~Hansen and H.~Leutwyler, \emph{{Charge correlations and topological
  susceptibility in QCD}},
  \href{http://dx.doi.org/10.1016/0550-3213(91)90259-Z}{\emph{Nucl. Phys.} {\bf
  B350} (1991) 201--227}.

\bibitem{GaLe:84}
J.~Gasser and H.~Leutwyler, \emph{{Chiral perturbation theory to one loop}},
  \href{http://dx.doi.org/10.1016/0003-4916(84)90242-2}{\emph{Ann. Phys.} {\bf
  158} (1984) 142}.

\bibitem{BiCoTa:98}
J.~Bijnens, G.~Colangelo and P.~Talavera, \emph{{The vector and scalar form
  factors of the pion to two loops}}, {\emph{JHEP} {\bf 05} (1998) 014},
  [\href{http://arxiv.org/abs/hep-ph/9805389}{{\tt hep-ph/9805389}}].

\bibitem{Ward:50}
J.~C. Ward, \emph{{An identity in quantum electrodynamics}},
  \href{http://dx.doi.org/10.1103/PhysRev.78.182}{\emph{Phys. Rev.} {\bf 78}
  (1950) 182}.

\bibitem{JLQCD:09}
{\scshape JLQCD} collaboration, S.~Aoki et~al., \emph{{Pion form factors from
  two-flavor lattice QCD with exact chiral symmetry}},
  \href{http://dx.doi.org/10.1103/PhysRevD.80.034508}{\emph{Phys. Rev.} {\bf
  D80} (2009) 034508}, [\href{http://arxiv.org/abs/0905.2465}{{\tt
  0905.2465}}].

\bibitem{BuJiTi:06}
T.~B. Bunton, F.-J. Jiang and B.~C. Tiburzi, \emph{{Extrapolations of lattice
  meson form factors}},
  \href{http://dx.doi.org/10.1103/PhysRevD.74.034514}{\emph{Phys. Rev.} {\bf
  D74} (2006) 034514}, [\href{http://arxiv.org/abs/hep-lat/0607001}{{\tt
  hep-lat/0607001}}].

\bibitem{HuJiTi:07}
J.~Hu, F.-J. Jiang and B.~C. Tiburzi, \emph{{Current renormalization in finite
  volume}}, \href{http://dx.doi.org/10.1016/j.physletb.2007.07.060}{\emph{Phys.
  Lett.} {\bf B653} (2007) 350--357},
  [\href{http://arxiv.org/abs/0706.3408}{{\tt 0706.3408}}].

\bibitem{Lu:83}
M.~L{\"u}scher, \emph{{On a relation between finite-size effects and elastic
  scattering processes}},  {1983}.

\bibitem{CoDuSo:02}
G.~Colangelo, S.~D{\"u}rr and R.~Sommer, \emph{{Finite-size effects on $M_\pi$
  in QCD from chiral perturbation theory}},
  \href{http://dx.doi.org/10.1016/S0920-5632(03)80450-4}{\emph{Nucl. Phys.
  Proc. Suppl.} {\bf 119} (2003) 254--256},
  [\href{http://arxiv.org/abs/hep-lat/0209110}{{\tt hep-lat/0209110}}].

\bibitem{CoWenWu:10}
G.~Colangelo, U.~Wenger and J.~M. Wu, \emph{{Twisted mass finite-volume
  effects}}, \href{http://dx.doi.org/10.1103/PhysRevD.82.034502}{\emph{Phys.
  Rev.} {\bf D82} (2010) 034502}, [\href{http://arxiv.org/abs/1003.0847}{{\tt
  1003.0847}}].

\bibitem{Ali:03}
{\scshape QCDSF-UKQCD} collaboration, A.~Ali~Khan et~al., \emph{{The nucleon
  mass in $N_{f}=2$ lattice QCD: finite size effects from chiral perturbation
  theory}},
  \href{http://dx.doi.org/10.1016/j.nuclphysb.2004.04.018}{\emph{Nucl. Phys.}
  {\bf B689} (2004) 175--194},
  [\href{http://arxiv.org/abs/hep-lat/0312030}{{\tt hep-lat/0312030}}].

\bibitem{Koma:04}
Y.~Koma and M.~Koma, \emph{{On the finite size mass shift formula for stable
  particles}},
  \href{http://dx.doi.org/10.1016/j.nuclphysb.2005.01.053}{\emph{Nucl. Phys.}
  {\bf B713} (2005) 575--597},
  [\href{http://arxiv.org/abs/hep-lat/0406034}{{\tt hep-lat/0406034}}].

\bibitem{Koma:05a}
Y.~Koma and M.~Koma, \emph{{Finite size mass shift formula for stable particles
  revisited}},
  \href{http://dx.doi.org/10.1016/j.nuclphysbps.2004.11.285}{\emph{Nucl. Phys.
  Proc. Suppl.} {\bf 140} (2005) 329--331},
  [\href{http://arxiv.org/abs/hep-lat/0409002}{{\tt hep-lat/0409002}}].

\bibitem{Koma:05b}
Y.~Koma and M.~Koma, \emph{{More on the finite size mass shift formula for
  stable particles}},  \href{http://arxiv.org/abs/hep-lat/0504009}{{\tt
  hep-lat/0504009}}.

\bibitem{CoFuHae:05}
G.~Colangelo, A.~Fuhrer and C.~H{\"a}feli, \emph{{The pion and proton mass in
  finite volume}},
  \href{http://dx.doi.org/10.1016/j.nuclphysbps.2006.01.004}{\emph{Nucl. Phys.
  Proc. Suppl.} {\bf 153} (2006) 41--48},
  [\href{http://arxiv.org/abs/hep-lat/0512002}{{\tt hep-lat/0512002}}].

\bibitem{CoFiUr:96}
G.~Colangelo, M.~Finkemeier and R.~Urech, \emph{{Tau decays and chiral
  perturbation theory}},
  \href{http://dx.doi.org/10.1103/PhysRevD.54.4403}{\emph{Phys. Rev.} {\bf D54}
  (1996) 4403--4418}, [\href{http://arxiv.org/abs/hep-ph/9604279}{{\tt
  hep-ph/9604279}}].

\bibitem{Va:15}
A.~Vaghi, \emph{{Finite-volume effects in chiral perturbation theory with
  twisted boundary conditions}}.
\newblock PhD thesis, {e-Dissertation (edbe) Universit{\"a}t Bern}, 2015.
\newblock \href{http://boris.unibe.ch/80733/}{10.7892/boris.80733}.

\bibitem{BiCoEcGaSa:97}
J.~Bijnens, G.~Colangelo, G.~Ecker, J.~Gasser and M.~Sainio, \emph{{Pion-pion
  scattering at low energy}},
  \href{http://dx.doi.org/10.1016/S0550-3213(97)00621-4,
  10.1016/S0550-3213(97)00621-4}{\emph{Nucl. Phys.} {\bf B508} (1997)
  263--310}, [\href{http://arxiv.org/abs/hep-ph/9707291}{{\tt
  hep-ph/9707291}}].

\bibitem{BeKaiMei:90}
V.~Bernard, N.~Kaiser and U.~G. Meissner, \emph{{$\pi K$-scattering in chiral
  perturbation theory to one loop}},
  \href{http://dx.doi.org/10.1016/0550-3213(91)90461-6}{\emph{Nucl. Phys.} {\bf
  B357} (1991) 129--152}.

\bibitem{BiCoGa:94}
J.~Bijnens, G.~Colangelo and J.~Gasser, \emph{{$K_{\ell4}$-decays beyond one
  loop}}, \href{http://dx.doi.org/10.1016/0550-3213(94)90634-3}{\emph{Nucl.
  Phys.} {\bf B427} (1994) 427--454},
  [\href{http://arxiv.org/abs/hep-ph/9403390}{{\tt hep-ph/9403390}}].

\bibitem{CoVa:16b}
G.~Colangelo and A.~Vaghi, \emph{{Work in progress}},  2016.

\bibitem{BeKaiMei:91}
V.~Bernard, N.~Kaiser and U.~G. Meissner, \emph{{$\pi\eta$-scattering in QCD}},
  \href{http://dx.doi.org/10.1103/PhysRevD.44.3698}{\emph{Phys. Rev.} {\bf D44}
  (1991) 3698--3701}.

\bibitem{CoGaLe:01}
G.~Colangelo, J.~Gasser and H.~Leutwyler, \emph{{$\pi\pi$-scattering}},
  \href{http://dx.doi.org/10.1016/S0550-3213(01)00147-X}{\emph{Nucl. Phys.}
  {\bf B603} (2001) 125--179}, [\href{http://arxiv.org/abs/hep-ph/0103088}{{\tt
  hep-ph/0103088}}].

\bibitem{FLAG:14}
S.~Aoki, Y.~Aoki, C.~Bernard, T.~Blum, G.~Colangelo et~al., \emph{{Review of
  lattice results concerning low-energy particle physics}},
  \href{http://dx.doi.org/10.1140/epjc/s10052-014-2890-7}{\emph{Eur. Phys. J.}
  {\bf C74} (2014) 2890}, [\href{http://arxiv.org/abs/1310.8555}{{\tt
  1310.8555}}].

\bibitem{BiJe:11}
J.~Bijnens and I.~Jemos, \emph{{A new global fit of the $L^{\textup{r}}_i$ at
  next-to-next-to-leading order in chiral perturbation theory}},
  \href{http://dx.doi.org/10.1016/j.nuclphysb.2011.09.013}{\emph{Nucl. Phys.}
  {\bf B854} (2012) 631--665}, [\href{http://arxiv.org/abs/1103.5945}{{\tt
  1103.5945}}].

\bibitem{MILC:09A}
{\scshape MILC} collaboration, A.~Bazavov et~al., \emph{{MILC results for light
  pseudoscalars}}, {\emph{PoS} {\bf CD09} (2009) 007},
  [\href{http://arxiv.org/abs/0910.2966}{{\tt 0910.2966}}].

\bibitem{Bor:12}
S.~Borsanyi, S.~Durr, Z.~Fodor, S.~Krieg, A.~Schafer et~al.,
  \emph{{$\mathrm{SU}(2)$ chiral perturbation theory low-energy constants from
  $(2+1)$-flavor staggered lattice simulations}},
  \href{http://dx.doi.org/10.1103/PhysRevD.88.014513}{\emph{Phys. Rev.} {\bf
  D88} (2013) 014513}, [\href{http://arxiv.org/abs/1205.0788}{{\tt
  1205.0788}}].

\bibitem{RBCUKQCD:12}
{\scshape RBC, UKQCD} collaboration, R.~Arthur et~al., \emph{{Domain wall QCD
  with near-physical pions}},
  \href{http://dx.doi.org/10.1103/PhysRevD.87.094514}{\emph{Phys. Rev.} {\bf
  D87} (2013) 094514}, [\href{http://arxiv.org/abs/1208.4412}{{\tt
  1208.4412}}].

\bibitem{NPLQCD:11}
S.~Beane, W.~Detmold, P.~Junnarkar, T.~Luu, K.~Orginos et~al.,
  \emph{{$\mathrm{SU}(2)$ low-energy constants from mixed-action lattice QCD}},
  \href{http://dx.doi.org/10.1103/PhysRevD.86.094509}{\emph{Phys. Rev.} {\bf
  D86} (2012) 094509}, [\href{http://arxiv.org/abs/1108.1380}{{\tt
  1108.1380}}].

\bibitem{MILC:10A}
A.~Bazavov, C.~Bernard, C.~DeTar, X.~Du, W.~Freeman et~al., \emph{{Staggered
  chiral perturbation theory in the two-flavor case and $\mathrm{SU}(2)$
  analysis of the MILC data}}, {\emph{PoS} {\bf LATTICE2010} (2010) 083},
  [\href{http://arxiv.org/abs/1011.1792}{{\tt 1011.1792}}].

\bibitem{PDG:14}
{\scshape Particle Data Group} collaboration, K.~Olive et~al., \emph{{Review of
  Particle Physics}},
  \href{http://dx.doi.org/10.1088/1674-1137/38/9/090001}{\emph{Chin. Phys.}
  {\bf C38} (2014) 090001}.

\bibitem{MILC:10}
{\scshape MILC} collaboration, A.~Bazavov et~al., \emph{{Results for light
  pseudoscalar mesons}}, {\emph{PoS} {\bf LATTICE2010} (2010) 074},
  [\href{http://arxiv.org/abs/1012.0868}{{\tt 1012.0868}}].

\bibitem{HPQCDUKQCD:07}
{\scshape HPQCD, UKQCD} collaboration, E.~Follana, C.~Davies, G.~Lepage and
  J.~Shigemitsu, \emph{{High precision determination of the $\pi$, $K$, $D$ and
  $D_{s}$ decay constants from lattice QCD}},
  \href{http://dx.doi.org/10.1103/PhysRevLett.100.062002}{\emph{Phys. Rev.
  Lett.} {\bf 100} (2008) 062002}, [\href{http://arxiv.org/abs/0706.1726}{{\tt
  0706.1726}}].

\bibitem{Hall:12}
J.~Hall, D.~Leinweber, B.~Owen and R.~Young, \emph{{Finite-volume corrections
  to charge radii}},
  \href{http://dx.doi.org/10.1016/j.physletb.2013.06.048}{\emph{Phys. Lett.}
  {\bf B725} (2013) 101--105}, [\href{http://arxiv.org/abs/1210.6124}{{\tt
  1210.6124}}].

\end{thebibliography}\endgroup

\end{document}